\documentclass[11pt]{article}

\usepackage[usenames,dvipsnames,table]{xcolor}
\usepackage[utf8]{inputenc}

\pdfoutput=1 
\usepackage{jheppub} 

\usepackage{graphicx}
\usepackage{amsmath, amssymb}
\usepackage{comment}

\usepackage{mathrsfs}

\usepackage{tikz}
\usetikzlibrary{arrows.meta,calc,positioning,decorations.pathmorphing,decorations.markings}

\tikzset{
 tube gauge/.style={circle,draw,line width=.75pt,minimum size=9mm,inner sep=0pt,outer sep=1pt,font=\normalsize},
 tube flavor/.style={rectangle,draw,line width=.75pt,minimum size=9mm,inner sep=0pt,outer sep=1pt,font=\normalsize},
 tube label/.style={inner sep=1pt,font=\normalsize},
 tube group/.style={inner sep=0pt,font=\small},
 tube edge/.style={line width=.75pt},
 tube cross/.style={fill=none,inner sep=0pt,outer sep=0pt,font=\normalsize}
}
\newcommand{\TubeLink}[4]{%
 \draw (#1)--node[tube label,above=5pt] {$#3$}
 node[tube cross,label={[tube label]below=5pt:$#4$}] {$\times$}(#2);}
\newcommand{\TubeAdjoint}[2]{%
 \draw (#1.55) .. controls +(60:1.05cm) and +(120:1.05cm) ..
 node[tube label,above=4pt] {$#2$}(#1.125);}
 
\usepackage{framed}
\definecolor{shadecolor}{rgb}{0.90,0.90,0.90}

\usepackage{mathtools}

\def\be{\begin{eqnarray}}
	\def\ee{\end{eqnarray}}
\def\bea{\begin{eqnarray}}
	\def\eea{\end{eqnarray}}

\newcommand{\Tr}{\mbox{Tr}}

\newcommand{\beal}{\begin{equation}
		\begin{aligned}}
		\newcommand{\eeal}{\end{aligned}
\end{equation}}
\newcommand{\bem}{\begin{multline}}
	\newcommand{\eem}{\end{multline}}

\def\SU{\text{SU}}

\def\E{\text{E}}
\def\g{{\frak g}}

\def\cS{{\cal S}}

\def\CN{\mathcal{N}}
\def\CF{\mathcal{F}}
\def\CJ{\mathcal{J}}

\def\su{{\frak{su}}}
\def\u1{{\frak{u}}_1}

\makeatletter
\gdef\@fpheader{\break}
\makeatother

\begin{document}

    \title{Irregularization of punctures:\\
    From $T_N$ to Argyres--Douglas via $\mathcal{N}=1$ flow}
    
	\medskip
	
	\author[1]{Ki-Hong Lee,\!}
    \author[2]{ Shlomo~S.~Razamat,\!}
    \author[3]{Jaewon Song,\!}
    \author[2]{and Thomas Waddleton}

	\medskip
	\affiliation[1]{Department of Physics and Haifa Center for Physics and Astrophysics, University of Haifa
    \\ Haifa 3498838, Israel}
	\affiliation[2]{Department of Physics, Technion, Haifa, 32000, Israel}
    \affiliation[3]{Department of Physics, Korea Advanced Institute of Science and Technology\\291 Daehak-ro, Yuseong-gu, Daejeon 34141, Republic of Korea}
    
    \emailAdd{khlee11812@gmail.com}
    \emailAdd{razamat@technion.ac.il}
    \emailAdd{jaewon.song@kaist.ac.kr}
\emailAdd{thomas.w@campus.technion.ac.il}


	\abstract{In this paper, we directly relate constructions of two large classes of $4d$  strongly coupled SCFTs: 1) compactifications of type $\mathfrak{g}$ $6d$ $(2,0)$ theories on spheres with irregular punctures leading to 
    the $D_p(\mathfrak{g})$ Argyres--Douglas theories, and 2) the compactifications of the same $6d$ $(2,0)$ theories on spheres with regular punctures. In particular, we show that the ${\cal N}=2$ $D_p(\mathfrak{g})$ Argyres--Douglas theories can be obtained starting from ${\cal N}=1$ compactifications of 6d $(2,0)$ theories of type $\mathfrak{g}$ on a sphere with two maximal punctures with flux linearly dependent on $p$ by gauging one of the maximal puncture symmetries, possibly with additional matter sectors. Importantly, part of the ${\cal N}=2$ R-symmetry of $D_p(\g)$ theories arises from the R-symmetry of the two-punctured sphere and the $\frak{u}_1$ geometric isometry symmetry. This construction gives in particular a geometric explanation of some of the ${\cal N}=1$ constructions of Argyres--Douglas theories, and also provides explicit new Lagrangian constructions (involving gauging of emergent symmetries in general) for $D_p({\mathfrak{su}_3})$ theories. More generally, we relate the problem of finding a Lagrangian for $D_p(\mathfrak{g})$ Argyres--Douglas theories (and thereby a large class of theories that are connected by Higgs branch RG flow) to the problem of finding Lagrangians for the $T_{\mathfrak{g}}$ theories. 
    }

	\maketitle

\section{Introduction}

The framework of supersymmetric quantum field theories with at least four supercharges in various dimensions is a very fertile ground for discovering and directly addressing strongly coupled physics. Utilizing the power of holomorphy~\cite{Seiberg:1994bp}, a web of interconnected, consistent, and well-motivated statements about the strong-coupling dynamics in these set-ups can be conjectured and tested. In particular, increasing the number of supercharges to eight allows for a complete IR solution of certain asymptotically free field theories~\cite{Seiberg:1994rs,Seiberg:1994aj,Nekrasov:2002qd}.

Early on, studying such solutions, and their avatars in the larger structure of string theory, led to the discovery of novel strongly coupled SCFTs. One such example is that of the 
Argyres--Douglas theories~\cite{Argyres:1995jj,Argyres:1995xn,Eguchi:1996vu}. A defining property of these theories is that they are
four-dimensional ${\cal N}=2$ superconformal field theories with Coulomb branch
operators of fractional dimensions.
Just by listing~\cite{Bhardwaj:2013qia} all possible manifestly ${\cal N}=2$ symmetric Lagrangians, one can show that no conventional ${\cal N}=2$
Lagrangian can describe them. In some sense, Argyres--Douglas theories are the simplest examples of intrinsically strongly
coupled ${\cal N}=2$ dynamics.  A large family of them, the $D_p(\g)$
theories~\cite{Cecotti:2012jx,Cecotti:2013lda}, labeled by a simply-laced Lie
algebra $\g$ and an integer $p\geq2$, has flavor symmetry at least $\g$ and central charges
that grow linearly with $p$.\footnote{There are many ways to label different Argyres--Douglas theories~\cite{Cecotti:2010fi}. For example, the $H_1 = (A_1, A_3)$ and $H_2 = (A_1, D_4)$ models, which were obtained on the Coulomb branches of $\SU(2)$ gauge theories in~\cite{Argyres:1995xn}, are $D_3(\su_2)$ and $D_4(\su_2)\cong D_2(\su_3)$, respectively. The original $H_0$ theory of~\cite{Argyres:1995jj} can be obtained via a Higgs branch flow from $D_5(\su(2))$~\cite{Beem:2019tfp}.}  These theories have a uniform
six-dimensional origin~\cite{Gaiotto:2009ma,Xie:2012hs,Wang:2015mra,Nanopoulos:2009uw}: they are
compactifications of the $6d$ $(2,0)$ theory of type $\g$ on a sphere with one
regular maximal puncture and one \emph{irregular} puncture.

Yet another class of early examples of interesting strongly coupled theories included ${\cal N}=2$ SCFTs with Coulomb branch operators of integer dimensions but with global symmetries which cannot be realized by manifestly ${\cal N}=2$ symmetric Lagrangians~\cite{Minahan:1996fg,Minahan:1996cj}.
In the framework of compactifications of $(2,0)$ six-dimensional SCFTs to four dimensions, these models are examples of compactifications on Riemann surfaces with \emph{regular} punctures~\cite{Gaiotto:2009we,Gaiotto:2009hg}.
A basic class of such theories is usually denoted by $T_\g$ and is obtained by compactifying a type $\g$ $(2,0)$ theory on a sphere with three \emph{regular} maximal punctures.

Although we do not have Lagrangian quantum field theories explicitly manifesting all the symmetries of the $D_p(\g)$ and $T_\g$ SCFTs (and their relatives), over time Lagrangian descriptions in the same IR or conformal universality class, manifesting only part of the (super)symmetry, were found for many of them.\footnote{By \emph{conformal universality class} we mean theories residing on the same conformal manifold, or flowing to the same conformal manifold.} For example, we have ${\cal N}=1$ Lagrangian constructions of $T_{{\frak{a}}_2}$ (\emph{aka} $T_3$, or the $\E_6$ MN theory)~\cite{Zafrir:2019hps,GarciaEtxebarria:2021deconf}, $T_{{\frak{a}}_3}$~\cite{Razamat:2019vfd}, and the $D_p({\frak{a}}_1)$ theories~\cite{Maruyoshi:2016tqk,Maruyoshi:2016aim,Agarwal:2016pjo,Agarwal:2017roi,Benvenuti:2017bpg,ChoSong:2026,Maruyoshi:2018nod}. There are more, rather sporadic, examples of Lagrangians
for some of the ${\cal N}=2$ SCFTs obtained by compactifications of $(2,0)$ theories~\cite{Razamat:2019vfd,GarciaEtxebarria:2021deconf,Razamat:2022gpm}.\footnote{In fact, one does not have to restrict to $(2,0)$ theories. We can start with any $(1,0)$ SCFT in six dimensions and consider its ${\cal N}=1$ compactifications to four dimensions. For many such compactifications, Lagrangian constructions directly in $4d$ are known; see {\it e.g.}~\cite{Razamat:2022gpm,Gaiotto:2015usa,Kim:2017toz,Kim:2018lfo,Razamat:2019vfd,Razamat:2019ukg,Razamat:2020bix,Kim:2023qbx}.} In addition to vanilla Lagrangian constructions, it is often useful to discuss sequences of flows that start from a vanilla Lagrangian but also involve gauging symmetries which are \emph{emergent} in the IR (or at strong coupling on the conformal manifold); see {\it e.g.}~\cite{Gadde:2015xta,Agarwal:2018ejn, Pasquetti:2019hxf, Hwang:2021xyw}.

A more general set of compactifications of the $(2,0)$ theories on a Riemann surface preserves only ${\cal N}=1$ supersymmetry~\cite{Benini:2009mz, Bah:2012dg,Bah:2011je,Bah:2011vv}. One can view the $(2,0)$ theory as a $(1,0)$ theory with an ${\frak{su}}_2$ global symmetry. Then turning on a flux $z$ for the Cartan generator of this global symmetry, supported on the Riemann surface, results in ${\cal N}=1$ supersymmetric theories in $4d$.
The ${\cal N}=2$ compactifications correspond to a value of $z$ related to the Euler characteristic of the surface. It is common to refer to theories obtained by compactifications of $(2,0)$ theories as theories of class ${\cal S}$.

\begin{figure}[t]
\centering
\begin{tikzpicture}[>=Stealth,scale=0.8]
\draw[very thick,fill=blue!5] (0,0) circle (1.35);
\fill (-1.35,0) circle (2.3pt);
\node[font=\small,right=3pt] at (-1.35,0) {maximal};
\draw[very thick,red!70!black,fill=red!15,decorate,decoration={zigzag,amplitude=1.5pt,segment length=4pt}] (1.35,0) circle (0.22);
\node[red!70!black,right,font=\small] at (1.65,0) {irregular};
\draw[thin,dashed,black!45] (-2.75,0) -- (-1.45,0);
\draw[->,thick,black!70] (-2.296,-0.583) arc (-110:200:0.28 and 0.62);
\node at (0,-1.85) {$D_p(\g)$};
\draw[<->,very thick] (4.45,0) -- (5.45,0);
\node[below=3pt] at (4.95,0) {$p=h_\g z+1$};
\begin{scope}[xshift=8.4cm]
\fill[green!8] (0,0.8) rectangle (4,-0.8);
\draw[very thick] (0,0.8) -- (4,0.8); \draw[very thick] (0,-0.8) -- (4,-0.8);
\draw[very thick,fill=blue!20] (0,0) ellipse (0.2 and 0.8);
\draw[very thick,fill=red!15] (4,0) ellipse (0.2 and 0.8);
\node[left,font=\small] at (-0.25,0) {$\g$};
\node[red!70!black,right,font=\small,align=left] at (4.3,0) {${\cal N}=2$\\vector};
\draw[thin,dashed,black!45] (-1.8,0) -- (-0.75,0);
\draw[->,thick,black!70] (-1.396,-0.583) arc (-110:200:0.28 and 0.62);
\foreach \x in {0.8,2,3.2}{\draw[->,thick,orange!85!black] (\x,0.8) -- (\x,1.3); \draw[->,thick,orange!85!black] (\x,-0.8) -- (\x,-1.3);}
\node[orange!85!black] at (3.55,1.12) {$z$};
\node at (2,-1.85) {${\cal T}^{(\g)}_z$};
\end{scope}
\end{tikzpicture}
\caption{Left: the $D_p(\g)$ theory as the compactification of the $6d$ $(2,0)$ theory of type $\g$ on a sphere with one maximal regular puncture and one irregular puncture. Right: the ${\cal N}=1$ theory ${\cal T}^{(\g)}_z$, the compactification on a sphere with two maximal punctures and flux $z$, with one of the puncture symmetries $\g$ gauged by an ${\cal N}=2$ vector multiplet. The circular arrows denote the $\u1$ isometries of the two surfaces, which rotate them around the axis through the punctures. The arrows pointing out of the tube represent the flux $z$.}
\label{fig:intro}
\end{figure}
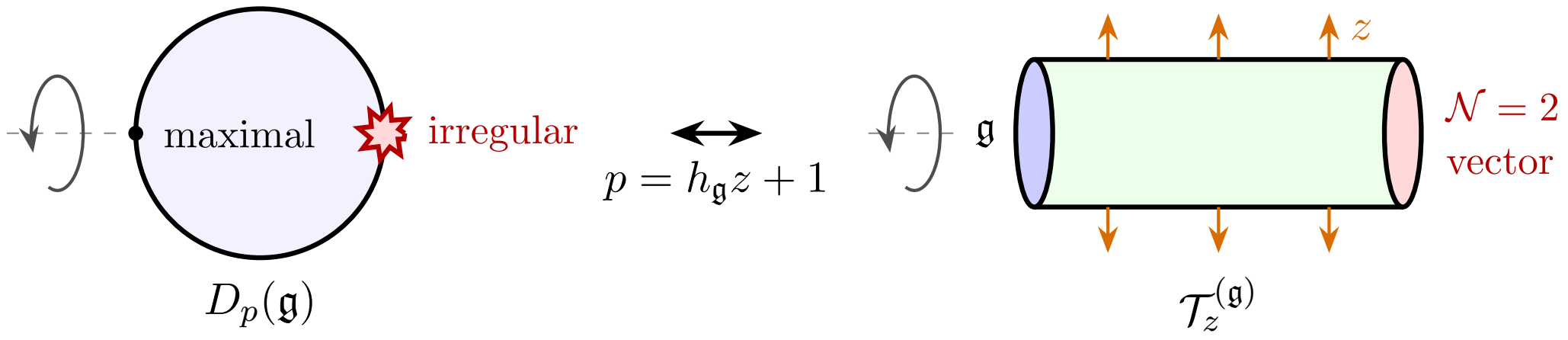

In this paper, we will relate the three ingredients above: we will argue that, starting from the $T_\g$ ${\cal N}=2$ theories, first building from them theories in ${\cal N}=1$ class ${\cal S}$, and then gauging some of their global symmetries, one can obtain ${\cal N}=2$ $D_p(\g)$ theories. In particular, we will claim that
\begin{shaded}
\begin{equation}
\mathcal{T}^{(\g)}_z\ /\,(\g\,{\oplus}\ V_{{\cal N}=2})\ \longrightarrow\
D_{h_\g z+1}(\g)\ \oplus\ \big(r_\g\,z\ \text{free chiral fields}\big)\,.
\label{eq:mainclaim}
\end{equation}
\end{shaded}\noindent  Here $h_\g$ and $r_\g$ are the Coxeter number and the rank of the Lie algebra $\g$ defining the $6d$ $(2,0)$ theory. The theory $\mathcal{T}^{(\g)}_z$ is the
${\cal N}=1$ class ${\cal S}$ model obtained by compactifying the type $\g$ $(2,0)$ theory on a sphere with two maximal punctures and flux $z$. Finally, $/\,(\g\,{\oplus}\ V_{{\cal N}=2})$ denotes gauging a maximal puncture symmetry $\g$ with an ${\cal N}=2$ vector multiplet. See Figure \ref{fig:intro} for an illustration. We will also discuss generalizations of this statement to other values of $p$. The gauging of the maximal puncture of the two-punctured sphere turns the maximal \emph{regular} puncture into an \emph{irregular} one. The type of the irregular puncture is determined by the value of the flux and the details of the gauging. 

An important ingredient of the construction which makes it work is the fact that the $\mathcal{T}^{(\g)}_z$ theory has a $\u1$ symmetry which neither comes from the global symmetry of the $6d$ SCFT nor is associated with the punctures: it is rather the geometric $\u1$ isometry of the two-punctured sphere.\footnote{In the description of the tube as a flux domain wall in the $5d$ Kaluza--Klein theory obtained by reducing the $6d$ SCFT on the circle of the tube, this symmetry descends from the $\u1$ Kaluza--Klein symmetry~\cite{Sabag:2022hyw} (see also~\cite{Braun:2026wao}); on a torus with flux it is broken to a discrete subgroup, and it was identified with the isometry of the surface in~\cite{Hwang:2021xyw}.} When gauging the puncture symmetry, the Cartan generator of the $6d$ $(1,0)$ $\su_2$ global symmetry and the isometry $\u1$ symmetry are anomalous, but a certain combination of these survives. This combination depends on the flux $z$ and on the details of the gauging. In particular, together with the non-anomalous ${\cal N}=1$ R-symmetry of the construction, it enhances to the $\su_2\times \u1$ ${\cal N}=2$ R-symmetry of the $D_p(\g)$ theory.

The derivation will proceed as follows. First, we will interpret the recent construction of $D_p(\su_2)$ theories discussed in~\cite{ChoSong:2026} as a geometric construction of \eqref{eq:mainclaim} in Section~\ref{sec:a1}. The crucial observation here is that the $\mathcal{T}^{(\g)}_z$ theories are already known for $\g=\su_2$ and are given by a simple linear ${\cal N}=1$ Lagrangian quiver of $\su_2$ gauge groups. Then, in Section~\ref{sec:a2}, we will explicitly generalize this claim to $D_p(\su_3)$. 
For $\g=\su_3$, we will first derive a Lagrangian for the ${{\cal T}^{(\su_3)}_{z=1}}$ theory, using one of the known Lagrangians for $T_{\su_3}$. 
The ${{\cal T}^{(\su_3)}_{z=1}}$ theory has an $\su_3\times \su_3$ global symmetry associated with the punctures. However, the Lagrangian that we will find manifests only the $\su_3\times (\su_2\times {\frak{u}_1})$ subgroup in the UV. We will argue that the full symmetry emerges in the IR. We will then use the Lagrangian and the statement \eqref{eq:mainclaim} to derive first a Lagrangian for $D_{3\times 1+1}(\su_3)$ in Section~\ref{sec:a2:lagp4}, and then generalize the construction to higher values of $p$. 
This generalization will involve the gauging of emergent symmetries. Once we have explicitly established the validity of \eqref{eq:mainclaim} in two concrete examples, we will discuss the generalization to any simply-laced $\g$ in Section~\ref{sec:g}.
We will give some evidence for this generalization by showing that it is consistent with certain anomaly computations.
In particular, we will perform in Section~\ref{sec:g} a field theoretic computation of expected anomalies of $4d$ $\mathcal{T}^{(\g)}_z$ starting from a type $\g$ $(2,0)$
 $6d$ SCFT. This will leave some of the anomaly coefficients undetermined.
 Even with the undetermined coefficients, the matching of the rest of the anomalies is a strong check of our proposal. We will show that the full consistency of our main conjecture \eqref{eq:mainclaim} sets the undetermined coefficients to particular values. We will also perform a holographic computation of the anomalies in the case of $\g=\su_N$ in Appendix~\ref{sec:holography}, which will independently set the ambiguous coefficients precisely to the values needed for \eqref{eq:mainclaim} to be consistent.
 Finally, in Section~\ref{sec:discussion}, we will further discuss the rationale behind our results and comment on additional checks and generalizations.

\section{Construction of \texorpdfstring{$D_p(\mathfrak{su}_2)$ theories from $T_2$}{Dp(su2) theories from T2}}\label{sec:a1}

In this section, we recast the construction of the $\mathcal{N}=1$ Lagrangians for the $D_p(\mathfrak{su}_2) = (A_1, D_p)$ theories of \cite{ChoSong:2026}. We note that this construction has a recurring building block, identical to the $\mathcal{N}=1$ class $\mathcal{S}$ theory corresponding to a tube (a two-punctured sphere) with non-trivial flux \cite{Agarwal:2015vla, Fazzi:2016eec}.  

\subsection{The tube of flux \texorpdfstring{$z$}{z}}

The tube theory can be obtained from the trinion theory $T_2$, which is just a theory of 8 free half-hypermultiplets. This is the class $\mathcal{S}$ theory of type $\mathfrak{a}_1$ on a three-punctured sphere, with manifest $\mathfrak{su}_2^3$ flavor symmetry. To obtain a tube, i.e.\ a two-punctured sphere with flux, we have to close a ``color-flipped'' puncture. In the $\mathcal{N}=1$ class $\mathcal{S}$ setup, one can assign $\mathbb{Z}_2$-valued ``colors'' to each pair of pants (three-punctured sphere) and to each puncture \cite{Bah:2012dg, Gadde:2013fma, Bah:2013aha, Agarwal:2013uga, Agarwal:2014rua, Agarwal:2015vla}. This represents the choice of two transverse directions (normal bundles). Flipping the color of a puncture (with respect to the color of the pair of pants) introduces a flip field $M$ for the moment map $\mu$, with the superpotential coupling $W=M\mu$. 

To close a puncture, we give a nilpotent vacuum expectation value $\rho(\sigma^+)$ to the moment map $\mu$ of an unflipped puncture, or to the flip field $M$ of a flipped puncture. Nilpotent vevs are classified by embeddings $\rho: \mathfrak{su}_2 \to \g$, and $\sigma^+ = (\sigma^1+i\sigma^2)/2$ is the usual raising operator written in terms of Pauli matrices. The commutant of $\rho(\sigma^+)$ in $\g$ gives the flavor symmetry associated with the puncture. For $\g = \mathfrak{a}_1$, there is only one type of puncture. The principal embedding corresponds to closing the puncture completely, breaking its flavor symmetry entirely. For an unflipped puncture, this simply amounts to removing the puncture. On the other hand, closing a flipped puncture affects the normal bundle. The degrees $(p, q)$ of the normal bundles satisfy the Euler number constraint $p+q=2g-2+n$, where $g$ and $n$ are the genus and the number of punctures, respectively. Closing a puncture decreases the total degree by one; depending on the color of the puncture, it decreases either $p$ or $q$. In our case, we start with $(p, q)=(1, 0)$ and close the color-flipped puncture to obtain $(p, q)=(1, -1)$. We denote the resulting theory by $\mathcal{T}_{z=1}^{(\g)}$ and call it the tube theory. 
When $\g = \mathfrak{a}_1 = \mathfrak{su}_2$, this procedure simply gives a theory of an $\su_2$ bifundamental $Q_{aa'}$ and a flip field $M$, with the coupling
\begin{align}
    W \sim M \epsilon^{ab}\epsilon^{a'b'} Q_{a a'} Q_{b b'} \sim M \det \, Q \ . 
\end{align}
The field content and the charges are given in Table~\ref{tab:Tsu2}, and the quiver is depicted in Figure~\ref{fig:su2-single-tube}. 
For general $\g$, this procedure involves closing a puncture of the non-Lagrangian $T_\g$ theory. 

\begin{table}[h] 
    \centering
    \renewcommand{\arraystretch}{1.25}
    \begin{tabular}{|c|cc|cc|}
        \hline
        field & $(\su_2)_L$ & $(\su_2)_R$ & $R_0$ & $\mathcal{F}$  \\
        \hline
        $Q$ & $\mathbf{2}$ & $\mathbf{2}$ & $0$ & $\phantom{-}1$ \\
        $M$ & $\mathbf{1}$ & $\mathbf{1}$ & $2$ & $-2$ \\
        \hline
    \end{tabular}
    \caption{The matter content of $\mathcal{T}_1^{(\su_2)}$.}    
    \label{tab:Tsu2}
\end{table}
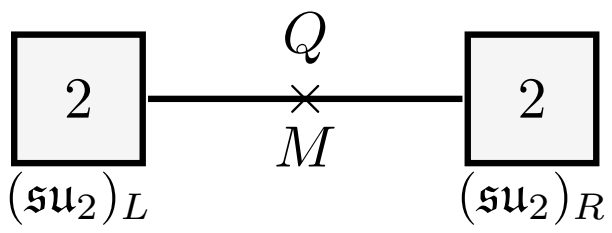
\begin{figure}
\begin{center}
\begin{tikzpicture}[>=Latex,tube edge,very thick,tube gauge/.append style={line width=1.2pt,minimum size=10mm,fill=blue!6},tube flavor/.append style={line width=1.2pt,minimum size=8.5mm,fill=black!4}]
\node[tube flavor,label={[tube group]below=10pt:$(\su_2)_L$}] (L) at (-1.5,0) {$2$};
\node[tube flavor,label={[tube group]below=10pt:$(\su_2)_R$}] (R) at (1.5,0) {$2$};
\TubeLink{L}{R}{Q}{M}
\end{tikzpicture}
\caption{One $\su_2$ tube. Both squares are flavor punctures;
the cross denotes the singlet coupling $M\,\det Q$. There is no gauge node.}
\label{fig:su2-single-tube}
\end{center}
\end{figure}

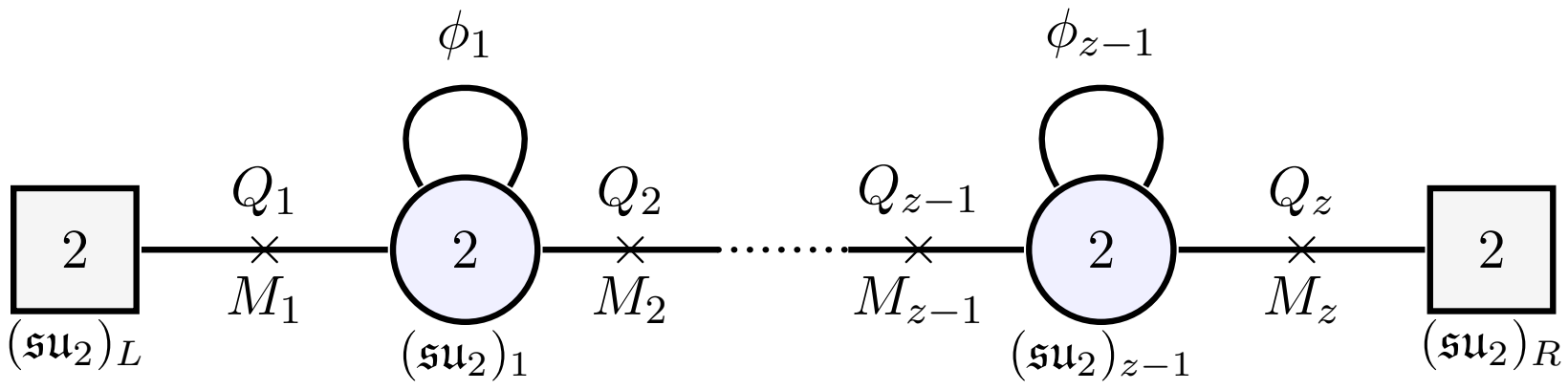
\begin{figure}
\begin{center}
\begin{tikzpicture}[>=Latex,tube edge,very thick,tube gauge/.append style={line width=1.2pt,minimum size=10mm,fill=blue!6},tube flavor/.append style={line width=1.2pt,minimum size=8.5mm,fill=black!4}]
\node[tube flavor,label={[tube group]below=10pt:$(\su_2)_L$}] (L) at (-4.9,0) {$2$};
\node[tube gauge,label={[tube group]below=10pt:$(\su_2)_1$}] (G1) at (-2.2,0) {$2$};
\node[tube gauge,label={[tube group]below=10pt:$(\su_2)_{z-1}$}] (Gn) at (2.2,0) {$2$};
\node[tube flavor,label={[tube group]below=10pt:$(\su_2)_R$}] (R) at (4.9,0) {$2$};
\coordinate (omitL) at (-.45,0);
\coordinate (omitR) at (.45,0);
\TubeLink{L}{G1}{Q_1}{M_1}
\TubeLink{G1}{omitL}{Q_2}{M_2}
\draw[dash pattern=on 0pt off 3pt,line cap=round] (omitL)--(omitR);
\draw (omitR)--node[tube label,pos=.4,above=5pt] {$Q_{z-1}$} node[tube cross,pos=.4,label={[tube label]below=5pt:$M_{z-1}$}] {$\times$}(Gn);
\TubeLink{Gn}{R}{Q_z}{M_z}
\TubeAdjoint{G1}{\phi_1}
\TubeAdjoint{Gn}{\phi_{z-1}}
\end{tikzpicture}
\caption{The $\mathcal{T}_z^{(\su_2)}$ theory, obtained by gluing $z$ copies of the elementary tube with $\mathcal{N}=2$ vector multiplets.}
\label{fig:Tsu2z}
\end{center}
\end{figure}

Starting from this tube, we can glue several cylinders together by gauging. Gluing $z \in \mathbb{Z}_{>0}$ cylinders with $\mathcal{N}=2$ vector multiplets, we obtain $\mathcal{T}_z^{(\mathfrak{su}_2)}$. This is an $(\su_2)^{z-1}$ linear quiver gauge theory, with a single bifundamental, together with its flip field, between each pair of adjacent nodes. The field content is given in Table~\ref{tab:Tsu2z}, and the quiver is depicted in Figure~\ref{fig:Tsu2z}. 
\begin{table}[h]
    \centering
    \renewcommand{\arraystretch}{1.25}
    \begin{tabular}{|c|c|cc|c|}
        \hline
        field & $(\su_2)_L \times (\su_2)^{z-1} \times (\su_2)_{R}$ &$R_0$ & $\mathcal{F}$ & $\mathcal{J}$ \\
        \hline
        $Q_{i=1, \cdots, z}$ & $(\mathbf{2}_{i-1}, \mathbf{2}_{i})$ & $0$ & $\phantom{-}1$ & $\frac{-z+2i-1}{2}$\\
        $\phi_{i=1, \cdots, z-1}$ & $\mathbf{3}_i$ & $1$ & $-1$ & $\frac{z-2i}{2}$\\ 
        $M_{i=1, \cdots, z}$ & $\mathbf{1}$ & $2$ & $-2$ & $z-2i+1$\\
        \hline
    \end{tabular}
    \caption{The matter content of $\mathcal{T}_z^{(\su_2)}$. Here $(\su_2)_0 = (\su_2)_L$ and $(\su_2)_{z} = (\su_2)_R$.}    
    \label{tab:Tsu2z}
\end{table}

The superpotential is
\begin{align}
    W = \sum_{i=1}^z M_i \det Q_i + \sum_{i=1}^{z-2} \phi_i Q_{i+1} \phi_{i+1} Q_{i+1} \ . 
\end{align}
Note that the superpotential contains quartic terms in the adjoints and the bifundamentals, and that the second sum runs only up to $z-2$. As a result, there is an additional $\u1$ symmetry, besides $\mathcal{F}$ (which comes from the normal direction of the UV curve), that we call $\mathcal{J}$ in Table~\ref{tab:Tsu2z}. It is an emergent symmetry that is not visible in the three-punctured sphere theory $T_\g$. Geometrically, it stems from the isometry of the cylinder \cite{Hwang:2021xyw}. The anomaly coefficients of this theory are
\begin{align}
\begin{split}
    & \Tr R_0 = \Tr R_0^3 = -3 \ , \quad \Tr \mathcal{F} = 3-z \ , \quad \Tr \mathcal{F}^3 = 3-7z \ , \\
    & \Tr R_0^2 \mathcal{F} = 2z \ , \quad \Tr \mathcal{F} \mathcal{J}^2 = -z(z-1)(7z-2)/12 \ , \\
    & \Tr \mathcal{J} = \Tr \mathcal{J}^3 = \Tr \mathcal{J} R_0^2 = \Tr \mathcal{J} \mathcal{F}^2 = \Tr \mathcal{J} R_0 \mathcal{F} = \Tr \mathcal{J}^2 R_0 = \Tr R_0 \mathcal{F}^2 =  0 \ , \\
    & \Tr R_0 (\su_2)^2_{L, R} = -1 \ , \quad \Tr \mathcal{F} (\su_2)_{L, R}^2 = 1 \ ,  \\ 
    & \Tr \mathcal{J} (\su_2)_L^2 = \frac{1-z}{2} \ , \quad \Tr \mathcal{J} (\su_2)_R^2 = \frac{z-1}{2} \ ,
\end{split}
\end{align}
with the normalization $\Tr T^a T^b = \frac{1}{2}\delta^{ab}$ for the generators $T^a$ of $\su_2$ in the fundamental representation. 
The superconformal R-symmetry is fixed by $a$-maximization \cite{Intriligator:2003jj}, and it is a mixture of $R_0$, $\mathcal{F}$ and $\mathcal{J}$:
\begin{align}
    R_{IR} = R_0 + \epsilon \mathcal{F} + \eta \mathcal{J}\ .
\end{align}
To determine $\epsilon$ and $\eta$, we compute the trial central charges,
\begin{align}
    a_{\text{trial}} &= \frac{3}{32} ( 3\Tr R^3 - \Tr R) = \frac{3}{32} \left( -6 + (19z-3)\epsilon + 3(3-7z)\epsilon^3 - 9 A_z \epsilon \eta^2 \right) \ , \\
    c_{\text{trial}} &= \frac{1}{32} ( 9\Tr R^3 - 5 \Tr R) = \frac{1}{32} \left( -12 + (59z- 15)\epsilon + 9(3-7z)\epsilon^3 - 27 A_z \epsilon \eta^2 \right) \ , 
\end{align}
where $A_z = z(z-1)(7z-2)/12$. The local maximum is at 
\begin{align}
    \eta = 0 \ , \quad \epsilon = \frac{1}{3}\sqrt{\frac{19z-3}{7z-3}} \ , 
\end{align}
so that the superconformal $R$-charge does not mix with $(\u1)_\mathcal{J}$. For $z=1$, we get $\epsilon=2/3$ and $(a, c)=(5/48, 5/24)$, which are the correct values for a theory of 5 free chiral multiplets. For $z\geq 2$, it is indeed an interacting $\CN=1$ SCFT with $(\su_2)^2\times (\u1)_\CF \times (\u1)_\CJ$ flavor symmetry. 

\subsection{The \texorpdfstring{$D_p(\mathfrak{su}_2)$}{Dp(su2)} theories}

\begin{figure}
\begin{center}
\begin{tikzpicture}[>=Latex,tube edge,very thick,tube gauge/.append style={line width=1.2pt,minimum size=10mm,fill=blue!6},tube flavor/.append style={line width=1.2pt,minimum size=8.5mm,fill=black!4}]
\node[tube gauge,label={[tube group]below=10pt:$(\su_2)_0$}] (L) at (-4.9,0) {$2$};
\node[tube gauge,label={[tube group]below=10pt:$(\su_2)_1$}] (G1) at (-2.2,0) {$2$};
\node[tube gauge,label={[tube group]below=10pt:$(\su_2)_{z-1}$}] (Gn) at (2.2,0) {$2$};
\node[tube flavor,label={[tube group]below=10pt:$(\su_2)_R$}] (R) at (4.9,0) {$2$};
\coordinate (omitL) at (-.45,0);
\coordinate (omitR) at (.45,0);
\TubeLink{L}{G1}{Q_1}{M_1}
\TubeLink{G1}{omitL}{Q_2}{M_2}
\draw[dash pattern=on 0pt off 3pt,line cap=round] (omitL)--(omitR);
\draw (omitR)--node[tube label,pos=.4,above=5pt] {$Q_{z-1}$} node[tube cross,pos=.4,label={[tube label]below=5pt:$M_{z-1}$}] {$\times$}(Gn);
\TubeLink{Gn}{R}{Q_z}{M_z}
\TubeAdjoint{L}{\phi_0}
\TubeAdjoint{G1}{\phi_1}
\TubeAdjoint{Gn}{\phi_{z-1}}
\end{tikzpicture}
\caption{The tube theory of flux $z$ with its left puncture gauged: the former flavor node $(\su_2)_L$ is now the gauge node $(\su_2)_0$, with the adjoint $\phi_0$.}
\label{fig:oddD-raw}
\end{center}
\end{figure}
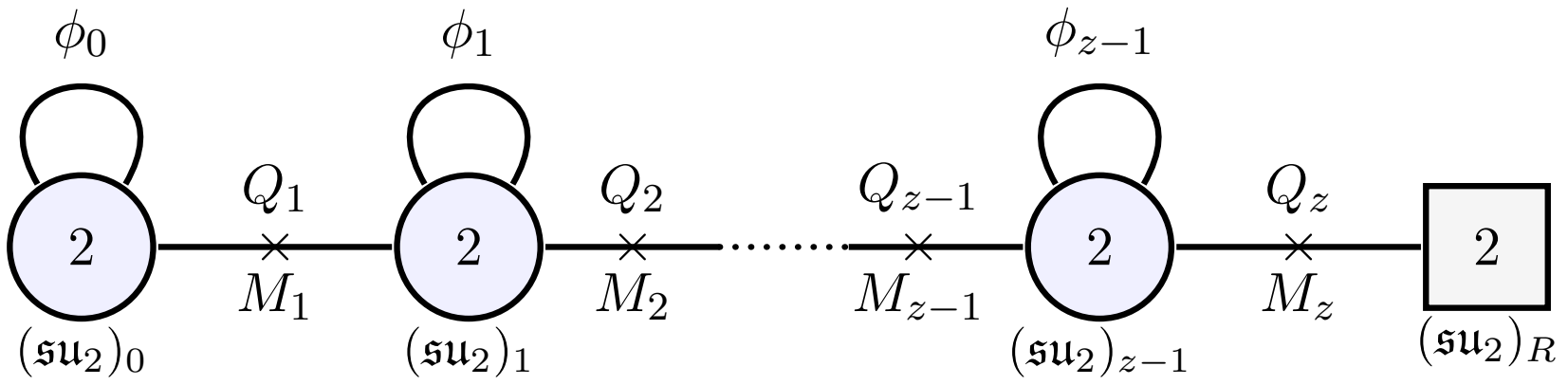

Let us now gauge one of the two flavor nodes, say $(\su_2)_L$, together with an adjoint chiral multiplet $\phi_0$ and the quartic coupling $W = \phi_0 Q_1 \phi_1 Q_1$, as depicted in Figure \ref{fig:oddD-raw}. In class $\mathcal{S}$ language, this corresponds to replacing one of the regular punctures with the irregular puncture of type $I_{2, -1}$ in the notation of \cite{Xie:2012hs}. This breaks the $(\u1)_\CF \times (\u1)_\CJ$ symmetry to $\u1$. Writing $R(\phi_0)=t$, the anomaly-free conditions and the superpotential give the relations
\begin{align}
\begin{split}
    2 + 2(t-1) + R(Q_1) -1 &= 0 \ , \\
    2 + 2(R(\phi_i)-1) + (R(Q_i) -1) + (R(Q_{i+1}) - 1) &= 0 \ , \quad (1 \le i \le z-1) \\
    R(\phi_{i-1}) + 2R(Q_i) + R(\phi_i) &= 2 \ , \quad (1 \le i \le z-1) \\
    R(M_i) + 2 R(Q_i) &= 2 \ . \quad (1 \le i \le z)
\end{split}
\end{align}
The solution is given as
\begin{align}
    R(\phi_i) = (2i+1)t \ , \quad R(Q_i) = 1 - 2i t \ , \quad R(M_i) = 4it \ , 
\end{align}
or, equivalently,
\begin{align}
    \eta = -2t \ , \quad \epsilon = 1-(z+1)t \ . 
\end{align}
The trial $a$-function is then
\begin{align}
    a(t) = \frac{3z}{32} \left( 3z(14z^2+16z+5) t^3 - 3(28z^2+24z+5)t^2 + 4(11z+5)t - 2 \right) \ .
\end{align}
We can determine the value of $t$ from $a$-maximization. However, we find that some of the gauge-invariant operators violate or saturate the unitarity bound $R(\mathcal{O}) \geq 2/3$. Such operators decouple along the RG flow and become free. This invalidates the naive $a$-maximization, since it introduces accidental $\u1$ symmetries that can mix with $R$ \cite{Kutasov:2003iy}. A convenient way to deal with this is to introduce a flip field $M$ with a coupling $W=M\mathcal{O}$ for each would-be-decoupled operator $\mathcal{O}$~\cite{Benvenuti:2017lle,Maruyoshi:2018nod}. We repeat this procedure until no operator violates the unitarity bound.\footnote{When several gauge-invariant operators violate the unitarity bound, we flip them one by one, starting from the one with the lowest (incorrect) $R$-charge. In this way, some of the would-be unitarity-violating operators may end up satisfying the bound. This prescription can be justified by the $a$-theorem \cite{Agarwal:2019crm, Agarwal:2020pol}.} 

\begin{figure}
\begin{center}
\begin{tikzpicture}[>=Latex,tube edge,very thick,tube gauge/.append style={line width=1.2pt,minimum size=10mm,fill=blue!6},tube flavor/.append style={line width=1.2pt,minimum size=8.5mm,fill=black!4}]
\node[tube gauge,label={[tube group]below=10pt:$(\su_2)_0$}] (G0) at (-4.9,0) {$2$};
\node[tube gauge,label={[tube group]below=10pt:$(\su_2)_1$}] (G1) at (-2.2,0) {$2$};
\node[tube gauge,label={[tube group]below=10pt:$(\su_2)_{z-1}$}] (Gn) at (2.2,0) {$2$};
\node[tube flavor,label={[tube group]below=10pt:$(\su_2)_R$}] (R) at (4.9,0) {$2$};
\coordinate (omitL) at (-.45,0);
\coordinate (omitR) at (.45,0);
\draw (G0)--node[tube label,above=5pt] {$Q_1$}(G1);
\draw (G1)--node[tube label,above=5pt] {$Q_2$}(omitL);
\draw[dash pattern=on 0pt off 3pt,line cap=round] (omitL)--(omitR);
\draw (omitR)--node[tube label,pos=.4,above=5pt] {$Q_{z-1}$} node[tube cross,pos=.4,label={[tube label]below=5pt:$M_{z-1}$}] {$\times$}(Gn);
\TubeLink{Gn}{R}{Q_z}{M_z}
\draw (G0.55) .. controls +(60:1.05cm) and +(120:1.05cm) ..
 node[tube cross,pos=.8,label={[tube label]left=8pt:$X_0$}] {$\times$}
 node[tube label,pos=.5,above=6pt] {$\phi_0$}(G0.125);
\draw (G1.55) .. controls +(60:1.05cm) and +(120:1.05cm) ..
 node[tube cross,pos=.8,label={[tube label]left=8pt:$X_1$}] {$\times$}
 node[tube label,pos=.5,above=6pt] {$\phi_1$}(G1.125);
\draw (Gn.55) .. controls +(60:1.05cm) and +(120:1.05cm) ..
 node[tube label,pos=.5,above=6pt] {$\phi_{z-1}$}(Gn.125);
\end{tikzpicture}
\caption{The $\CN=1$ Lagrangian for the $D_{2z+1}(\su_2)$ theory \cite{ChoSong:2026}.}
\label{fig:DoddSU2}
\end{center}
\end{figure}
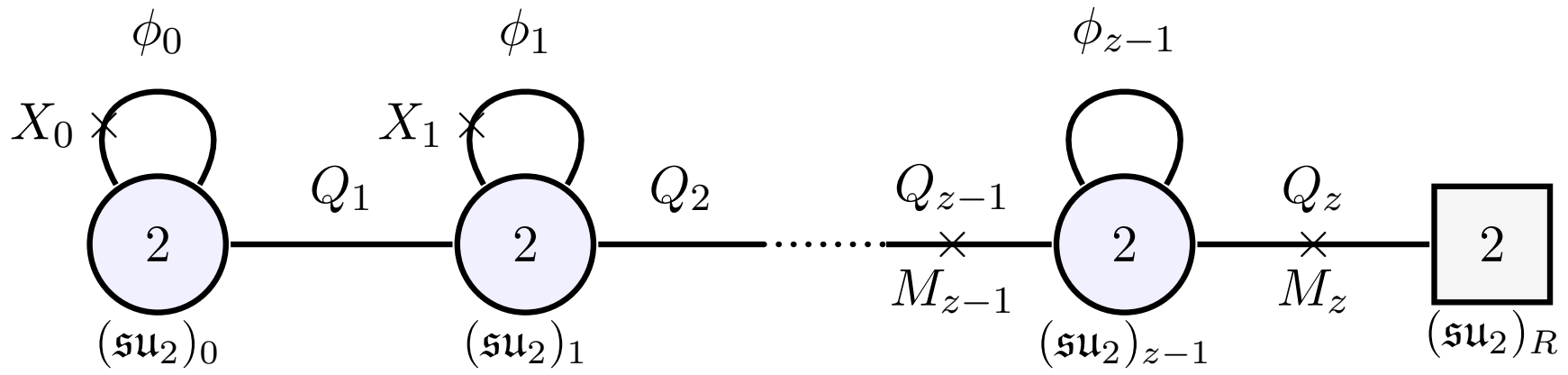

We find that the following operators decouple along the RG flow:
\begin{align}
M_i \quad (1 \le i \le \lfloor z/2 \rfloor) \ , \qquad \Tr \phi_j^2 \quad (0 \le j \le \lceil z/2 \rceil -1 )\,.   
\end{align}
There are $z$ decoupled operators in total. After flipping them, the interacting part at the fixed point is described by the superpotential
\begin{align}
    W = \sum_{i=1}^{z-1} \phi_{i-1} Q_{i} \phi_i Q_{i} + \sum_{i= \lfloor z/2 \rfloor +1}^z M_i \det Q_i + \sum_{j=0}^{\lceil z/2 \rceil -1} X_j \Tr \phi_j^2 \ ,
\end{align}
and its matter content is summarized in Figure~\ref{fig:DoddSU2}. The mixing parameters are now
\begin{align}
 t = \frac{2}{3(2z+1)},\qquad
 \epsilon = \frac{4z+1}{3(2z+1)},\qquad
 \eta = -\frac{4}{3(2z+1)} \ ,
\end{align}
from which we obtain the central charges
\begin{align}
    a = \frac{z(8z+3)}{8(2z+1)} \ , \quad c = \frac{z}{2} \ , \quad k_{(\su_2)_R} = -6\Tr R (\su_2)_R^2 = \frac{8z}{2z+1} \ . 
\end{align}
These values precisely match those of the $\CN=2$ $(A_1, D_{2z+1}) = D_{2z+1}(\su_2)$ theory~\cite{Xie:2012hs,Cecotti:2013lda,Shapere:2008zf}. We also find that the Coulomb branch operators are the $M_i$ and the $\Tr \phi_j^2$, with scaling dimensions
\begin{align}
    \Delta(M_i) = \frac{4i}{2z+1}  \quad (\lfloor z/2 \rfloor < i \le z)\ , \qquad \Delta(\Tr \phi_j^2) = \frac{2(2j+1)}{2z+1} \quad (\lceil z/2 \rceil \le j < z) \ .
\end{align}
These also agree with the dimensions of the Coulomb branch operators of the Argyres--Douglas theory.

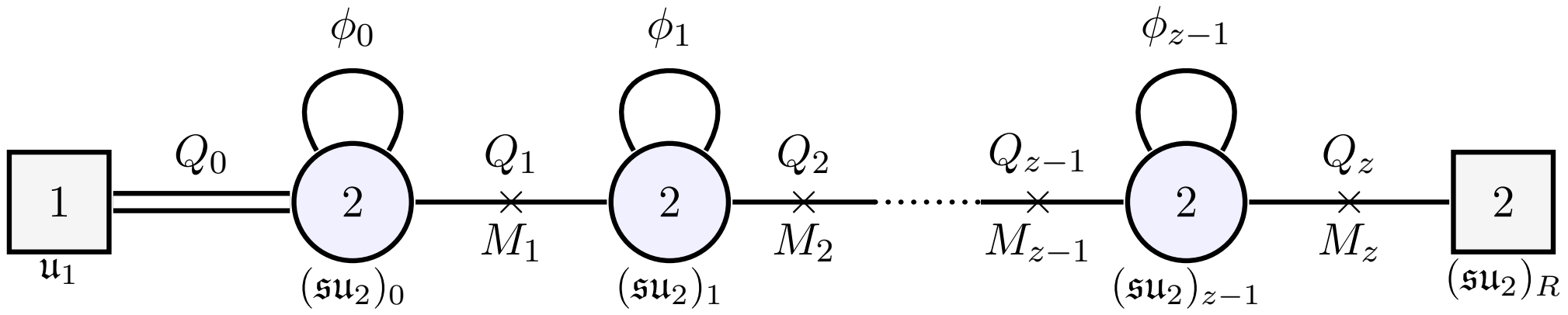
\begin{figure}
\begin{center}
\begin{tikzpicture}[>=Latex,tube edge,very thick,tube gauge/.append style={line width=1.2pt,minimum size=10mm,fill=blue!6},tube flavor/.append style={line width=1.2pt,minimum size=8.5mm,fill=black!4}]
\node[tube flavor,label={[tube group]below=10pt:$\mathfrak{u}_1$}] (LL) at (-7.4,0) {$1$};
\node[tube gauge,label={[tube group]below=10pt:$(\su_2)_0$}] (L) at (-4.9,0) {$2$};
\node[tube gauge,label={[tube group]below=10pt:$(\su_2)_1$}] (G1) at (-2.2,0) {$2$};
\node[tube gauge,label={[tube group]below=10pt:$(\su_2)_{z-1}$}] (Gn) at (2.2,0) {$2$};
\node[tube flavor,label={[tube group]below=10pt:$(\su_2)_R$}] (R) at (4.9,0) {$2$};
\coordinate (omitL) at (-.45,0);
\coordinate (omitR) at (.45,0);
\draw[double,double distance=3pt] (LL)--node[tube label,above=5pt] {$Q_0$}(L);
\TubeLink{L}{G1}{Q_1}{M_1}
\TubeLink{G1}{omitL}{Q_2}{M_2}
\draw[dash pattern=on 0pt off 3pt,line cap=round] (omitL)--(omitR);
\draw (omitR)--node[tube label,pos=.4,above=5pt] {$Q_{z-1}$} node[tube cross,pos=.4,label={[tube label]below=5pt:$M_{z-1}$}] {$\times$}(Gn);
\TubeLink{Gn}{R}{Q_z}{M_z}
\TubeAdjoint{L}{\phi_0}
\TubeAdjoint{G1}{\phi_1}
\TubeAdjoint{Gn}{\phi_{z-1}}
\end{tikzpicture}
\caption{The tube theory of flux $z$ with its left puncture gauged together with $D_2(\su_2)$, the quiver tail: the former flavor node $(\su_2)_L$ is now the gauge node $(\su_2)_0$, with the adjoint $\phi_0$ and two fundamental chiral multiplets.}
\label{fig:evenD-raw}
\end{center}
\end{figure}

To obtain the $(A_1, D_{2z+2}) = D_{2z+2}(\su_2)$ theory, we repeat the same procedure, except that the gauged $\su_2$ node is also coupled to an additional fundamental hypermultiplet $Q_0$. The gauged theory is described in Figure \ref{fig:evenD-raw}. This can be thought of as attaching a quiver tail of the form $(2)-[1]$, or, in class $\cS$ language, as introducing an irregular puncture of type $I_{2, 0}$. The quiver tail contains an extra $\mathfrak{u}_1$ flavor symmetry, under which the two half-hypers in $Q_0$ have charges $\pm1$.

\begin{figure}
\begin{center}
\begin{tikzpicture}[>=Latex,tube edge,very thick,tube gauge/.append style={line width=1.2pt,minimum size=10mm,fill=blue!6},tube flavor/.append style={line width=1.2pt,minimum size=8.5mm,fill=black!4}]
\node[tube flavor,label={[tube group]below=10pt:$\mathfrak{u}_1$}] (L) at (-7.4,0) {$1$};
\node[tube gauge,label={[tube group]below=10pt:$(\su_2)_0$}] (G0) at (-4.9,0) {$2$};
\node[tube gauge,label={[tube group]below=10pt:$(\su_2)_1$}] (G1) at (-2.2,0) {$2$};
\node[tube gauge,label={[tube group]below=10pt:$(\su_2)_{z-1}$}] (Gn) at (2.2,0) {$2$};
\node[tube flavor,label={[tube group]below=10pt:$(\su_2)_R$}] (R) at (4.9,0) {$2$};
\coordinate (omitL) at (-.45,0);
\coordinate (omitR) at (.45,0);
\draw[double,double distance=3pt] (L)--node[tube label,above=5pt] {$Q_0$}(G0);
\draw (G0)--node[tube label,above=5pt] {$Q_1$}(G1);
\draw (G1)--node[tube label,above=5pt] {$Q_2$}(omitL);
\draw[dash pattern=on 0pt off 3pt,line cap=round] (omitL)--(omitR);
\draw (omitR)--node[tube label,pos=.4,above=5pt] {$Q_{z-1}$} node[tube cross,pos=.4,label={[tube label]below=5pt:$M_{z-1}$}] {$\times$}(Gn);
\TubeLink{Gn}{R}{Q_z}{M_z}
\draw (G0.55) .. controls +(60:1.05cm) and +(120:1.05cm) ..
 node[tube cross,pos=.8,label={[tube label]left=8pt:$X_0$}] {$\times$}
 node[tube label,pos=.5,above=6pt] {$\phi_0$}(G0.125);
\draw (G1.55) .. controls +(60:1.05cm) and +(120:1.05cm) ..
 node[tube cross,pos=.8,label={[tube label]left=8pt:$X_1$}] {$\times$}
 node[tube label,pos=.5,above=6pt] {$\phi_1$}(G1.125);
\draw (Gn.55) .. controls +(60:1.05cm) and +(120:1.05cm) ..
 node[tube label,pos=.5,above=6pt] {$\phi_{z-1}$}(Gn.125);
\end{tikzpicture}
\caption{The $\CN=1$ Lagrangian for the $D_{2z+2}(\su_2)$ theory \cite{ChoSong:2026}.}
\label{fig:DevenSU2}
\end{center}
\end{figure}
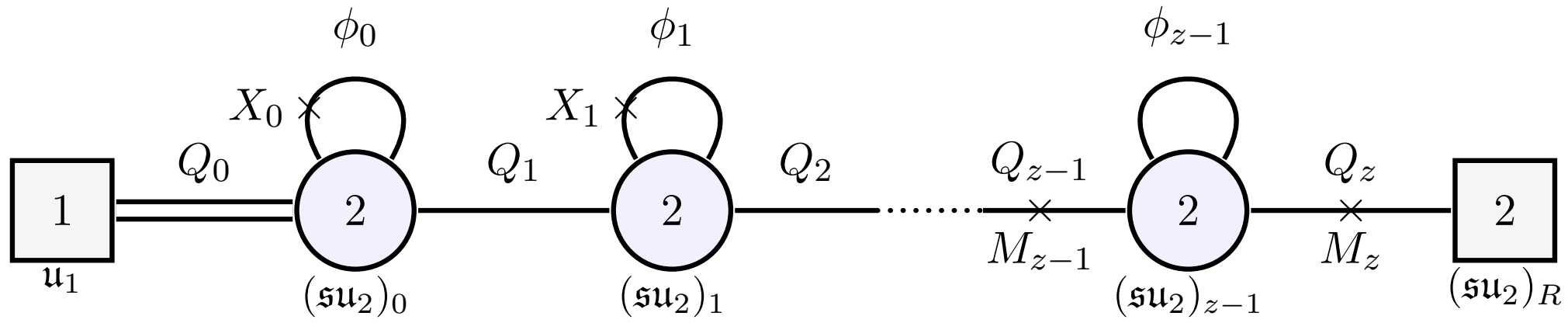

In this case, we turn on an additional cubic superpotential $\Delta W=Q_0\phi_0Q_0$:
\begin{align}
    W=\sum_{i=1}^{z-1}\phi_{i-1}Q_i\phi_iQ_i+Q_0\phi_0Q_0+\sum_{i=1}^z M_i\det Q_i\,.
\end{align}
Notice that the extra $\mathfrak{u}_1$ flavor symmetry does not mix with the $R$-charge, because it is baryonic. Denoting the $R$-charge of $\phi_0$ by $R(\phi_0)=t$, we get the relations between the $R$-charges of the fields from the anomaly-free conditions and the superpotential as follows.
\begin{align}
\begin{split}
    2R(Q_0)+t=2\,,&\\
    2+2(R(\phi_i)-1)+(R(Q_i)-1)+(R(Q_{i+1})-1)=0\,,&\quad(0\leq i\leq z-1)\\
    R(\phi_{i-1})+2R(Q_i)+R(\phi_i)=2\,,&\quad(1\leq i\leq z-1)\\
    R(M_i)+2R(Q_i)=2\,.&\quad (1\leq i\leq z)
    \end{split}
\end{align}
Then, the solution is
\begin{align}
    R(\phi_i)=(i+1)t\,,\quad R(Q_i)=1-\frac{2i+1}{2}t\,,\quad R(M_i)=(2i+1)t\,.
\end{align}
With this parametrization, the trial $a$ central charge is written as
\begin{align}
\begin{split}
    a(t)=\frac{3}{128}\Big(&3(7z^4+22z^3+25z^2+12z-2)t^3-6z(14z^2+33z+25)t^2\\
    &+8(11z^2+16z+1)t-8z\Big)\,.
\end{split}
\end{align}
However, it turns out that several operators decouple, invalidating the $a$-maximization from $a(t)$. In the end, one finds that the following operators decouple:
\begin{align}
    M_i\,,\quad(1\leq i\leq \lfloor z/2\rfloor)\,,\qquad\Tr\,\phi_j^2\quad(0\leq j\leq \lceil z/2\rceil-1)\,.
\end{align}
The matter content of the interacting part of the IR SCFT is depicted in Figure \ref{fig:DevenSU2}. Then, the superpotential describing the interacting part of the IR SCFT is
\begin{align}
    W=\sum_{i=1}^{z-1}\phi_{i-1}Q_i\phi_iQ_i+Q_0\phi_0Q_0+\sum_{i=\lfloor z/2\rfloor+1}^zM_i\det Q_i+\sum_{j=0}^{\lceil z/2\rceil-1}X_j\Tr\,\phi_j^2\,.
\end{align}
In the end, we find $t$ and the central charges to be
\begin{align}
    t=\frac{2}{3(z+1)}\,,\quad a=\frac{z}{2}+\frac{1}{12}\,,\quad c=\frac{z}{2}+\frac{1}{6}\,,\quad k_{(\su_2)_R}=-6\Tr\,R(\su_2)_R^2=\frac{2(2z+1)}{z+1}\,.
\end{align}
The anomaly coefficients perfectly match those of the $D_{2z+2}(\su_2)$ theory~\cite{Xie:2012hs,Cecotti:2013lda,Shapere:2008zf}. Furthermore, we find the Coulomb branch operators from the remaining $M_i$ and $\Tr\,\phi_j^2$, whose scaling dimensions are
\begin{align}
    \Delta(M_i)=\frac{2i+1}{z+1}\quad(\lfloor z/2\rfloor+1\leq i\leq z)\,,\qquad\Delta(\Tr\,\phi_j^2)=\frac{2(j+1)}{z+1}\quad(\lceil z/2\rceil\leq j\leq z-1)\,.
\end{align}
The Coulomb branch spectrum also coincides with that of $D_{2z+2}(\su_2)$.

\section{Construction of \texorpdfstring{$D_p(\mathfrak{su}_3)$ theories from $T_3$}{Dp(su3) theories from T3}}\label{sec:a2}
As our second explicit construction, we will relate the $T_{\su_3}$ theory and ${\cal N}=1$ two-punctured spheres constructed from it to the $D_p(\mathfrak{su}_3)$ Argyres--Douglas theories.  We start by constructing a Lagrangian for the $z=1$ two-punctured sphere $\mathcal{T}^{(\su_3)}_1$.

\subsection{The tube of flux \texorpdfstring{$z$}{z}}\label{sec:a2:tubea2}
We start from a Lagrangian for the $T_{\su_3}$ theory (\emph{aka} the $\E_6$ MN theory, or the $T_3$ trinion) derived in~\cite{GarciaEtxebarria:2021deconf}. The Lagrangian of~\cite{GarciaEtxebarria:2021deconf} is an ${\cal N}=1$ $\su_2\times \su_3$ gauge theory with only a subgroup, $\u1^3\times \su_5\subset \u1\times {\frak{e}}_6$, of the expected global symmetry of $T_{\su_3}$ (in ${\cal N}=1$ language) visible in the UV. The fields of the Lagrangian are listed in Table~\ref{tab:T3}.\footnote{We denote by $\tilde Y$, $\tilde\Phi$ and $\tilde P$ the fields called $Y_1$, $\Phi_1$ and $P_1$ in~\cite{GarciaEtxebarria:2021deconf}, to avoid confusion with the components of the $\su_5$ multiplets $Y$, $\Phi$ and $P$ ({\it e.g.}\ $Y_m$, $Y_4$, $\Phi^m$).} The superpotential preserving the symmetries listed in the table is
\begin{equation}
W_{T_{\su_3}}={\rm Tr}\big(A_1A_2Z+A_2Y\Phi+A_1\tilde Y\tilde\Phi+PQY+\tilde PQ\tilde Y\big)+A_1^2Q\,\bar\phi\,.
\end{equation}
Two of the three puncture $\su_3$ symmetries are emergent. We can explicitly write the moment map operators, both for the emergent and for the manifest symmetries. For example, the moment map $\mu_A$ of one of the emergent puncture $\su_3$ symmetries is the octet of operators
\be
\mu_A\;:\;\;\;\;\;\; P^{4,5},\;\;\;  Y{}_{4,5}\cdot Z,\;\;\;  \epsilon_{\alpha\beta}\Phi^{\alpha 4,5}A_1^\beta \cdot Y_{4,5}\,.
\ee Here $\cdot$ represents the contraction of $(\su_3)_g$ gauge indices, and Greek indices are $(\su_2)_g$ gauge indices. All the components of the moment map have $(R_0,{\cal F})=(1,1)$. We will call this puncture $(A)$ and the other emergent puncture $(B)$, and we will denote the manifest puncture, $\su_3\subset\su_5$, by $(C)$.

\begin{table}[h]
\centering
\renewcommand{\arraystretch}{1.25}
\begin{tabular}{|c|ccc|cc|cc|}
\hline
field & $(\su_3)_g$ & $(\su_2)_g$ & $\su_5$ & $R_0$ & ${\cal F}$ & $(\u1)_y$ & $(\u1)_z$ \\
\hline
$A_1$ & $\bar{\bf 3}$ & ${\bf 2}$ & ${\bf 1}$ & $\phantom{-}\frac16$ & $\phantom{-}\frac16$ & $-\frac56$ & $-\frac13$\\
$A_2$ & $\bar{\bf 3}$ & ${\bf 2}$ & ${\bf 1}$ & $\phantom{-}\frac76$ & $-\frac56$ & $-\frac56$ & $-\frac13$\\
$Y$ & ${\bf 3}$ & ${\bf 1}$ & $\bar{\bf 5}$ & $\phantom{-}\frac13$ & $\phantom{-}\frac13$ & $-\frac16$ & $\phantom{-}\frac13$\\
$\tilde Y$ & ${\bf 3}$ & ${\bf 1}$ & ${\bf 1}$ & $\phantom{-}\frac13$ & $\phantom{-}\frac13$ & $\phantom{-}\frac56$ & $-\frac53$\\
$Z$ & $\bar{\bf 3}$ & ${\bf 1}$ & ${\bf 1}$ & $\phantom{-}\frac23$ & $\phantom{-}\frac23$ & $\phantom{-}\frac53$ & $\phantom{-}\frac23$\\
$P$ & ${\bf 1}$ & ${\bf 1}$ & ${\bf 5}$ & $\phantom{-}1$ & $\phantom{-}1$ & $-\frac32$ & $-1$\\
$\tilde P$ & ${\bf 1}$ & ${\bf 1}$ & ${\bf 1}$ & $\phantom{-}1$ & $\phantom{-}1$ & $-\frac52$ & $\phantom{-}1$\\
$Q$ & $\bar{\bf 3}$ & ${\bf 1}$ & ${\bf 1}$ & $\phantom{-}\frac23$ & $-\frac43$ & $\phantom{-}\frac53$ & $\phantom{-}\frac23$\\
$\Phi$ & ${\bf 1}$ & ${\bf 2}$ & ${\bf 5}$ & $\phantom{-}\frac12$ & $\phantom{-}\frac12$ & $\phantom{-}1$ & $\phantom{-}0$\\
$\tilde\Phi$ & ${\bf 1}$ & ${\bf 2}$ & ${\bf 1}$ & $\phantom{-}\frac32$ & $-\frac12$ & $\phantom{-}0$ & $\phantom{-}2$\\
$\bar\phi$ & ${\bf 1}$ & ${\bf 1}$ & ${\bf 1}$ & $\phantom{-}1$ & $\phantom{-}1$ & $\phantom{-}0$ & $\phantom{-}0$\\
\hline
\end{tabular}
\caption{The ${\cal N}=1$ Lagrangian for $T_{\su_3}$~\cite{GarciaEtxebarria:2021deconf}.  The
$R_0$ and ${\cal F}$ charges are those of the ${\cal N}=2$ theory,
$R_0=I_3+\frac r2$ and ${\cal F}=I_3-\frac r2$, with $I_3$ the Cartan generator of the $\su_2$ R-symmetry and $r$ the generator of the $\u1$ R-symmetry. One finds $a=\frac{41}{24}$,
$c=\frac{13}{6}$ and $k_{\E_6}=6$, the values of the $T_{\su_3}\cong \E_6$ MN theory. The $(\u1)_y$ and $(\u1)_z$ charges are such that $\su_5\times(\u1)_y\subset{\frak{so}}_{10}$ and ${\frak{so}}_{10}\times(\u1)_z\subset{\frak{e}}_6$.}
\label{tab:T3}
\end{table}

\ 

Next, we close the maximal puncture $(A)$ by first flipping it and then giving a principal nilpotent vev to the flippers~\cite{Agarwal:2015vla,Maruyoshi:2016aim,Gadde:2013fma}. This closes the puncture
and shifts the flux by $+\frac12$~\cite{Beem:2012yn, Gaiotto:2015usa, Razamat:2022gpm}. The
trinion $T_{\su_3}$ has flux $z=+\frac12$, so the resulting two-punctured sphere (the tube $\mathcal{T}^{(\su_3)}_{z=1}$) has flux $z=+1$. Let us work this out in more detail.

\begin{figure}[t]
\centering
\begin{tikzpicture}[>=Stealth,
 gauge/.style={circle,draw,very thick,minimum size=10mm,fill=blue!6},
 flav/.style={rectangle,draw,very thick,minimum size=8.5mm,fill=black!4},
 flip/.style={font=\Large,inner sep=0pt},
 midarrow/.style={postaction={decorate},decoration={markings,mark=at position 0.55 with {\arrow{Stealth[length=2.6mm]}}}}]
\node[gauge] (L) at (0,0) {$2$};
\node[gauge] (R) at (4.6,0) {$2$};
\node[flav] (C) at (2.3,2.7) {$3$};
\node[flav] (U) at (2.3,-2.7) {$1$};
\node[flav] (X) at (-2.6,0) {$2$};
\draw[very thick] (L) to[bend left=18] node[above]{$A_2$} (R);
\draw[very thick] (L) to[bend right=18] coordinate[pos=0.62] (fA) node[pos=0.38,below]{$A_1$} (R);
\node[flip] at (fA) {$\times$};
\node[below=3pt] at (fA) {$b_a$};
\draw[very thick,midarrow] (L) to[bend left=10] node[above left]{$Q_L$} (C);
\draw[very thick,midarrow] (C) to[bend left=10] node[above right]{$Q_R$} (R);
\draw[very thick] (L) to[bend right=10] node[below left]{$\tilde Q_L$} (U);
\draw[very thick] (R) to[bend left=10] node[below right]{$\tilde Q_R$} (U);
\draw[very thick] (X) -- coordinate[pos=0.5] (fS) node[pos=0.5,above=5pt]{$q_i$} (L);
\node[flip] at (fS) {$\times$};
\node[below=3pt] at (fS) {$b_q$};
\end{tikzpicture}
\caption{The tube ${{\cal T}^{(\su_3)}_{z=1}}$. Circles are the gauge nodes $(\su_2)_L$ (left) and
$(\su_2)_R$ (right).  The square labeled $3$ is the manifest puncture symmetry
$(\su_3)_C$. The squares labeled $2$ and $1$ are the manifest subgroup
$(\su_2)_{B_1-B_2}\times (\frak{u}_1)_{B_1+B_2}$ of the emergent puncture symmetry $(\su_3)_B$. Crosses
denote flip fields: $b_a$ flips $\det A_1$, and $b_q$ flips the meson $\det q$.
The superpotential is given in Eq.~\eqref{eq:Wtube}.}
\label{fig:tube}
\end{figure}
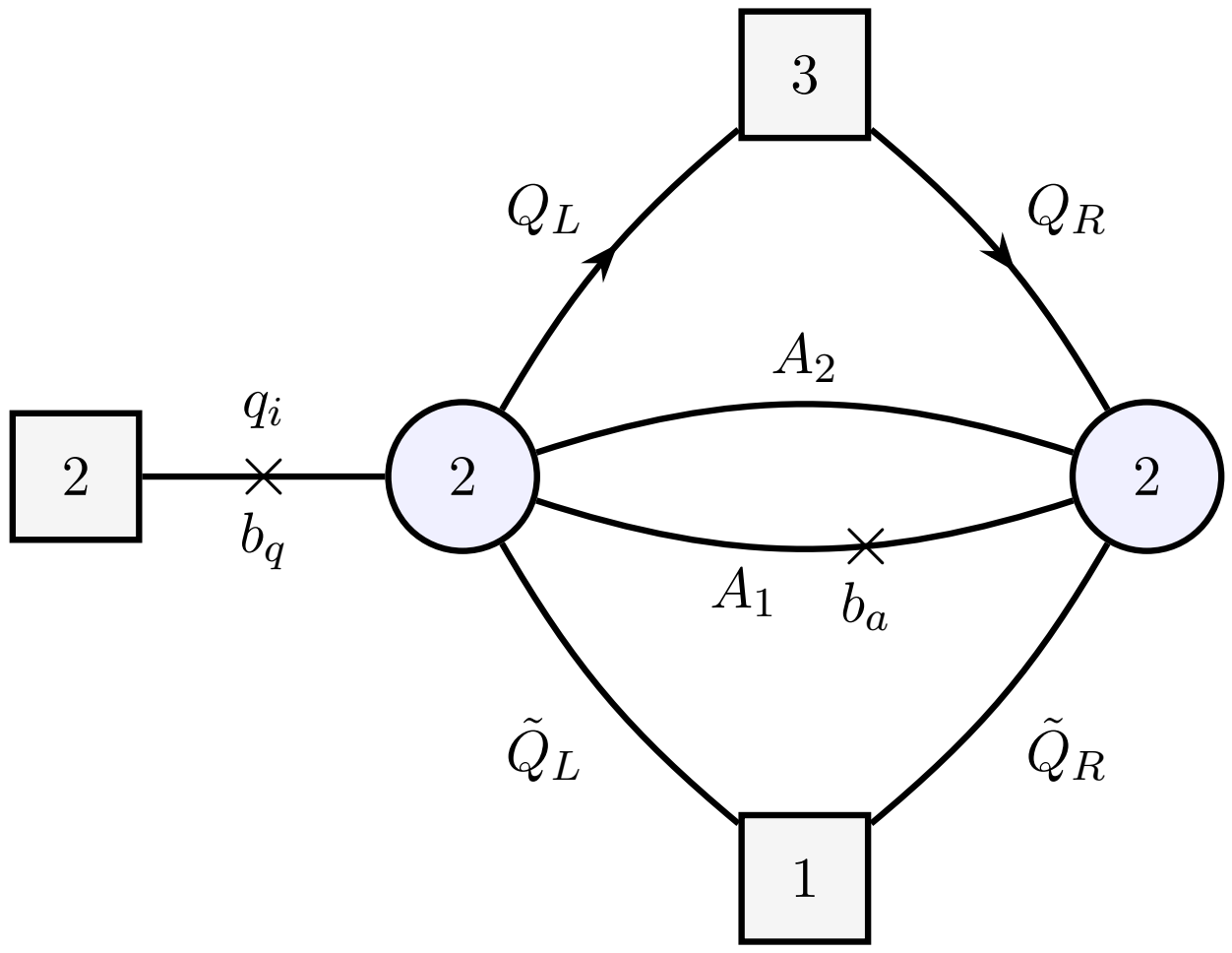

We deform the theory by adding a chiral field $M_A$ in the adjoint representation of the puncture $(A)$ symmetry and coupling it through a flipping superpotential term,
\be {\rm Tr}\,M_A\mu_A\,,
\ee 
and we give it the principal vacuum expectation value
$\langle M_A\rangle=\rho(\sigma^+)$. The R-symmetry preserved by this deformation is shifted by the
charge of the Cartan generator $\rho(\sigma_3)$. Only the two lowest components of the flip field $M_A$
survive, one for each Casimir of $\frak a_2$. We will denote these two singlets by $S_2$ and $S_3$. These operators have $(R_0,{\cal F})=(2,-2)$ and
$(3,-3)$, respectively. In the deformed Lagrangian, the vacuum expectation value produces a \emph{linear} term $P^5$ in the superpotential. Its
F-term triggers expectation values for $Y$ and $Q$ that Higgs $(\su_3)_g\to \su_2$. We then integrate out the massive fields. The Higgs mechanism and the mass terms
remove $Q$, $Y_5$, $P^I$ and $\tilde P$, and they pair up
the third-color components of $A_{1}$ and $A_{2}$ 
with $\Phi^4$ and $\Phi^5$, and $S_2$ with $Z_3$. We are left with the theory
$\mathcal{T}^{(\su_3)}_1$ of Table~\ref{tab:tube} and Figure~\ref{fig:tube}, with the superpotential
\begin{align} \label{eq:Wtube}
\begin{split}
W_{\mathcal{T}^{(\su_3)}_1}&=A_2\,Y_m\,\Phi^m+A_1\,\tilde Y\,\tilde\Phi+\bar\phi\,\det A_1+S_3\,Y_4Z\,,\\
&=A_2\,Q_L\,Q_R+A_1\,\tilde Q_L\,\tilde Q_R+b_a\,\det\,A_1+b_q\,\det\, q\,.
\end{split}
\end{align}
Here we wrote the superpotential both in terms of the field names inherited from the Lagrangian of $T_{\su_3}$ and in terms of a more natural set of field names for ${{\cal T}^{(\su_3)}_{z=1}}$.
The gauge group is $(\su_2)_L\times (\su_2)_R$, where $(\su_2)_L$ is the unbroken part of
$(\su_3)_g$ and $(\su_2)_R\equiv (\su_2)_g$.

\begin{table}[h]
\centering
\renewcommand{\arraystretch}{1.25}
\begin{tabular}{|c|c|ccc|cc|cc|c|}
\hline
Old field & New field & $(\su_2)_L$ & $(\su_2)_R$ & $(\su_3)_C$ & $B_1$ & $B_2$ & $R_0$ & ${\cal F}$ & ${\cal J}$ \\
\hline
$A_1$ & $A_1$ & ${\bf 2}$ & ${\bf 2}$ & ${\bf 1}$ & $\phantom{-}0$ & $\phantom{-}0$ & $\phantom{-}\frac12$ & $-\frac12$ & $-\frac14$\\
$A_2$ & $A_2$ & ${\bf 2}$ & ${\bf 2}$ & ${\bf 1}$ & $\phantom{-}0$ & $\phantom{-}0$ & $\phantom{-}\frac32$ & $-\frac32$ & $\phantom{-}\frac14$\\
$Y_m$ & $Q_L$ & ${\bf 2}$ & ${\bf 1}$ & $\bar{\bf 3}$ & $\phantom{-}\frac12$ & $\phantom{-}\frac12$ & $\phantom{-}\frac12$ & $\phantom{-}\frac12$ & $-\frac14$\\
$Y_4$ & $q_1$ & ${\bf 2}$ & ${\bf 1}$ & ${\bf 1}$ & $-\frac12$ & $\phantom{-}\frac12$ & $-\frac12$ & $\phantom{-}\frac32$ & $\phantom{-}\frac14$\\
$Z$ & $q_2$ & ${\bf 2}$ & ${\bf 1}$ & ${\bf 1}$ & $\phantom{-}\frac12$ & $-\frac12$ & $-\frac12$ & $\phantom{-}\frac32$ & $\phantom{-}\frac14$\\
$\tilde Y$ & $\tilde Q_L$ &  ${\bf 2}$ & ${\bf 1}$ & ${\bf 1}$ & $-\frac32$ & $-\frac32$ & $\phantom{-}\frac32$ & $-\frac12$ & $\phantom{-}\frac14$\\
$\Phi^m$ & $Q_R$ &  ${\bf 1}$ & ${\bf 2}$ & ${\bf 3}$ & $-\frac12$ & $-\frac12$ & $\phantom{-}0$ & $\phantom{-}1$ & $\phantom{-}0$\\
$\tilde\Phi$ & $\tilde Q_R$ & ${\bf 1}$ & ${\bf 2}$ & ${\bf 1}$ & $\phantom{-}\frac32$ & $\phantom{-}\frac32$ & $\phantom{-}0$ & $\phantom{-}1$ & $\phantom{-}0$\\
$\bar\phi$ & $b_a$ & ${\bf 1}$ & ${\bf 1}$ & ${\bf 1}$ & $\phantom{-}0$ & $\phantom{-}0$ & $\phantom{-}1$ & $\phantom{-}1$ & $\phantom{-}\frac12$\\
$S_3$ & $b_q$ & ${\bf 1}$ & ${\bf 1}$ & ${\bf 1}$ & $\phantom{-}0$ & $\phantom{-}0$ & $\phantom{-}3$ & $-3$ & $-\frac12$\\
\hline
\end{tabular}
\caption{The flux-one $\frak a_2$ tube ${{\cal T}^{(\su_3)}_{z=1}}$. $B_{1,2}$ are the Cartan generators
of the emergent puncture symmetry $(\frak{su}_3)_B$. These two symmetries are certain linear combinations of $(\u1)_y$, $(\u1)_z$, and a $\u1$ appearing in the decomposition $\u1\times \su_3\times \su_2\subset\su_5$.
The symmetry ${\cal J}$ is emergent here in the IR. It will be identified with the geometric isometry of the two-punctured sphere. The leftmost column gives the field names inherited from the $T_{\su_3}$ Lagrangian, while the next column gives field names which are more natural for ${{\cal T}^{(\su_3)}_{z=1}}$.}
\label{tab:tube}
\end{table}

The UV Lagrangian of the theory ${{\cal T}^{(\su_3)}_{z=1}}$ has four non-anomalous non-R $\u1$'s. Two of them, $B_{1,2}$, are the Cartan generators of the emergent puncture symmetry $(\su_3)_B$. We also have the symmetry ${\cal F}$, the Cartan generator of the $6d$ $(1,0)$ $\su_2$ global symmetry. The last symmetry, which is denoted by ${\cal J}$ in Table~\ref{tab:tube}, is emergent after the closure of the puncture. This symmetry is not present in the three-punctured sphere. We identify it with the geometric isometry symmetry of the two-punctured sphere. This identification will be further motivated by anomaly computations in Section~\ref{sec:g}. For later reference, let us quote here the 't~Hooft anomalies of the various symmetries,
\be
&&{\rm Tr}\,R_0={\rm Tr}\,R_0^3=-8\,,\quad {\rm Tr}\,{\cal F}=6\,,\quad {\rm Tr}\,{\cal F}^3=-18\,,\quad
{\rm Tr}\,R_0^2{\cal F}=8\,,\quad {\rm Tr}\,R_0{\cal F}^2=0\,,\nonumber\\
&&{\rm Tr}\,{\cal J}^2R_0=0\,,\qquad {\rm Tr}\,{\cal J}^2{\cal F}=-\frac12\,,\qquad
{\rm Tr}\,{\cal J}\,(\su_3)_B^2=-{\rm Tr}\,{\cal J}\,(\su_3)_C^2=\frac14\,,\\
&&{\rm Tr}\,{\cal J}={\rm Tr}\,{\cal J}^3={\rm Tr}\,{\cal J}\,R_0\,{\cal F}={\rm Tr}\,{\cal J}R_0^2={\rm Tr}\,{\cal J}{\cal F}^2=0\,.\nonumber
\ee 
The moment map operators for the two punctures are as follows. First, for the puncture with the symmetry manifest in the UV,
\begin{equation}\label{eq:muC}
(\mu_C)^m_n=(Q_R)^m\,A_1\,(Q_L)_n-\tfrac13\,\delta^m_n\,(Q_R)^r\,A_1\,(Q_L)_r\,,
\end{equation}
and we note that $\mu_C$ has ${\cal J}=-\frac12$. The eight components of $\mu_B$ are
\begin{equation}
(A_1\,q_{(i})(A_2\,q_{j)})\,,\qquad (A_2\,q_i)\tilde Q_R\,,\qquad \tilde Q_L\,q_i\,, \qquad b_a\,,
\end{equation}
and all have ${\cal J}=+\frac12$. 

In Appendix \ref{sec:N4} we will perform yet another check of the tube by relating it to ${\cal N}=4$ $\su_3$ SYM.\footnote{Let us also comment here that a construction manifesting {\it both} puncture symmetries can be engineered by the same procedure of closing a puncture, but starting with~\cite{Gadde:2015xta}. However, the construction of the $T_{\su_3}$ theory itself in~\cite{Gadde:2015xta} involves gauging of emergent symmetries.}

\subsection{Lagrangian of \texorpdfstring{$D_4(\mathfrak{su}_3)$}{D4(su3)}}\label{sec:a2:lagp4}

Having derived the theory for the two-punctured sphere, we can now proceed to check our general claim \eqref{eq:mainclaim} for the case $\g=\su_3$. We will do so in three steps. First, in this subsection we will discuss the special case of $p=4$.
In the next subsection we will generalize to $p=3z+1$, and in the last subsection to other values of $p$.

In the case of $p=4=3\times 1+1$, \eqref{eq:mainclaim} implies that gauging the $\su_3$ symmetry of a maximal puncture will produce the $D_4(\su_3)$ Argyres--Douglas theory and a pair of free fields ($r_\g\times z=2\times 1=2$). Let us work this out in detail. 

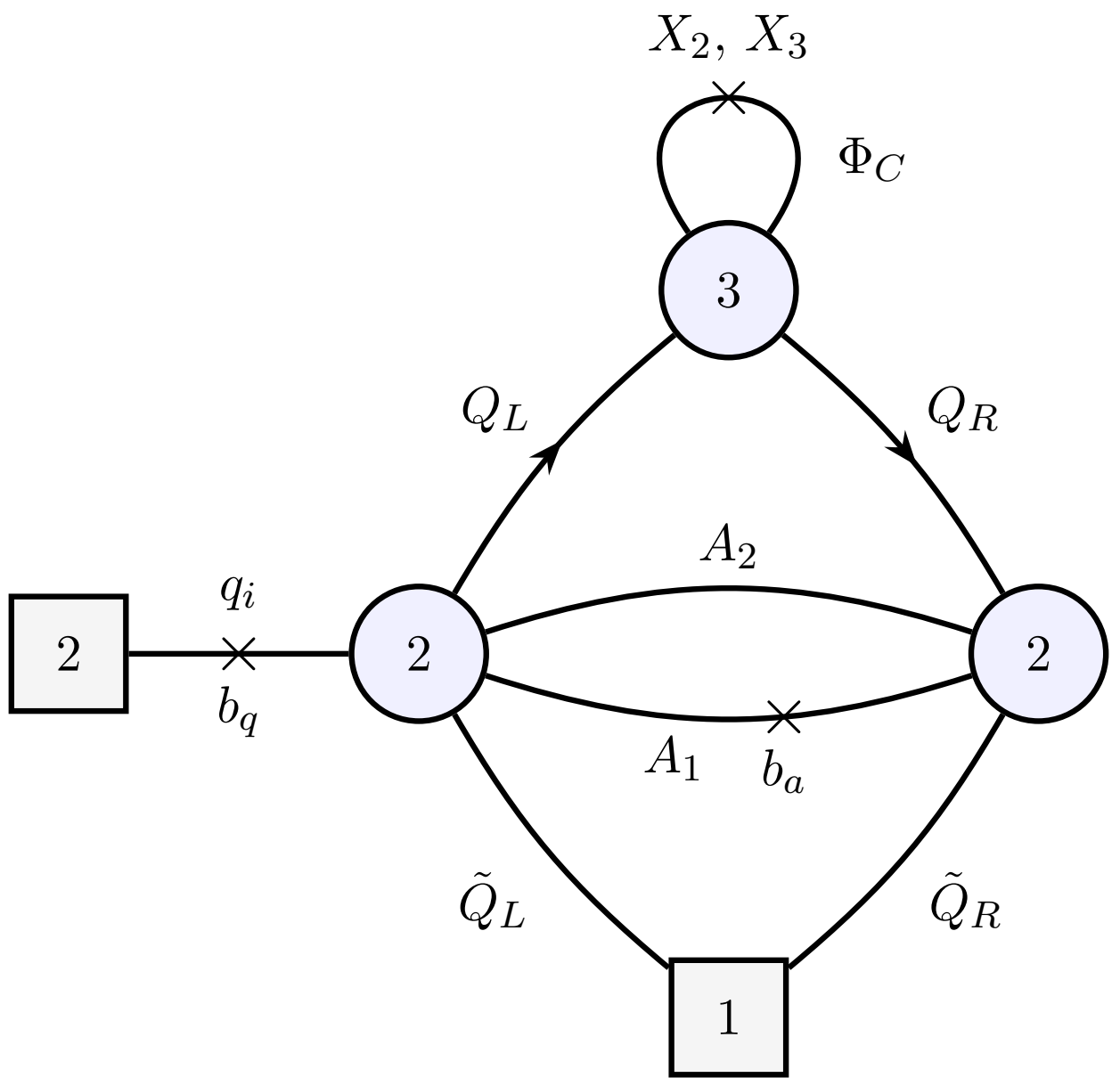
\begin{figure}[t]
\centering
\begin{tikzpicture}[>=Stealth,
 gauge/.style={circle,draw,very thick,minimum size=10mm,fill=blue!6},
 flav/.style={rectangle,draw,very thick,minimum size=8.5mm,fill=black!4},
 flip/.style={font=\Large,inner sep=0pt},
 midarrow/.style={postaction={decorate},decoration={markings,mark=at position 0.55 with {\arrow{Stealth[length=2.6mm]}}}}]
\node[gauge] (L) at (0,0) {$2$};
\node[gauge] (R) at (4.6,0) {$2$};
\node[gauge] (C) at (2.3,2.7) {$3$};
\node[flav] (U) at (2.3,-2.7) {$1$};
\node[flav] (X) at (-2.6,0) {$2$};
\draw[very thick] (L) to[bend left=18] node[above]{$A_2$} (R);
\draw[very thick] (L) to[bend right=18] coordinate[pos=0.62] (fA) node[pos=0.38,below]{$A_1$} (R);
\node[flip] at (fA) {$\times$};
\node[below=3pt] at (fA) {$b_a$};
\draw[very thick,midarrow] (L) to[bend left=10] node[above left]{$Q_L$} (C);
\draw[very thick,midarrow] (C) to[bend left=10] node[above right]{$Q_R$} (R);
\draw[very thick] (C) to[out=125,in=55,looseness=7] coordinate[pos=0.5] (fC) (C);
\node[flip] at (fC) {$\times$};
\node[above=4pt] at (fC) {$X_2,\,X_3$};
\node[right=9pt] at ($(C.north)+(0.35,0.45)$) {$\Phi_C$};
\draw[very thick] (L) to[bend right=10] node[below left]{$\tilde Q_L$} (U);
\draw[very thick] (R) to[bend left=10] node[below right]{$\tilde Q_R$} (U);
\draw[very thick] (X) -- coordinate[pos=0.5] (fS) node[pos=0.5,above=5pt]{$q_i$} (L);
\node[flip] at (fS) {$\times$};
\node[below=3pt] at (fS) {$b_q$};
\end{tikzpicture}
\caption{The ${\cal N}=1$ Lagrangian for $D_4(\su_3)$: the tube ${{\cal T}^{(\su_3)}_{z=1}}$ of
Figure~\ref{fig:tube} with its manifest puncture $(\su_3)_C$ gauged by an ${\cal N}=2$
vector multiplet. All circles are gauge nodes; the loop on the $\su_3$ node is
its adjoint $\Phi_C$, coupled through $W\supset{\rm Tr}\,\Phi_C\mu_C$. Crosses
denote flip fields: $b_a$ flips $\det A_1$, $b_q$ flips the meson $\det q$, and $X_2$,
$X_3$ flip ${\rm Tr}\,\Phi_C^2$ and ${\rm Tr}\,\Phi_C^3$. The squares labeled $2$
and $1$ are the manifest subgroup $(\su_2)_{B_1-B_2}\times (\u1)_{B_1+B_2}$ of the flavor
symmetry $(\su_3)_B$, which is emergent. The superpotential is given in Eq.~\eqref{eq:WD4}.}
\label{fig:D4}
\end{figure}

Take the Lagrangian for ${{\cal T}^{(\su_3)}_{1}}$ that we have derived and gauge the \emph{manifest} puncture $(\su_3)_C$ with an ${\cal N}=2$
vector multiplet. The ${\cal N}=2$ vector multiplet contains an ${\cal N}=1$ vector multiplet and an ${\cal N}=1$ chiral field, which we will denote by $\Phi_C$.
The resulting theory is a
Lagrangian with gauge group $(\su_2)_L\times (\su_2)_R\times (\su_3)_C$. We will discuss this shortly using $a$-maximization, but the operators which fall below the unitarity bound and become free are $\Tr\,\Phi_C^2$ and $\Tr\,\Phi_C^3$. After removing these operators by flipping them with chiral fields $X_2$ and $X_3$, the claim \eqref{eq:mainclaim} is thus that the Lagrangian for $D_4(\su_3)$ is the gauge theory described above, with the superpotential
\begin{equation}
W=W_{{{\cal T}^{(\su_3)}_{1}}}+{\rm Tr}\,\Phi_C\,\mu_C+X_2\,{\rm Tr}\,\Phi_C^2+X_3\,{\rm Tr}\,\Phi_C^3\,,
\label{eq:WD4}
\end{equation}
where $\mu_C$ is defined in \eqref{eq:muC}.
This theory is also depicted in Figure~\ref{fig:D4}.
The flavor symmetry is the emergent $(\su_3)_B$.
Let us analyze the rest of the symmetries in detail. The field $\Phi_C$ has
$(R_0,{\cal F},{\cal J})=(1,-1,\frac12)$. The $(\su_3)_C$ gauge anomaly kills
one combination of $R_0$, ${\cal F}$, ${\cal J}$. We denote the two non-anomalous combinations by
\begin{equation}
2I_3=R_0+{\cal F}\,,\qquad r=R_0-\frac14\,{\cal F}-\frac32\,{\cal J}\,.
\label{eq:D4map}
\end{equation}
These in fact become the Cartan generators of the $\su_2\times \u1$ ${\cal N}=2$ R-symmetry of the $D_4(\su_3)$ Argyres--Douglas theory. The 't~Hooft anomalies are
\begin{equation}
{\rm Tr}\,(2I_3)\,(\su_3)_B^2=0\,,\qquad {\rm Tr}\,r\,(\su_3)_B^2=-\frac94\,,
\end{equation}
{\it i.e.}\ $k_{\su_3}=\frac92$, the flavor central charge of $D_4(\su_3)$. The
charges of all the fields under the non-anomalous symmetries are listed in Table~\ref{tab:D4}.

\begin{table}[h]
\centering
\renewcommand{\arraystretch}{1.3}
\begin{tabular}{|c|ccccccccccccc|}
\hline
 & $A_1$ & $A_2$ & $Q_L$ & $q_1$ & $q_2$ & $\tilde Q_L$ & $Q_R$ & $\tilde Q_R$ & $b_a$ & $b_q$ & $\Phi_C$ & $X_2$ & $X_3$\\
\hline
$2I_3$ & $0$ & $0$ & $1$ & $1$ & $1$ & $1$ & $1$ & $1$ & $2$ & $0$ & $0$ & $2$ & $2$\\
$r$ & $1$ & $\frac32$ & $\frac34$ & $-\frac54$ & $-\frac54$ & $\frac54$ & $-\frac14$ & $-\frac14$ & $0$ & $\frac92$ & $\frac12$ & $1$ & $\frac12$\\
$R_{\rm IR}$ & $\frac13$ & $\frac12$ & $\frac{11}{12}$ & $\frac14$ & $\frac14$ & $\frac{13}{12}$ & $\frac7{12}$ & $\frac7{12}$ & $\frac43$ & $\frac32$ & $\frac16$ & $\frac53$ & $\frac32$\\
\hline
\end{tabular}
\caption{The charges of the $D_4(\su_3)$ Lagrangian (scalar components).
$R_{\rm IR}=\frac43I_3+\frac13r$ is the superconformal R-symmetry, which
$a$-maximization selects.}
\label{tab:D4}
\end{table}

Let us now perform $a$-maximization~\cite{Intriligator:2003jj} for the theory without the flip fields $X_{2,3}$, namely with $W=W_{{{\cal T}^{(\su_3)}_{1}}}+{\rm Tr}\,\Phi_C\,\mu_C$. The non-anomalous R-symmetries form the one-parameter family $R_x=x\,(2I_3)+(1-x)\,r$, with $2I_3$ and $r$ given in Table~\ref{tab:D4}; the ${\cal N}=2$ value is $x=\frac23$. The trial central charge is
\begin{equation}
a(x)=\frac{1}{256}\Big(-2907\,x^3+1971\,x^2+1131\,x-291\Big)\,,
\label{eq:aD4}
\end{equation}
which is maximized at
\begin{equation}
x_*=\frac{219+2\sqrt{42433}}{969}\simeq 0.651\,,\qquad a(x_*)\simeq1.869\,.
\end{equation}
At this point $R(\Phi_C)=\frac{1-x_*}{2}\simeq0.174$, so that
\begin{equation}
R\big({\rm Tr}\,\Phi_C^2\big)\simeq0.349\,,\qquad R\big({\rm Tr}\,\Phi_C^3\big)\simeq0.523\,,
\end{equation}
{\it i.e.}\ $\Delta=\frac32R\simeq0.52$ and $0.78$. Both operators fall strictly below the unitarity bound $\Delta=1$ ($R=\frac23$). Thus, in the IR, these operators become free fields decoupled from the interacting theory. The other low-lying gauge-invariant operators remain above the bound; for example, $R({\rm Tr}\,A_1A_2)\simeq0.87$, $R(\det A_2)\simeq1.05$ and $R(b_q)\simeq1.57$. Following~\cite{Kutasov:2003iy}, we remove the two free operators and repeat the maximization. Equivalently, we flip them with $X_2$ and $X_3$~\cite{Benvenuti:2017lle}, which leads to the superpotential~\eqref{eq:WD4}. The $a$-maximization after the removal of the free fields gives exactly $x=\frac23$, and no further operator violates the bound. At $x=\frac23$ the two decoupled operators have $R=\frac13$ and $\frac12$, {\it i.e.}\ $\Delta=\frac12$ and $\frac34$. The conformal anomalies after removing the free fields are
\begin{equation}
a=\frac{15}{8}\,,\qquad c=2\,,
\end{equation}
the central charges of $D_4(\su_3)$~\cite{Cecotti:2013lda}.
The Coulomb branch of $D_4(\su_3)$ has operators of dimensions
$\{2-\frac34,\ 3-\frac34,\ 3-\frac32\}=\{\frac54,\frac94,\frac32\}$. The
Lagrangian has natural candidates for these operators with $2I_3=0$:
\begin{equation}
{\rm Tr}\,A_1A_2:\ \Delta=\tfrac54\,,\qquad b_q:\ \Delta=\tfrac94\,,\qquad
\det A_2:\ \Delta=\tfrac32\,.
\end{equation}
The roots of the $(\su_3)_B$ moment map have $R_{\rm IR}=\frac43$, {\it i.e.}\ $\Delta=2$,
as required for ${\cal N}=2$ moment map operators. It is thus natural to conjecture that the Lagrangian~\eqref{eq:WD4} flows to
$D_4(\su_3)$, with the flavor symmetry enhanced to $\su_3$.

A sharp test of this conjecture is the computation of indices, and in particular the Schur limit~\cite{Gadde:2011ik,Gadde:2011uv} of the
superconformal index. For Argyres--Douglas theories, a lot is known about the Schur index~\cite{Buican:2015ina,Cordova:2015nma,Buican:2015tda,Song:2015wta,Buican:2017uka} because of the relation of these theories to chiral algebras through the VOA/${\cal N}=2$ SCFT correspondence of~\cite{Beem:2013sza}.
In particular, the chiral algebra associated with $D_4(\su_3)$ is
$(\widehat{\frak{su}}_3)_{-9/4}$~\cite{Xie:2016evu,Song:2017oew,Creutzig:2017qyf}.  This is a boundary
admissible level, $k+h^\vee=h^\vee/p$, for which the formula of \cite{Song:2017oew}, proved in \cite{Kac:2017}, gives the vacuum character in closed form.  The expected Schur index is therefore
\begin{equation}
{\cal I}_{\rm Schur}={\rm PE}\left[\frac{q-q^4}{(1-q)(1-q^4)}\,\chi_{\bf 8}(b)\right]
=1+8q+44q^2+192q^3+718q^4+2400q^5+\dots\,,
\label{eq:SchurD4}
\end{equation}
where $b$ are the $(\su_3)_B$ fugacities and the expansion is at $b=1$. Since $2I_3$ and $r$ are symmetries of the Lagrangian, the full ${\cal N}=2$
superconformal index ${\cal I}(p,q,t)$ can be computed from it explicitly~\cite{Maruyoshi:2016tqk,Maruyoshi:2016aim,Agarwal:2016pjo}. For example, a chiral
multiplet with charges $(2I_3,\,r)$ contributes $\Gamma_e\big((pq/t)^{r/2}t^{I_3}z;q,p\big)$,
where $z$ are the gauge and flavor fugacities and $\Gamma_e(z;q,p)$ is the standard elliptic Gamma function~\cite{Dolan:2008qi}. Setting $t=q$, one obtains the Schur index. For
an ${\cal N}=2$ SCFT, the result must be independent of $p$ and equal to the Schur index, even though the contribution of each field might depend on $p$.
We have evaluated the Schur limit both in an expansion in $p$ and $q$ and numerically, and found the result to be
indeed independent of $p$ and equal to~\eqref{eq:SchurD4}.  This also
holds for generic values of the fugacities $b$.

We can also compute the Macdonald index to get (here $t=qT$)
\begin{align}
\begin{split}
\mathcal{I}_{\textrm{Mac}} ={}&1+8qT
 +q^2(9T+35T^2)
 +q^3(9T+72T^2+111T^3)\\
&+q^4(9T+117T^2+306T^3+286T^4)\\
&+q^5(9T+153T^2+665T^3+936T^4+637T^5)\\
&+q^6(9T+198T^2+1118T^3+2422T^4
                +2331T^5+1274T^6)
 +O(q^7).
\end{split}
\label{eq:MacD4sur}
\end{align}
Our result also agrees with the conjectural expression for the Macdonald index ($p \to 0$ limit) computed in \cite{Kang:2026ona, Kang:2026ney}, which is obtained by computing the Hilbert series of the arc space of a certain graded filtered affine scheme \cite{Bhargava:2023hsc, Kang:2025zub, Kang:2026nge, Andrews:2025krn}. This method is essentially a more efficient version of \cite{Song:2016yfd, Agarwal:2018zqi}, where the Macdonald index is computed by introducing a specific filtration of the associated VOA.\footnote{In fact, for the case of $D_p(\g)$ theory (and their Higgsed ones), the bifiltered scheme of \cite{Kang:2026ona} is identical to the associated \emph{scheme} (which contains the nilpotency structure) \cite{Arakawa:2010ni, Beem:2017ooy, RastelliTalk} of the VOA upon forgetting the filtration. However, they are not in general identical.} See also \cite{ArabiArdehali:2025fad} on the filtration for the associated VOA. This provides further evidence for both of the conjectures.

\subsection{The \texorpdfstring{$D_{3z+1}(\mathfrak{su}_3)$}{D3z+1(su3)} theories}

The construction of the $D_p(\su_3)$ Argyres--Douglas theories with $p=3\,z+1$, $z\in {\mathbb N}$, is now straightforward following \eqref{eq:mainclaim}. 
A tube of flux $z$, $\mathcal{T}^{(\su_3)}_z$, is obtained by gluing $z$ copies of $\mathcal{T}^{(\su_3)}_1$ with
${\cal N}=2$ vector multiplets.  Gauging one of the two maximal punctures of $\mathcal{T}^{(\su_3)}_z$ then gives
$D_{3z+1}(\su_3)$ plus $2\,z$ free fields according to \eqref{eq:mainclaim}.

\begin{figure}[t]
\centering
\begin{tikzpicture}[>=Stealth,scale=0.95]
\def\tube#1#2#3#4#5{%
 \fill[green!8] (#1,0.7) rectangle (#2,-0.7);
 \draw[very thick] (#1,0.7) -- (#2,0.7); \draw[very thick] (#1,-0.7) -- (#2,-0.7);
 \node at ({(#1+#2)/2},0) {#3};
 \node[font=\footnotesize] at ({#1+0.35},-1.05) {#4};
 \node[font=\footnotesize] at ({#2-0.35},-1.05) {#5};}
\tube{0}{2.4}{$\mathcal{T}$}{$C$}{$B$}
\tube{2.4}{5.0}{$\mathcal{T}$}{$B$}{$C$}
\tube{5.0}{7.4}{$\mathcal{T}$}{$C$}{$B$}
\node at (8.0,0) {$\cdots$};
\tube{8.6}{11.0}{$\mathcal{T}$}{$B$}{$C$}
\draw[very thick,blue!70!black,fill=blue!70!black!25] (0,0) ellipse (0.2 and 0.7);
\draw[very thick,blue!70!black,fill=blue!70!black!25] (11.0,0) ellipse (0.2 and 0.7);
\foreach \x/\c in {2.4/red!70!black,5.0/blue!70!black,8.6/red!70!black}{\draw[very thick,\c,fill=\c!25] (\x,0) ellipse (0.2 and 0.7);}
\node[blue!70!black,font=\footnotesize,align=center] at (-0.95,0) {${\cal N}=2$\\vector};
\node[blue!70!black,font=\footnotesize,align=center] at (11.85,0) {flavor\\$(\su_3)_C$};
\node[red!70!black,font=\footnotesize,align=center] at (2.4,1.25) {$B$--$B$\\emergent};
\node[blue!70!black,font=\footnotesize,align=center] at (5.0,1.25) {$C$--$C$\\Lagrangian};
\node[red!70!black,font=\footnotesize,align=center] at (8.6,1.25) {$B$--$B$\\emergent};
\node[font=\footnotesize] at (-0.95,-1.75) {${\cal J}$:};
\node[font=\footnotesize] at (1.2,-1.75) {${\cal J}_1-\frac{z-1}2\,{\cal F}_1$};
\node[font=\footnotesize] at (3.7,-1.75) {$-{\cal J}_2-\frac{z-3}2\,{\cal F}_2$};
\node[font=\footnotesize] at (6.2,-1.75) {${\cal J}_3-\frac{z-5}2\,{\cal F}_3$};
\node[font=\footnotesize] at (8.0,-1.75) {$\cdots$};
\node[font=\footnotesize] at (9.8,-1.75) {$-{\cal J}_z+\frac{z-1}2\,{\cal F}_z$};
\end{tikzpicture}
\caption{The tube $\mathcal{T}^{(\su_3)}_z$ built from $z$ copies of $\mathcal{T}=\mathcal{T}^{(\su_3)}_1$, glued alternately
along their emergent punctures $B$ and their manifest punctures $C$. The manifest puncture $C$ at the left end is gauged with an ${\cal N}=2$ vector multiplet. The figure shows the case of even $z$, in which the chain ends with a $C$ puncture, the flavor symmetry $(\su_3)_C$; for odd $z$ it ends with a $B$ puncture. Below each copy we show how the isometry ${\cal J}$ of Eq.~\eqref{eq:Jchain} acts on it, in terms of the symmetries ${\cal J}_k$ and ${\cal F}_k$ of the $k$-th copy.}
\label{fig:chain3l}
\end{figure}
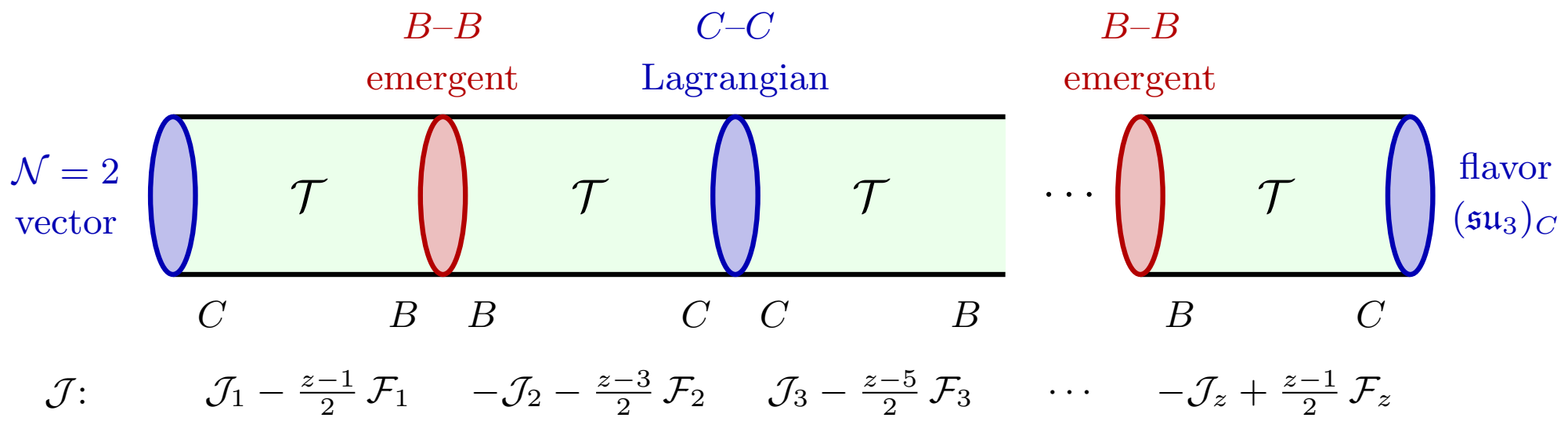

Since one of the puncture symmetries of the Lagrangian of $\mathcal{T}^{(\su_3)}_1$ is manifest and the other one is emergent, for $z>1$ we necessarily need to gauge emergent symmetries in order to obtain a description of the Argyres--Douglas theory. An example of a gluing is depicted in Figure~\ref{fig:chain3l}.
The ${\cal F}$ and $R_0$ symmetries are non-anomalous for each of the gaugings: these symmetries of the two glued blocks are simply identified. The geometric isometry ${\cal J}$ of the glued copies, on the other hand, has to be identified properly. Let us work this out in detail.

Let $\mathcal{T}\equiv \mathcal{T}^{(\su_3)}_1$ be the tube of Table~\ref{tab:tube} and Figure~\ref{fig:tube}. Consider first two copies of $\mathcal{T}$ glued along their $B$ punctures with an ${\cal N}=2$ vector multiplet: we gauge the diagonal $(\su_3)_B$ with an ${\cal N}=1$ vector multiplet, and add the adjoint chiral field $\Phi$ with superpotential coupling $W\supset{\rm Tr}\,\Phi\,(\mu_B-\mu'_B)$. Before the gluing, each copy has its own symmetries ${\cal F}_i$ and ${\cal J}_i$ ($i=1,2$), with the charges of Table~\ref{tab:tube}. The moment maps $\mu_B$ have $({\cal F},{\cal J})=(1,\frac12)$, and each $B$ puncture contributes ${\rm Tr}\,{\cal F}\,\su_3^2=\frac32$ and ${\rm Tr}\,{\cal J}\,\su_3^2=\frac14$. Consider a general combination of the abelian symmetries,
\begin{equation}
U=a_1{\cal J}_1+a_2{\cal J}_2+f_1{\cal F}_1+f_2{\cal F}_2\,.
\end{equation}
The superpotential requires $U(\mu_B)=U(\mu'_B)=-U(\Phi)$, {\it i.e.}\ $\frac{a_1}2+f_1=\frac{a_2}2+f_2$. Requiring $U$ to be non-anomalous then gives
\begin{equation}
0={\rm Tr}\,U\,\su_3^2=\frac{a_1+a_2}4+\frac32\,(f_1+f_2)-3\Big(\frac{a_1}2+f_1\Big)=-\frac{a_1+a_2}2\,,
\end{equation}
which requires $a_2=-a_1$. Thus, besides $R_0$ and the diagonal ${\cal F}={\cal F}_1+{\cal F}_2$, exactly one combination of the two isometries survives the gluing. Up to the addition of a multiple of ${\cal F}$, it is
\begin{equation}
{\cal J}=\Big({\cal J}_1-\frac12{\cal F}_1\Big)-\Big({\cal J}_2-\frac12{\cal F}_2\Big)\,,\qquad {\cal J}(\Phi)=0\,.
\end{equation} The combination ${\cal J}$ has the property that exchanging the two puncture symmetries together with ${\cal J}\to -{\cal J}$ is an outer automorphism of the flavor symmetry.
We will refer to ${\cal J}$ as the isometry symmetry of the glued theory. 

Next, we glue the $C$ puncture of the second copy to the $C$ puncture of a third one, and continue alternating between $B$ and $C$ punctures:
\begin{equation}
\mathcal{T}^{(\su_3)}_z=\mathcal{T}\,\cup_B\,\mathcal{T}\,\cup_C\,\mathcal{T}\,\cup_B\,\mathcal{T}\,\cup_C\,\cdots\qquad(z\ \text{copies})\,.
\label{eq:chain}
\end{equation}
The seams thus alternate between gaugings of the emergent symmetry ($B$--$B$) and Lagrangian gaugings ($C$--$C$). At a $C$--$C$ seam, $\mu_C$ has ${\cal J}=-\frac12$ and each puncture contributes ${\rm Tr}\,{\cal J}\,\su_3^2=-\frac14$. The same computation again requires opposite coefficients for the two isometries. At every seam the coefficient of ${\cal F}$ increases by one unit. Hence, besides $R_0$ and ${\cal F}$, there is exactly one more non-anomalous $\u1$. Again up to a multiple of ${\cal F}$, it is
\begin{equation}
{\cal J}=\sum_{k=1}^z\Big(\epsilon_k\,{\cal J}_k+\kappa_k\,{\cal F}_k\Big)\,,\qquad \epsilon_k=(-1)^{k+1}\,,\qquad \kappa_k=-\frac z2+k-\frac12\,,
\label{eq:Jchain}
\end{equation}
where ${\cal J}_k$ and ${\cal F}_k$ act on the fields of the $k$-th copy of ${\cal T}$, counted from the left, with the charges of Table~\ref{tab:tube}. See Figure~\ref{fig:chain3l} for an illustration. The moment maps meeting at the $m$-th seam ($m=1,\dots,z-1$) have ${\cal J}=-\frac z2+m$, and the adjoint chiral field of the vector multiplet there has
\begin{equation}
(R_0,{\cal F},{\cal J})(\Phi_m)=\Big(1,\,-1,\,\frac z2-m\Big)\,.
\label{eq:Jseam}
\end{equation}
The symmetry \eqref{eq:Jchain} is naturally identified with the $\u1$ isometry of the tube of flux $z$, which rotates it around the axis through the two punctures. On each copy it acts as the isometry of that copy, up to a sign, combined with ${\cal F}$ with a coefficient $\kappa_k$ that grows by one unit for each unit of flux along the tube, from $-\frac{z-1}{2}$ at the left end to $\frac{z-1}{2}$ at the right end.\footnote{Such a mixing of the isometry with the flux mimics the equivariant treatment of the flux when the surface is fibered over spacetime~\cite{Bah:2019rgq,Hosseini:2020vgl}.} We will show in Section~\ref{sec:g} that the anomalies of ${\cal J}$ are indeed consistent with those of the tube theory of flux $z$, $\mathcal{T}^{(\su_3)}_z$.
 
Finally, we gauge the manifest puncture $C$ at the left end of the chain with an ${\cal N}=2$ vector multiplet, exactly as for $D_4(\su_3)$ in Section~\ref{sec:a2:lagp4} (see Figure~\ref{fig:chain3l}). The adjoint chiral field $\Phi_0$ of this vector multiplet has the charges \eqref{eq:Jseam} with $m=0$. The remaining flavor symmetry is $(\su_3)_B$ for odd $z$ and $(\su_3)_C$ for even $z$.
In particular, ${\cal J}(\Phi_0)=\frac z2$, and the $(\su_3)_C$ gauge anomalies of the symmetries are
\begin{equation}
{\rm Tr}\,R_0\,(\su_3)_C^2=\frac32\,,\qquad {\rm Tr}\,{\cal F}\,(\su_3)_C^2=-\frac32\,,\qquad
{\rm Tr}\,{\cal J}\,(\su_3)_C^2=\frac{3z+2}4\,.
\end{equation}
The non-anomalous combinations are
\begin{equation}
2I_3=R_0+{\cal F}\,,\qquad r=R_0-\frac{1}{3z+1}\,{\cal F}-\frac{6}{3z+1}\,{\cal J}\,,
\label{eq:D3z1map}
\end{equation}
which reduce to \eqref{eq:D4map} for $z=1$. For the remaining flavor symmetry, ${\rm Tr}\,(2I_3)\,\su_3^2=0$ and ${\rm Tr}\,r\,\su_3^2=-\frac{9z}{3z+1}$, {\it i.e.}\ $k_{\su_3}=\frac{18z}{3z+1}$, the flavor central charge of $D_{3z+1}(\su_3)$~\cite{Cecotti:2012jx,Cecotti:2013lda}. We will now argue that the construction indeed describes $D_{3z+1}(\su_3)$ together with $2\,z$ decoupled free chiral fields.

To give evidence for this statement, let us first discuss the Coulomb branch operators. We assume for now that the R-symmetries \eqref{eq:D3z1map} are the superconformal ones; we will justify this below. 
Identifying $I_3$ with the Cartan generator of the $\su_2$ part of the ${\cal N}=2$ R-symmetry, the Coulomb branch operators should have $I_3=0$. The natural operators with $2I_3=0$ are the Casimirs ${\rm Tr}\,\Phi_m^2$ and ${\rm Tr}\,\Phi_m^3$ of the $z$ vector multiplets ($m=0,\dots,z-1$), and, in each copy, the three operators ${\rm Tr}\,A_1A_2$, $\det A_2$ and $b_q$. With $\Delta\equiv\frac r2$, and labeling the copies of ${\cal T}$ by $k=1,\dots,z$, Eqs.~\eqref{eq:Jchain}--\eqref{eq:D3z1map} give
\begin{equation}
\begin{split}
&\Delta\big({\rm Tr}\,\Phi_m^j\big)=\frac{j\,(3m+1)}{3z+1}\,,\qquad
\Delta\big({\rm Tr}\,A_1A_2\big)_k=\frac{6k-1}{3z+1}\,,\\
&\big\{\Delta(\det A_2)_k,\,\Delta(b_q)_k\big\}=\Big\{\frac{9k-3}{3z+1},\,\frac{9k}{3z+1}\Big\}\,,
\end{split}
\label{eq:dec_3z+1}
\end{equation}
where for odd $k$ the first value belongs to $\det A_2$ and the second to $b_q$, and for even $k$ the two are exchanged. The numerators $2+6m$ and $6k-1$ exhaust the integers $\equiv2$ mod $3$ between $2$ and $6z-1$, while $3+9m$, $9k-3$ and $9k$ exhaust the integers $\equiv0$ mod $3$ between $3$ and $9z$. Since $3z+1\equiv1$ mod $3$, no operator has $\Delta=1$, and the $5z$ operators split into two sets:
\begin{itemize}
\item $3z$ operators with $\Delta>1$, with dimensions $\{2-\frac{3s}{3z+1}\}_{s=1}^{z}\cup\{3-\frac{3s}{3z+1}\}_{s=1}^{2z}$. This is precisely the Coulomb branch spectrum of $D_{3z+1}(\su_3)$~\cite{Cecotti:2012jx,Cecotti:2013lda}.
\item $2z$ operators with dimensions $\{\frac{2+3n}{3z+1},\frac{3+3n}{3z+1}\}_{n=0}^{z-1}$, below the unitarity bound. These become free, in agreement with the $r_\g z=2z$ free fields of \eqref{eq:mainclaim}.
\end{itemize}
The decoupled operators are the ones localized close to the gauged end: the Casimirs of $\Phi_m$ with small $m$ and the operators of the copies with small $k$. For example, for $z=2$ they are ${\rm Tr}\,\Phi_0^2$, ${\rm Tr}\,\Phi_0^3$, ${\rm Tr}\,A_1A_2$ and $\det A_2$ of the first copy, with $\Delta=\frac27,\frac37,\frac57,\frac67$, while the Coulomb branch of $D_7(\su_3)$, $\{\frac87,\frac{11}7,\frac97,\frac{12}7,\frac{15}7,\frac{18}7\}$, comes from ${\rm Tr}\,\Phi_1^2$, ${\rm Tr}\,A_1A_2$ of the second copy, $b_q$ of the first copy, ${\rm Tr}\,\Phi_1^3$, and $b_q$ and $\det A_2$ of the second copy. As for $z=1$, we flip the $2z$ decoupled operators with chiral fields $X$.

After decoupling the free fields, the anomalies of the R-symmetries are
\begin{equation}
{\rm Tr}\,I_3={\rm Tr}\,I_3^3={\rm Tr}\,I_3\,r^2=0\,,\qquad {\rm Tr}\,r={\rm Tr}\,r^3=48(a-c)\,,\qquad {\rm Tr}\,(2I_3)^2r=16a-8c\,,
\end{equation}
with
\begin{equation}
a=\frac{3z(4z+1)}{2(3z+1)}\,,\qquad c=2z\,.
\end{equation}
These are the 't~Hooft anomalies of an ${\cal N}=2$ SCFT~\cite{Shapere:2008zf} with the central charges of $D_{3z+1}(\su_3)$~\cite{Cecotti:2013lda}. Consistently, summing over the Coulomb branch operators found above gives $2a-c=\frac14\sum(2\Delta-1)=\frac{z(6z+1)}{3z+1}$.

 Let us now show that the $2z$ operators listed above indeed decouple, by performing $a$-maximization~\cite{Intriligator:2003jj} in the theory without the flip fields. We define the non-anomalous trial R-symmetry as $R_x=x\,(2I_3)+(1-x)\,r$. Denote by ${\cal O}_i$, $i=1,\dots,2z$, the operators claimed to decouple. They have $2I_3=0$ and $r=2\Delta_i$, with $\Delta_i<1$ the dimensions found above, so that $R_x({\cal O}_i)=2\Delta_i\,(1-x)$. Since the anomalies of the theory without the ${\cal O}_i$ are those of $D_{3z+1}(\su_3)$, the trial central charge of the full theory is
\begin{equation}
a(x)=a_{{\cal N}=2}(x)+\sum_{i=1}^{2z}a_\chi\big(2\Delta_i(1-x)\big)\,,\qquad a_\chi(R)=\frac3{32}\Big(3(R-1)^3-(R-1)\Big)\,,
\label{eq:atrial}
\end{equation}
where $a_\chi(R)$ is the contribution of a chiral field of R-charge $R$, and
\begin{equation}
a_{{\cal N}=2}(x)=\frac{9z\,(1-x)}{4(3z+1)}\Big[3(6z+1)\,x^2-3(1-x)^2+1\Big]
\end{equation}
is the trial central charge computed from the anomalies above. It is maximized at $x=\frac23$, where $a_{{\cal N}=2}(\frac23)=a=\frac{3z(4z+1)}{2(3z+1)}$. The dimensions of the decoupling operators are $\Delta_i=\frac{n_i}{3z+1}$ with $n_i\in\{2,3,5,6,\dots,3z-1,3z\}$, so that the second term in \eqref{eq:atrial} is $\sum_{n}a_\chi\big(\frac{2n\,(1-x)}{3z+1}\big)$. For $z=1$, \eqref{eq:atrial} reduces to the polynomial~\eqref{eq:aD4}. The structure of the iterative maximization~\cite{Kutasov:2003iy} then follows from three simple observations.
\begin{itemize}
\item At $x=\frac23$ each ${\cal O}_i$ has $R=\frac23\Delta_i<\frac23$, where $a_\chi'(R)>0$. Therefore $a'(\frac23)<0$, and as long as some of the ${\cal O}_i$ have not been removed the maximum lies at $x<\frac23$.
\item For $x\leq\frac23$ the Coulomb branch operators have $R_x=2\Delta(1-x)\geq\frac23\Delta>\frac23$, so they never violate the unitarity bound.
\item At the maximum, at least one of the remaining ${\cal O}_i$ violates the bound. Indeed, if all of them had $\frac23\leq R\leq\frac43$, their contributions to $a'(x)$, $-2\Delta_i\,a_\chi'(R)$, would be non-negative. Since $a_{{\cal N}=2}'(x)>0$ for $\frac13<x<\frac23$, which is the relevant range, the point $x$ could not be a maximum.
\end{itemize}
The violating operators become free; we remove them by flipping them, and repeat the maximization. The procedure stops only once all the ${\cal O}_i$ have been removed, at which point the maximum is exactly at $x=\frac23$. Thus iterative $a$-maximization decouples precisely the $2z$ operators ${\cal O}_i$ and lands on $R_{\rm IR}=\frac43I_3+\frac13r$. We have also checked that the other low-lying gauge-invariant operators, such as the moment maps and $b_a$, stay above the bound throughout. As an illustration, for $z=2$ the first maximization gives $x_1\simeq0.652$. At this point ${\rm Tr}\,\Phi_0^2$, ${\rm Tr}\,\Phi_0^3$, and ${\rm Tr}\,A_1A_2$ and $\det A_2$ of the first copy have $R\simeq0.20,\,0.30,\,0.50,\,0.60$, while all other operators have $R\gtrsim0.80$. After flipping these four operators the maximum is at $x=\frac23$. For $z=3$, $x_1\simeq0.652$, and the six operators ${\rm Tr}\,\Phi_0^2$, ${\rm Tr}\,\Phi_0^3$, ${\rm Tr}\,\Phi_1^2$, and ${\rm Tr}\,A_1A_2$, $\det A_2$ and $b_q$ of the first copy have $R\simeq0.14,\,0.21,\,0.56,\,0.35,\,0.42,\,0.63$. All other operators have $R\gtrsim0.76$, and again a single step suffices. For larger values of $z$ (we have checked up to $z=40$), a second step is in general needed to decouple the few operators with $\Delta_i$ close to one.

Finally, let us mention that
although the construction involves an
emergent gauging, its superconformal index is well defined and can be computed.  The index of each copy of $\mathcal{T}=\mathcal{T}^{(\su_3)}_1$ is
invariant under the Weyl group of $(\su_3)_B$, so it can be
integrated against the $\su_3$ Haar measure together with an ${\cal N}=2$ vector multiplet. 
As in Section~\ref{sec:a2:lagp4}, we compute the Schur index by setting $t=q$ in the contribution of each field. For example, for $z=2$ the result is
independent of $p$, as expected, and equals
\begin{equation}
{\cal I}_{\rm Schur}={\rm PE}\left[\frac{q-q^7}{(1-q)(1-q^7)}\,\chi_{\bf 8}\right]\,,
\end{equation}
the vacuum character of $(\widehat{\frak{su}}_3)_{-18/7}$~\cite{Kac:2017}, as expected~\cite{Song:2015wta,Xie:2016evu,Song:2017oew,Creutzig:2017qyf}.

\subsection{The \texorpdfstring{$D_{3z+2}(\mathfrak{su}_3)$ and $D_{3z+3}(\mathfrak{su}_3)$}{D3z+2(su3) and D3z+3(su3)} theories}

Now we generalize our construction of the $D_{3z+1}(\mathfrak{su}_3)$ theory to the generic $D_p(\mathfrak{su}_3)$ theories with arbitrary $p$. We start with $D_{3z+3}(\su_3)$. The construction is almost identical. We glue the tubes with unit flux to build $\mathcal{T}_z^{(\mathfrak{su}_3)}$, and gauge one of the two maximal punctures together with the $D_3(\mathfrak{su}_3)$ theory, which is the $\mathcal{N}=2$ quiver tail $[\mathfrak{su}_3]-(\mathfrak{su}_2)-[\mathfrak{su}_1]$. We claim that the theory flows to the $D_{3z+3}(\mathfrak{su}_3)$ theory with extra decoupled free chirals under the renormalization group flow.
\begin{align}
\mathcal{T}_z^{(\mathfrak{su}_3)}/(\mathfrak{su}_3\,\oplus\,(\mathfrak{su}_3)-(\mathfrak{su}_2)-[1]\,\oplus V_{\mathcal{N}=2})\,\longrightarrow\,D_{3z+3}(\mathfrak{su}_3)\,\oplus\,(2z\text{ free chiral fields})\,.
\end{align}\noindent
In the class $\mathcal{S}$ language, we are replacing a regular puncture with an irregular puncture $I_{3,0}$. Conversely, one can also think that gluing a $z$-flux tube alters $I_{3,0}$ to $I_{3,3z}$.

The $D_3(\mathfrak{su}_3)$ theory is the $\mathfrak{su}_2$ gauge theory with eight fundamental half-hypermultiplets that are in the vector representation of the flavor $\mathfrak{so}_8$. We split them into six $B\,,\widetilde{B}$ that are in $(\mathbf{3}_{1,0},\overline{\mathbf{3}}_{-1,0})$ of $(\mathfrak{su}_3)_C\times (\mathfrak{u}_1)_{\mathcal{C}}\times(\mathfrak{u}_1)_\mathcal{A}\subset\mathfrak{so}_8$ among which $(\su_3)_C$ will be gauged, and the remaining two $h\,,\widetilde{h}$ that have non-trivial charges under $(\mathfrak{u}_1)_{\mathcal{A}}$. The matter content, along with their charges under the relevant symmetry groups, are summarized in Table \ref{tab:D3su3}. The relevant non-trivial anomaly coefficients of the $D_3(\su_3)$ theory are
\begin{equation}
\begin{gathered}
    \Tr\,R_0^3=1\,,\quad \Tr\,R_0^2\mathcal{F}=2\,,\quad \Tr\,R_0\mathcal{F}^2=-2\,,\quad \Tr\,R_0\mathcal{C}^2=-6\,,\quad \Tr\,R_0\mathcal{A}^2=-2\,,\\
    \Tr\,R_0(\su_3)_C^2=-1\,,\quad\Tr\,\mathcal{F}(\su_3)_C^2=1\,,\quad\Tr\,\mathcal{F}^3=-1\,,\quad\Tr\,R_0=-5\,,\quad \Tr\,\mathcal{F}=5\,.
    \end{gathered}
    \label{eq:anomD3A2}
\end{equation}

\begin{table}[h] 
    \centering
    \renewcommand{\arraystretch}{1.25}
    \begin{tabular}{|c|cc|cccc|}
        \hline
        field & $(\mathfrak{su}_3)_C$ & $(\mathfrak{su}_2)_\g$ & $R_0$ & $\mathcal{F}$ & $\mathcal{C}$ & $\mathcal{A}$  \\
        \hline
        $B$ & $\mathbf{3}$ & $\mathbf{2}$ & $\frac{1}{2}$ & $\phantom{-}\frac{1}{2}$ & $\phantom{-}1$ & $\phantom{-}0$ \\
        $\widetilde{B}$ & $\overline{\mathbf{3}}$ & $\mathbf{2}$ & $\frac{1}{2}$ & $\phantom{-}\frac{1}{2}$ & $-1$ & $\phantom{-}0$ \\
        $h$ & $\mathbf{1}$ & $\mathbf{2}$ & $\frac{1}{2}$ & $\phantom{-}\frac{1}{2}$ & $\phantom{-}0$ & $\phantom{-}1$\\
        $\widetilde{h}$ & $\mathbf{1}$ & $\mathbf{2}$ & $\frac{1}{2}$ & $\phantom{-}\frac{1}{2}$ & $\phantom{-}0$ & $-1$\\
        $\phi$ & $\mathbf{1}$ & $\mathbf{3}$ & $1$ & $-1$ & $\phantom{-}0$ & $\phantom{-}0$\\
        \hline
        $\Phi_C$ & $\mathbf{8}$ & $\mathbf{1}$ & $1$ & $-1$ & $\phantom{-}0$ & $\phantom{-}0$\\
        \hline
    \end{tabular}
    \caption{Matter content of $D_3(\su_3)$. $\phi$ is the $\mathcal{N}=1$ adjoint chiral in the $\mathcal{N}=2$ vector multiplet of $\su_2$. $\Phi_C$, the $\mathcal{N}=1$ chiral in the $\mathcal{N}=2$ vector multiplet of $(\su_3)_C$, has exactly the same charges under the global symmetries as $\phi$ due to the superpotential \eqref{eq:WD_3z+3}.}    
    \label{tab:D3su3}
\end{table}
\noindent After gauging, the full superpotential is
\begin{align}
    W=W_{\mathcal{T}_z^{(\mathfrak{su}_3)}}+\widetilde{B}\Phi_CB-B\phi\widetilde{B}+\widetilde{h}\phi h+\Tr\,\Phi_C\mu_C\,,
    \label{eq:WD_3z+3}
\end{align}
where $\phi$ is the $\mathcal{N}=1$ chiral multiplet in the $\mathcal{N}=2$ vector multiplet of the $\mathfrak{su}_2$ gauge group in $D_3(\mathfrak{su}_3)$, $\Phi_C$ is that of the $(\mathfrak{su}_3)_C$ vector, and $\mu_C$ is the moment map operator in $\mathcal{T}_z^{(\mathfrak{su}_3)}$ that is specified in \eqref{eq:WD4}. After the gauging, four anomaly-free $\mathfrak u_1$'s survive:
\begin{align}
    R_0+\mathcal{F}_\mathcal{T}+\mathcal{F}\,,\quad \mathcal{J}+\frac{z+2}{2}\mathcal{F}_\mathcal{T}+\mathcal{F}\,,\quad \mathcal{C}\,,\quad \mathcal{A}\,.
\end{align}
Here $\mathcal{F}_{\mathcal{T}}$ is the $\mathcal{F}$ flavor symmetry of the $\mathcal{T}_z^{(\su_3)}$, while $\mathcal{F}$ is that of the quiver tail.
Then, the $\mathcal{N}=1$ superconformal $R$-charge at the IR fixed point should be a linear combination of the $\mathfrak{u}_1$'s
\begin{align}
    R_{\text{IR}}=R_0+\mathcal{F}_\mathcal{T}+\mathcal{F}-x\left(\mathcal{J}+\frac{z+2}{2}\mathcal{F}_\mathcal{T}+\mathcal{F}\right)\,,
\end{align}
whose coefficient $x$ is fixed by $a$-maximization, while $\mathcal{C}$ and $\mathcal{A}$ do not mix because they are baryonic. The trial central charge $a$ is given by
\begin{align}
    a(x)=&\frac{9}{64}(13z^4+44z^3+55z^2+30z+2)x^3-\frac{3}{32}(78z^3+198z^2+165z+27)x^2\nonumber\\
    &+\frac{3}{32}(80z^2+128z+32)x-\frac{3z}{8}\,.
\end{align}
However, several operators decouple in the process of $a$-maximization. Candidate operators are similar to \eqref{eq:dec_3z+1}:
\begin{align}
    \Tr\,\Phi_k^2\,,\quad\Tr\,\Phi_k^3\,,\quad(\Tr\,A_1A_2)_k\,,\quad (\det\,A_2)_k\,,\quad (b_q)_k\,,\quad \Tr\,\Phi_C^2\,,\quad \Tr\,\Phi_C^3\,,\quad \Tr\,\phi^2\,.
    \label{eq:CoulombOps_3z+3}
\end{align}
The $R_{IR}$-charges of  these operators are
\begin{equation}
\begin{gathered}
    \{R(\det\,A_2)_{k\in 2\mathbb{Z}+1},R(b_q)_{k\in2\mathbb{Z}}\}=\{(3k+1)x\,,\quad k=1,\cdots,z\}\,,\\
    \{R(\det\,A_2)_{k\in 2\mathbb{Z}},R(b_q)_{k\in2\mathbb{Z}+1}\}=\{(3k+2)x\,,\quad k=1,\cdots,z\}\,,\\    R(\Tr\,A_1A_2)_k=(2k+1)x\,,\quad k=1,\cdots, z\,,\\
    R(\Tr\,\Phi_m^2)=(2m+2)x\,,\quad m=1,\dots,z-1\,,\\
    R(\Tr\,\Phi_m^3)=(3m+3)x\,,\quad m=1,\dots,z-1\,,\\
    R(\Tr\,\Phi_C^2)=R(\Tr\,\phi^2)=2x\,,\quad R(\Tr\,\Phi_C^3)=3x\,.
    \end{gathered}
    \label{eq:CoulombR_3z+3}
\end{equation}
It turns out that $x>0$ and operators in \eqref{eq:CoulombR_3z+3} with $R_{IR}\leq (z+1)x$ decouple along the iterative $a$-maximization. Then, the trial $a$-central charge after decoupling is 
\begin{align}
    a'(x)=a(x)-\frac{3}{16}\sum_{k=2}^{z+1}\left(3(kx-1)^3-(kx-1)\right)\,.
\end{align}
The resulting mixing coefficient $x$ and the central charges of the interacting part of the IR SCFT are
\begin{align}
    x=\frac{2}{3(z+1)},\quad a=\frac{48z+23}{24}\,,\quad c=\frac{12z+7}{6}\,,
\end{align}
which exactly match the central charges of $D_{3z+3}(\su_3)$~\cite{Cecotti:2013lda}. Furthermore, one can easily find that the remaining operators in \eqref{eq:CoulombOps_3z+3} form the Coulomb branch operators of $D_{3z+3}(\su_3)$. First, their scaling dimensions are exactly those of the Coulomb branch operators. Second, their scaling dimensions are exactly proportional to $\left(\mathcal{J}+\frac{z+2}{2}\mathcal{F}_\mathcal{T}+\mathcal{F}\right)$, which is the only remaining $\mathcal{N}=1$ abelian flavor symmetry that can be identified with $\mathcal{F}_{\mathcal{N}=2}=I_3-r/2$ (up to normalization) in the enhanced $\mathcal{N}=2$ symmetry at the IR fixed point. More precisely, the scaling dimensions of Coulomb branch operators in the IR are given by
\begin{align}
    \Delta_{\text{Coulomb}}=\frac{3}{2}R_{\text{IR}}=-\frac{1}{z+1}\left(\mathcal J+\frac{z+2}{2}\mathcal{F}_\mathcal{T}+\mathcal F\right)\,.
\end{align}
For Coulomb branch operators $\Delta=\frac{r}{2}=R_0=-\mathcal{F}_{\mathcal{N}=2}$, so the $\mathcal{F}_{\mathcal{N}=2}$ in the IR is \begin{align}
    \mathcal{F}_{IR}=\frac{1}{z+1}\left(\mathcal{J}+\frac{z+2}{2}\mathcal{F}_\mathcal{T}+\mathcal{F}\right)\,.
\end{align}

To provide further evidence for our construction, we compute the Schur index of this theory. For example, at $z=2$, we obtain the unrefined Schur index
\begin{align}
    I_{D_9(\su_3)}=1+10q+65q^2+348q^3+1610q^4+\mathcal{O}(q^5)\,,
\end{align}
which exactly matches the Schur index that can be obtained from the 2d TQFT formalism \cite{Buican:2017uka}.

Now we construct the $D_{3z+2}(\su_3)$ theory. We repeat the construction of $D_{3z+3}(\su_3)$, and glue the $D_2(\su_3)$ theory instead of the $D_3(\su_3)$ quiver tail to the regular puncture of the $\mathcal{T}_z^{(\su_3)}$ by gauging $(\su_3)_C$. 
\begin{align}
\mathcal{T}_z^{(\mathfrak{su}_3)}/(\mathfrak{su}_3\,\oplus\,D_2(\su_3)\,\oplus V_{\mathcal{N}=2})\,\longrightarrow\,D_{3z+2}(\mathfrak{su}_3)\,\oplus\,(2z\text{ free chiral fields})\,.
\end{align}\noindent
One can take it the other way around. We start from the $D_{2}(\su_3)$ theory with one regular puncture and one irregular puncture $I_{3,-1}$. Then, by gluing $\mathcal{T}_z^{(\su_3)}$ on the regular puncture of $D_2(\su_3)$, the ``irregularity'' of the irregular puncture deepens to $I_{3,3z-1}$. 

$D_2(\su_3)$ is a non-Lagrangian theory that has $\mathcal{N}=2$ supersymmetry and $\su_3$ flavor symmetry. Still, a UV $\mathcal{N}=1$ Lagrangian that flows to it under the renormalization group flow is known \cite{Agarwal:2016pjo}, which enables detailed analysis including the computation of the superconformal index. The anomaly coefficients of $D_2(\su_3)$ are given as follows.
\begin{equation}
    \begin{gathered}
        \Tr\,R_0^3=1\,,\quad \Tr\,R_0^2\mathcal{F}=1\,,\quad\Tr\,R_0\mathcal{F}^2=-1\,,\quad\Tr\,\mathcal{F}^3=-1\\
        \Tr\,R_0(\su_3)^2=-\frac{3}{4}\,,\quad \Tr\,\mathcal{F}(\su_3)^2=\frac{3}{4}\,,\quad\Tr\,R_0=-2\,,\quad\Tr\,\mathcal{F}=2\,.
    \end{gathered}
\end{equation}

After $\mathcal{N}=2$ gauging, there are two anomaly-free symmetries other than $(\su_3)_B$ or $(\su_3)_C$:
\begin{align}
R_0+\mathcal{F}_\mathcal{T}+\mathcal{F}\,,\quad\mathcal{J}+\frac{3z+4}{6}\mathcal{F}_\mathcal{T}+\frac{2}{3}\mathcal{F}\,,
\end{align}
where we take the $\mathcal{N}=2$ vector multiplet that we couple to $(\su_3)_C$ to be charged under $\mathcal{F}$. Since the gauging is not conformal, the theory undergoes a renormalization group flow. Then, the superconformal $R$-symmetry at the IR fixed point is a linear combination
\begin{align}
    R_{\text{IR}}=R_0+\mathcal{F}_\mathcal{T}+\mathcal{F}-x\left(\mathcal{J}+\frac{3z+4}{6}\mathcal{F}_\mathcal{T}+\frac{2}{3}\mathcal{F}\right)\,.
\end{align}
Again, the mixing coefficient $x$ is fixed by maximizing the trial $a$ central charge
\begin{align}
    a(x)=&\,\frac{1}{192}(351z^4+720z^3+531z^2+164z+16)x^3-\frac{3}{32}(78z^3+120z^2+59z+8)x^2\nonumber\\
    &+\frac{1}{4}(30z^2+28z+5)x-\frac{3z}{8}\,.
\end{align}
As in the cases of $D_{3z+1}(\su_3)$ and $D_{3z+3}(\su_3)$, several operators decouple along the flow, which invalidates the naive $a$-maximization. The candidate decoupling operators are the following:
\begin{equation}
    \begin{gathered}
        \Tr\,\Phi_k^2\,,\quad\Tr\,\Phi_k^3\,,\quad(\Tr\,A_1A_2)_k\,,\quad (\det\,A_2)_k\,,\quad (b_q)_k\,,\quad \Tr\,\Phi_C^2\,,\quad \Tr\,\Phi_C^3\,,\quad u\,.
        \label{eq:CoulombOps_3z+2}
    \end{gathered}
\end{equation}
Here, all operators are exactly the same as those in \eqref{eq:CoulombOps_3z+3}, except that $\phi^2$ is replaced with $u$, the $\mathcal{N}=1$ chiral multiplet that includes the Coulomb branch operator of $D_2(\su_3)$. $R_{\text{IR}}$-charges of the above operators are
\begin{equation}
    \begin{gathered}
            \{R(\det\,A_2)_{k\in 2\mathbb{Z}},R(b_q)_{k\in2\mathbb{Z}+1}\}=\{(3k+1)x\,,\quad k=1,\cdots,z\}\,,\\
    \{R(\det\,A_2)_{k\in 2\mathbb{Z}+1},R(b_q)_{k\in2\mathbb{Z}}\}=\{3kx\,,\quad k=1,\cdots,z\}\,,\\    R(\Tr\,A_1A_2)_k=\left(2k+\frac{1}{3}\right)x\,,\quad k=1,\cdots, z\,,\\
    R(\Tr\,\Phi_m^2)=\left(2m+\frac{4}{3}\right)x\,,\quad m=1,\dots,z-1\,,\\
    R(\Tr\,\Phi_m^3)=(3m+2)x\,,\quad m=1,\dots,z-1\,,\\
    R(\Tr\,\Phi_C^2)=\frac{4}{3}x\,,\quad R(\Tr\,\Phi_C^3)=2x\,,\quad R(u)=x\,.
    \end{gathered}
\end{equation}
In the sequence of $a$-maximization, it turns out that $x>0$, and operators with $R_{IR}\leq\left(z+\frac{2}{3}\right)x$ decouple. The trial $a$ central charge that is successfully maximized at the true IR $R$ charge is
\begin{equation}
    \begin{split}
        a'(x) = a(x)-\frac{3}{32}\Bigg(&\sum_{k=1}^{z}\left(3\left(\left(k+\frac{1}{3}\right)x-1\right)^3-\left(\left(k+\frac{1}{3}\right)x-1\right)\right)\\
        &+\sum_{k=1}^{z}\left(3\left(kx-1\right)^3-\left(kx-1\right)\right)\Bigg)\,.
    \end{split}
\end{equation}
The mixing coefficient and the central charges at the local maximum of $a'(x)$ are
\begin{align}
    x=\frac{2}{3z+2}\,,\quad a=\frac{(3z+1)(12z+7)}{6(3z+2)}\,,\quad c=\frac{2(3z+1)}{3}\,.
\end{align}
The central charges match exactly those of $D_{3z+2}(\su_3)$. Moreover, the remaining operators in \eqref{eq:CoulombOps_3z+2} exactly reproduce the Coulomb branch operators of $D_{3z+2}(\su_3)$. The match of central charges and Coulomb branch spectrum strongly supports our claim. 

Another piece of evidence supporting our construction is the Schur index. For simplicity, we unrefine the flavor $\su_3$ fugacities, and compute the Schur index of $D_8(\su_3)$. Using the letter indices from our Lagrangian, we obtain the following Schur index:
\begin{align}
    \mathcal{I}^{D_8(\su 3)}_{\text{Schur}}=\text{PE}\left[\frac{q-q^8}{(1-q)(1-q^8)}\chi_{\mathbf{8}}\right]=1+8q+44q^2+192q^3+726q^4+2464q^5+\dots\,,
\end{align}
where $\chi_\mathbf{8}$ is the character of the adjoint representation of the flavor $\su_3$ symmetry that the ungauged regular puncture of $\mathcal{T}_z^{(\su_3)}$ carries. The coefficients match perfectly with the vacuum character of $(\widehat{\su}_3)_{-21/8}$ \cite{Kac:2017} that is expected \cite{Xie:2016evu,Song:2017oew}.
We can also compute the Macdonald index. For the case of $D_5(\su 3)$, we obtain
\begin{align}
\begin{split}
\mathcal{I}_{\textrm{Mac}}^{D_5(\su 3)} ={}&1+8qT
 +q^2(9T+35T^2)
 +q^3(9T+72T^2+111T^3)\\
&+q^4(9T+117T^2+314T^3+286T^4)\\
&+q^5(9T+153T^2+674T^3+983T^4+637T^5)\\
&+q^6(9T+198T^2+1127T^3+2540T^4
                +2492T^5+1274T^6)
 +O(q^7) \ ,
\end{split}
\label{eq:MacD5su3}
\end{align}
which reduces to the known Schur index as we take $T \to 1$. This result also agrees with that of \cite{Kang:2026ona, Kang:2026ney}.

\section{Construction of \texorpdfstring{$D_p(\mathfrak{su}_N)$ theories from $T_N$}{DpsuN theories from TN}}\label{sec:g}

In the previous two sections we have discussed two concrete examples of our main claim \eqref{eq:mainclaim}: the examples of $\g=\su_2$ and $\g=\su_3$.
In these examples explicit Lagrangians for the tube theories ${\cal T}^{(\g)}_{z=1}$ are known. For other choices of $\g$, at least at the moment,  explicit Lagrangians for ${\cal T}^{(\g)}_{z=1}$ are not known.  In particular, the ``trinion'' theories $T_\g$ do not have a known Lagrangian description.\footnote{There is, however, an exception: $T_{\su_4}$~\cite{Razamat:2019vfd}. That Lagrangian, however, does not manifest enough symmetry to be useful for us here.} Nevertheless, we can test the consistency of the claim \eqref{eq:mainclaim} for {\it any} simply-laced $\g$ at the level of some of the 't~Hooft anomalies. The anomalies of ${\cal T}^{(\g)}_z$ can be computed directly from six dimensions, and the anomalies of the $D_p(\g)$ theories are known~\cite{Cecotti:2012jx,Cecotti:2013lda}. In this section we perform this comparison. We will work with a general simply-laced $\g$, and specialize to $\g=\su_N$ only at the end.\footnote{ In more general compactification scenarios starting with $6d$ $(1,0)$ theories the anomaly computations of the sort we will perform here were very useful to nail down $4d$ descriptions. By now there is a vast literature on this subject. We refer the reader to \cite{Razamat:2022gpm} for a review and to \cite{Razamat:2016dpl,Kim:2017toz,Bah:2017gph} for many detailed worked-out examples.}
We will proceed in two steps. In Section~\ref{sec:tubeanomalies} we compute the anomalies of the tube ${\cal T}^{(\g)}_z$ from six dimensions, \emph{including} the anomalies of the geometric isometry symmetry ${\cal J}$. In Section~\ref{sec:gauging} we gauge one of the two punctures with an ${\cal N}=2$ vector multiplet, and compare with the anomalies of $D_p(\g)$ plus free fields. We will see that the anomalies that we will compute match if $p=h_\g z+1$. In Section~\ref{sec:suN} we summarize the results for $\g=\su_N$. The goal of this section is twofold. First, we want to phrase the statement of the construction of the Argyres--Douglas theories of the two preceding sections purely in geometric terms using the anomalies. Second, the same geometric language will apply for any algebra $\g$ and thus will motivate the general form of the conjecture \eqref{eq:mainclaim}. 

\ 

\noindent Let us first set up the notations. Throughout this section $\g$ is a simply-laced Lie algebra with
\begin{equation}
r_\g\ (\text{rank}),\qquad d_\g\ (\text{dimension}),\qquad
 h_\g\ (\text{Coxeter number}),\qquad K\equiv r_\g+d_\g\,h_\g\,,
\end{equation}
and $d_j$, $j=1,\dots,r_\g$, are the degrees of its Casimir invariants, with $d_1=2$ and $d_{r_\g}=h_\g$. For a simply-laced algebra $\g$ the Coxeter number equals the dual Coxeter number, and $d_\g=r_\g\,(h_\g+1)$. To avoid confusion, we always keep the subscript on $r_\g$ and $d_\g$: a bare $r$ denotes the $(\u1)_r$ charge, and $d_j$ a Casimir degree. For $\g=\su_N$ we have $r_\g=N-1$, $d_\g=N^2-1$, $h_\g=N$, $d_j=j+1$ and $K=N^3-1$.

The symmetries of the tube that we will use, along with their characteristic classes, are collected in Table~\ref{tab:notation6d}. As in the rest of the paper, an ${\cal N}=2$ theory viewed as an ${\cal N}=1$ theory has R-symmetry $R_0=I_3+\frac12 r$ and global symmetry ${\cal F}=I_3-\frac12 r$, where $I_3$ is the Cartan generator of the $\su_2$ R-symmetry and $r$ is the $(\u1)_r$ charge, normalized so that the adjoint chiral field of a vector multiplet has $r=2$. In particular, the adjoint chiral field of an ${\cal N}=2$ vector multiplet has $(R_0,{\cal F})=(1,-1)$, a hypermultiplet scalar has $(\frac12,\frac12)$, and a moment map operator has $(1,1)$. We  use the anomaly polynomial  to encode the various 't~Hooft anomalies.\footnote{We are using the normalization for the Dynkin index $I_\g(\mathbf{R})$ of a representation $\mathbf{R}$ such that $I_\g(\text{Adj.})=h_\g$ and $I_{\su_N}({\bf N})=\frac12$.}
\begin{table}[h]
\centering
\renewcommand{\arraystretch}{1.3}
\begin{tabular}{|c|c|c|}
\hline
Symmetry & Background & Origin \\
\hline
$(\u1)_{R_0}$ & $x=c_1(R_0)$ & Cartan of $\su_2\subset{\frak{so}}_5$ R-symmetry, twisted along the surface \\
$(\u1)_{\cal F}$ & $y=c_1({\cal F})$ & Cartan of the other $\su_2\subset{\frak{so}}_5$; carries the flux $z$ \\
$(\u1)_{\cal J}$ & $\epsilon=c_1({\cal J})$ & Rotation of the surface around the axis through the punctures \\
$\g_{\rm L},\,\g_{\rm R}$ & $c_2(\g_{\rm L,R})$ & The two maximal puncture symmetries \\
\hline
\end{tabular}
\caption{The symmetries of the tube ${\cal T}^{(\g)}_z$. $R_0$ is an R-symmetry, while ${\cal F}$, ${\cal J}$ and $\g_{\rm L,R}$ are flavor symmetries.}
\label{tab:notation6d}
\end{table}

\subsection{'t~Hooft anomalies of \texorpdfstring{${\cal T}^{(\g)}_z$}{Tgz} theories from six dimensions}
\label{sec:tubeanomalies}

Let us first derive the 't~Hooft anomalies of ${\cal T}^{(\g)}_z$. We compute them by placing the $6d$ $(2,0)$ theory of type $\g$ on a two-punctured sphere, integrating the $6d$ anomaly polynomial over this Riemann surface, and taking into account the contributions of the punctures. For the symmetries of the $6d$ theory this procedure is by now standard~\cite{Benini:2009mz}. Here, however, we also have the geometric isometry. The technology to compute its anomalies for closed Riemann surfaces was developed in~\cite{Hosseini:2020vgl}, see also~\cite{Bah:2019rgq,Ferrero:2021wvk,Bah:2021mzw,Bah:2021hei}, and it was used in compactification computations in~\cite{Hwang:2021xyw}. We will derive the contributions of the geometric isometry to the anomalies by a {\it cutting procedure}. A sphere ${\mathbb S}^2$ with flux $z$ can be constructed by gluing together two discs with fluxes $+\frac12$ and $-\frac12$ and a tube with flux $z$~\cite{Kim:2025ziz}. The gluing is done by gauging the diagonal combination of the symmetries $\g$ associated to the boundaries of the two discs and the tube. In this way we derive the anomalies of the once-punctured sphere and the tube. The latter is the object we are after. This procedure fixes the anomalies of the geometric isometry up to ambiguities which cancel when a tube is closed with caps. Let us proceed with this computation.

We start with the anomaly polynomial of the $6d$ $(2,0)$ theory of type $\g$~\cite{Harvey:1998bx,Benini:2009mz,Ohmori:2014kda,Intriligator:2000eq,Yi:2001bz},
\begin{equation}
I_8=\frac{r_\g}{48}\Big[p_2(N)-p_2(T)+\tfrac14\big(p_1(T)-p_1(N)\big)^2\Big]+\frac{d_\g\,h_\g}{24}\,p_2(N)\,.
\label{eq:I8}
\end{equation}
Here $N$ is the normal bundle, whose structure group is the ${\frak{so}}_5$ R-symmetry. We place the theory on ${\mathbb S}^2$ and twist to preserve ${\cal N}=1$ supersymmetry. For this twist we decompose the $6d$ R-symmetry as ${\frak{so}}_5\supset\su_2\times\su_2\supset(\u1)_{R_0}\times(\u1)_{\cal F}$. We denote by $\hat x$ and $\hat y$ the $6d$ Chern roots of the Cartan generators of the two $\su_2$ factors, so that $p_1(N)=2(\hat x^2+\hat y^2)$ and $p_2(N)=(\hat x^2-\hat y^2)^2$, and we write $I_8(\hat x,\hat y)$ for \eqref{eq:I8} with these substitutions. On a Riemann surface $\Sigma$ with Euler class $e$, the twist and the flux $z$ for ${\cal F}$ are implemented by~\cite{Benini:2009mz,Bah:2012dg},\footnote{When $\chi(\Sigma)=0$ one defines and uses a 2-form $\omega$ so that $\int \omega=1$ on the surface instead of $e$.}
\begin{equation}
\hat x=x-\tfrac{1}{2}\,e\,,\qquad \hat y=y-\tfrac{z}{\chi(\Sigma)}\,e\,,\qquad
p_1(T_6)=p_1(T_4)+e^2\,.
\label{eq:twist}
\end{equation}

Next, we compute the anomaly polynomial $I_6$ of the compactification on $\Sigma$ by integrating $I_8$ over the Riemann surface and adding the contributions localized at the punctures. 
As $\int_\Sigma e=\chi(\Sigma)$, and as long as the background for the isometry of $\Sigma$ is turned off (we will turn it on below), only the part of $I_8$ which is linear in $e$ survives the integration over $\Sigma$. We can thus think of the integral as a first-order differential operator acting on $I_8(x,y)$. For a sphere with $s$ maximal punctures of the same sign,\footnote{The sign of the puncture is determined by the sign of the charge of the moment map under $(\u1)_{\cal F}$.} $\chi=2-s$, we thus find
\begin{equation}
I_6=-\frac{\chi}{2}\,\partial_xI_8(x,y)-z\,\partial_yI_8(x,y)-\frac{s}{2}\,I_{\text{Vec.}}(x,y)\,.
\label{eq:genred}
\end{equation}
Here $-\frac{s}{2}\,I_{\text{Vec.}}$ is the contribution of the punctures.
Each maximal puncture contributes minus half of an ${\cal N}=2$ vector multiplet~\cite{Bah:2012dg,Bah:2018gwc,Chacaltana:2010ks,Bah:2019jts}. The basic logic is again that pairs of punctures can be removed by gluing them together. In field theory, the gluing is implemented by gauging the diagonal $\g$ of the two punctures with an ${\cal N}=2$ vector multiplet~\cite{Gaiotto:2009we}. The $4d$ anomaly polynomial of the ${\cal N}=2$ vector multiplet is
\be
I_{\text{Vec.}}=d_\g\,\left(\Big(\frac{x^3}{6}-\frac{x\,p_1}{24}\Big)-\Big(\frac{y^3}{6}-\frac{y\,p_1}{24}\Big)\right)\,.
\label{eq:Vpoly}
\ee
Two special cases are important for us: the sphere without punctures ($\chi=2$, $s=0$) and the sphere with two punctures ($\chi=0$, $s=2$). For the latter we obtain
\begin{equation}
\begin{aligned}
I_6[{\cal T}^{(\g)}_z]&=-z\,\partial_yI_8(x,y)-I_{\text{Vec.}}(x,y)\\
&=-\frac{d_\g}{6}x^3+\frac{d_\g-Kz}{6}\,y^3+\frac{d_\g h_\g\,z}{6}\,x^2y
+\frac{d_\g}{24}\,x\,p_1+\frac{r_\g z-d_\g}{24}\,y\,p_1\,,
\end{aligned}
\label{eq:tubeI6}
\end{equation}
or, in terms of 't~Hooft anomalies,
\begin{equation}
\boxed{\begin{gathered}\label{eq:nonJtubeanomalies}
{\rm Tr}\,R_0={\rm Tr}\,R_0^3=-d_\g\,,\qquad {\rm Tr}\,{\cal F}=d_\g-r_\g z\,,\qquad {\rm Tr}\,{\cal F}^3=d_\g-Kz\,,\\
{\rm Tr}\,R_0^2{\cal F}=\frac{d_\g h_\g\,z}{3}\,,\qquad {\rm Tr}\,R_0{\cal F}^2=0\,.
\end{gathered}}
\end{equation}
The anomalies involving the puncture symmetry $\g$ follow from the requirement that the ${\cal N}=2$ gauging is not anomalous:
\begin{equation}
\boxed{{\rm Tr}\,\g^2R_0=-\frac{h_\g}{2}\,,\qquad {\rm Tr}\,\g^2{\cal F}=\frac{h_\g}{2}\,.}
\label{eq:punctanom}
\end{equation}
Let us check these expressions against the theories we have already encountered.
\begin{itemize}
\item The theory $T_{\su_3}$ corresponds to the three-punctured sphere ($\chi=-1$, $s=3$), and it is the $\E_6$ Minahan--Nemeschansky theory~\cite{Minahan:1996fg}. The anomalies derived here match the values in Table~\ref{tab:T3} if we take $z=\frac12$. For instance, ${\rm Tr}\,R_0^2{\cal F}=\frac13d_\g h_\g\,z=4$, ${\rm Tr}\,R_0^3=1$ and ${\rm Tr}\,{\cal F}^3=-1$. Thus, in our conventions, the three-punctured sphere has flux $\frac12$.

\item For $\g=\su_2$ and $z=1$, \eqref{eq:tubeI6} is the anomaly polynomial of a single bifundamental chiral field $Q$ with $(R_0,{\cal F})=(0,1)$ plus a singlet flipping $\det Q$, the ${\cal T}^{(\su_2)}_{z=1}$ tube of Section~\ref{sec:a1}~\cite{Agarwal:2015vla,Nardoni:2016ffl,Fazzi:2016eec,Razamat:2019sea}.
\item For $\g=\su_3$ and $z=1$, \eqref{eq:tubeI6} and \eqref{eq:punctanom} give ${\rm Tr}\,R_0={\rm Tr}\,R_0^3=-8$, ${\rm Tr}\,{\cal F}=6$, ${\rm Tr}\,{\cal F}^3=-18$, ${\rm Tr}\,R_0^2{\cal F}=8$, ${\rm Tr}\,R_0{\cal F}^2=0$ and ${\rm Tr}\,\su_3^2(R_0,{\cal F})=(-\frac32,\frac32)$, exactly the anomalies of the Lagrangian ${\cal T}^{(\su_3)}_{z=1}$ of Table~\ref{tab:tube}.
\end{itemize}

\ 

We have thus derived the anomalies of ${\cal T}^{(\g)}_z$ for all the symmetries except the geometric isometry ${\cal J}$. Equivalently, we have set $\epsilon=c_1({\cal J})$ of Table~\ref{tab:notation6d} to zero. The geometric isometry is not a symmetry of the $6d$ SCFT, and thus it was not covered by the analysis so far. We now fill this gap and consider $\epsilon\neq0$.

First, consider the closed sphere ${\mathbb S}^2$ without punctures. Its isometry is ${\frak{so}}_3\cong\su_2$. In order to include the background field for the isometry, one has to fiber the UV curve $\Sigma$ over spacetime~\cite{Bah:2019rgq,Hosseini:2020vgl}. The Euler class $e$ of the sphere is then replaced by the corresponding class of the fibration, which has components along spacetime as well. As a result, also the terms in $I_8$ which are cubic in $e$ contribute to the integral over the surface. By the Bott--Cattaneo formula~\cite{BottCattaneo:1997}, the powers of $e$ integrate over the sphere as
\begin{equation}
\int_{{\mathbb S}^2}e^{2k}=0\,,\qquad \int_{{\mathbb S}^2}e^{2k+1}=2\,p_1({\frak{so}}_3)^k\,,
\label{eq:BC}
\end{equation}
where $p_1({\frak{so}}_3)$ is the first Pontryagin class of the ${\frak{so}}_3$ background. For $k=0$ this is just $\int_{{\mathbb S}^2}e=\chi=2$, the Gauss--Bonnet theorem. Importantly for us, the case $k=1$ is also non-trivial, and it contributes to the four-dimensional anomaly polynomial. Keeping only the Cartan $\u1$ of ${\frak{so}}_3$, generated by ${\cal J}$, we have $p_1({\frak{so}}_3)\to\epsilon^2$. We normalize ${\cal J}$ so that the triplet of ${\frak{so}}_3$ has charges $0,\pm1$. Since only the odd powers of $e$ survive, \eqref{eq:BC} can be summarized as follows: for any polynomial $\omega(e)$,
\begin{equation}
\int_{{\mathbb S}^2}\omega(e)=\frac{\omega(\epsilon)-\omega(-\epsilon)}{\epsilon}\,.
\label{eq:localization}
\end{equation}
The two terms have a simple geometric meaning: $+\epsilon$ and $-\epsilon$ can be thought of as the values of $e$ at the north and south poles, the two points which are fixed by the rotation~\cite{Hosseini:2020vgl,Martelli:2023oqk,BenettiGenolini:2023yfe,Couzens:2025loc}. 

We now apply these formulas to ${\mathbb S}^2$ with flux $z$ for ${\cal F}$. All the backgrounds in the twist \eqref{eq:twist} are polynomials in $e$: $\hat x=x-\frac12e$, $\hat y=y-\frac z2e$ and $p_1(T_6)=p_1(T_4)+e^2$. Thus,
\begin{equation}
I_6[{\mathbb S}^2_z]=\frac{\widehat I_8\big(x-\tfrac{\epsilon}{2},y-\tfrac{z\epsilon}{2}\big)-\widehat I_8\big(x+\tfrac{\epsilon}{2},y+\tfrac{z\epsilon}{2}\big)}{\epsilon}\,,
\label{eq:finitediff}
\end{equation}
where the hat means that also $p_1(T_6)\to p_1(T_4)+\epsilon^2$. Extracting the anomalies of the ${\frak{so}}_3$ isometry we find, for every $\g$,
\begin{equation}
{\rm Tr}\,{\frak{so}}_3^2\,R_0=\frac{d_\g\,h_\g}{12}\,(z^2-1)\,,\qquad
{\rm Tr}\,{\frak{so}}_3^2\,{\cal F}=-\frac{K}{12}\,z\,(z^2-1)\,.
\label{eq:sphereISO}
\end{equation}
For $\g=\su_N$ this is the result of~\cite{Hwang:2021xyw}, which was first obtained from anomaly inflow in M-theory in~\cite{Bah:2019rgq}. There, the result was checked for $\g=\su_2$ against explicit constructions of spheres with flux, using the Lagrangians which can be built from the tubes discussed in Section~\ref{sec:a1}.\footnote{In~\cite{Hwang:2021xyw} the same technology was used to check the constructions of spheres with flux from compactifications of the rank $Q$ E-string theory. The tubes used there were derived in~\cite{Pasquetti:2019hxf}. As for $\g=\su_N$ with $N>2$, the basic tube Lagrangian there manifests only one of the puncture symmetries, while the other one is emergent. Thus, also for the rank $Q$ E-string, general tube theories involve the gauging of emergent symmetries.}

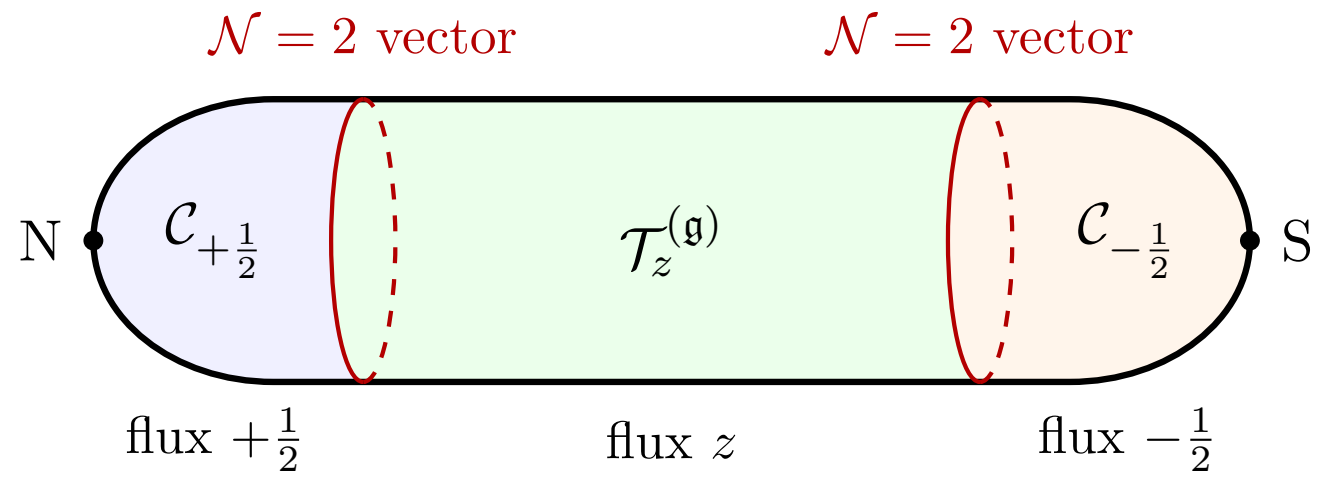
\begin{figure}[t]
\centering
\begin{tikzpicture}[scale=0.85,>=Stealth]
\fill[blue!6] (-4.5,0) arc (180:90:1.4 and 1.1) -- (-2.4,1.1) arc (90:270:0.25 and 1.1) -- (-3.1,-1.1) arc (270:180:1.4 and 1.1) -- cycle;
\fill[green!8] (-2.4,1.1) -- (2.4,1.1) arc (90:270:0.25 and 1.1) -- (-2.4,-1.1) arc (270:90:0.25 and 1.1) -- cycle;
\fill[orange!8] (2.4,1.1) -- (3.1,1.1) arc (90:0:1.4 and 1.1) arc (0:-90:1.4 and 1.1) -- (2.4,-1.1) arc (270:90:0.25 and 1.1) -- cycle;
\draw[very thick] (-4.5,0) arc (180:90:1.4 and 1.1) -- (3.1,1.1) arc (90:0:1.4 and 1.1) arc (0:-90:1.4 and 1.1) -- (-3.1,-1.1) arc (270:180:1.4 and 1.1);
\draw[thick,dashed,red!70!black] (-2.4,-1.1) arc (-90:90:0.25 and 1.1);
\draw[thick,red!70!black] (-2.4,1.1) arc (90:270:0.25 and 1.1);
\draw[thick,dashed,red!70!black] (2.4,-1.1) arc (-90:90:0.25 and 1.1);
\draw[thick,red!70!black] (2.4,1.1) arc (90:270:0.25 and 1.1);
\fill (-4.5,0) circle (2.2pt) node[left=2pt] {N};
\fill (4.5,0) circle (2.2pt) node[right=2pt] {S};
\node at (-3.55,0) {${\cal C}_{+\frac12}$};
\node at (3.55,0) {${\cal C}_{-\frac12}$};
\node at (0,0) {${\cal T}^{(\g)}_z$};
\node[font=\small] at (0,-1.55) {flux $z$};
\node[red!70!black,font=\small] at (-2.4,1.6) {${\cal N}=2$ vector};
\node[red!70!black,font=\small] at (2.4,1.6) {${\cal N}=2$ vector};
\node[font=\small] at (-3.55,-1.55) {flux $+\frac12$};
\node[font=\small] at (3.55,-1.55) {flux $-\frac12$};
\end{tikzpicture}
\caption{The decomposition of a sphere with flux $z$ into two discs with fluxes $\pm\frac12$ and a two-punctured sphere with flux $z$. The two cuts carry ${\cal N}=2$ vector multiplets. The southern disc is the image, under a rotation which reverses ${\cal J}$, of a northern disc with flux $-\frac12$.}
\label{fig:disc}
\end{figure}

Having derived the anomalies of the closed sphere with flux, our second step is to cut it open and derive the anomalies of the once- and twice-punctured spheres. The idea is that 't~Hooft anomalies are {\it topological}. The anomalies associated to a surface depend only on its topology, on the degrees of freedom localized at the punctures, and on the total flux through it. For example, they do not depend on how the flux is distributed on the surface. We may therefore choose a distribution of the flux on ${\mathbb S}^2$ in which flux $+\frac12$ is supported in a disc around the north pole, flux $-\frac12$ in a disc around the south pole, and flux $z$ on the cylinder between them, see Figure~\ref{fig:disc}. The anomalies of the cylinder  are what we are after.\footnote{In general, we expect the flux to be half-integer on surfaces with an odd number of punctures, and integer on surfaces with an even number of punctures~\cite{Gaiotto:2015usa,Kim:2017toz,Razamat:2016dpl}.}

The symmetry ${\cal F}$ comes from six dimensions, and thus it is easy to identify. The geometric symmetry ${\cal J}$, on the other hand, is identified only up to a possible mixing with ${\cal F}$, ${\cal J}\to{\cal J}+\lambda\,{\cal F}$. We call the choice of the mixing $\lambda$ the frame of the theory. We can choose the frame of each building block of the geometry separately, in a natural way, but we will need to identify the symmetries properly when gluing the building blocks together.
Changing the frame amounts to the substitution $y\to y+\lambda\,\epsilon$ in the anomaly polynomial.

We fix the frame of each piece in the decomposition of the sphere into two discs (caps) and a tube as follows.
\begin{itemize}
\item For a cap, we use the frame in which the moment map $\mu$ of the puncture symmetry $\g$ is not charged under ${\cal J}$.
\item For the tube, we use the {\it symmetric frame}, in which the two moment maps have opposite charges, ${\cal J}(\mu_{\rm R})=-{\cal J}(\mu_{\rm L})\equiv m_z$. The value of $m_z$ remains to be fixed.
\item For a closed sphere, we use the frame fixed by the requirement that ${\cal J}$ is the Cartan generator of the ${\frak{so}}_3$ symmetry. The anomalies are then given by \eqref{eq:finitediff}, and in particular the anomaly polynomial is even in $\epsilon$.
\end{itemize}
When two pieces of the geometry are glued together by gauging the diagonal combination of the puncture symmetries $\g$ with an ${\cal N}=2$ vector multiplet, the ${\cal N}=2$ superpotential ${\rm Tr}\,\Phi\,(\mu-\mu')$ locks the isometry charges of the glued pieces. In particular, \be \label{eq:superpotentialJ}{\cal J}(\mu)={\cal J}(\mu')=-{\cal J}(\Phi)\equiv\lambda\,, \ee for every gluing. Since $\mu$ and $\mu'$ have ${\cal F}$ charge $+1$, a piece whose moment map has charge $j$ under ${\cal J}$ in its own frame enters the glued surface in the frame shifted by $\lambda-j$. Since $\Phi$ has ${\cal F}$ charge $-1$, the vector multiplet on the cut contributes $I_{\text{Vec.}}(x,y+\lambda\epsilon)$ to the anomaly polynomial.

We denote by ${\cal C}_{\pm\frac12}(x,y,\epsilon)$ the anomaly polynomial of the cap theory with flux $\pm\frac12$, and by ${\cal D}_\pm(x,y,\epsilon)$ the anomaly polynomial of the cap with flux $\pm\frac12$ together with the vector multiplet for its puncture:
\begin{equation}
{\cal D}_\pm(x,y,\epsilon)\equiv {\cal C}_{\pm\frac12}(x,y,\epsilon)+I_{\text{Vec.}}(x,y)\,.
\end{equation}
Let $\lambda_{z}$ be the frame of the tube within the sphere of flux $z$. Then the moment maps at the two cuts have charges $\lambda_{z}\mp m_z$, and the decomposition of Figure~\ref{fig:disc} reads
\begin{equation}
I_6[{\mathbb S}^2_z](x,y,\epsilon)={\cal D}_+\big(x,y+(\lambda_{z}-m_z)\epsilon,\epsilon\big)+I_6\left[{\cal T}^{(\g)}_z\right]\big(x,y+\lambda_{z}\epsilon,\epsilon\big)+{\cal D}_-\big(x,y+(\lambda_{z}+m_z)\epsilon,-\epsilon\big)\,.
\label{eq:discdecomp}
\end{equation}
Here $I_6\left[{\cal T}^{(\g)}_z\right]$ is the anomaly polynomial of ${\cal T}^{(\g)}_z$ in its symmetric frame. Note that the two caps enter with opposite signs of $\epsilon$, since one sits at the north pole and the other at the south pole: the two caps are rotated {\it geometrically} by $\pi$ relative to each other.
Shifting $y\to y-\lambda_{z}\,\epsilon$, we can solve for the anomaly polynomial of the tube,
\begin{equation}
I_6\left[{\cal T}^{(\g)}_z\right](x,y,\epsilon)=I_6[{\mathbb S}^2_z]\big(x,y-\lambda_{z}\epsilon,\epsilon\big)-{\cal D}_+\big(x,y-m_z\epsilon,\epsilon\big)-{\cal D}_-\big(x,y+m_z\epsilon,-\epsilon\big)\,.
\label{eq:tubedirectgen}
\end{equation}

\

\noindent Let us now fix $\lambda_z$, $m_z$ and ${\cal D}_\pm$ by consistency requirements. The caps do not depend on $z$. We determine them from three closed spheres, each of which is glued from two caps (Figure~\ref{fig:disc} without the tube). In each case, both moment maps at the cut are neutral in the frames of their caps, so both caps enter with the same value of $\lambda$.
\begin{itemize}
\item ${\mathbb S}^2_{+1}$ consists of two caps with flux $+\frac12$, the southern one being the reflection of the northern one:
\begin{equation}
I_6[{\mathbb S}^2_{+1}](x,y,\epsilon)={\cal D}_+(x,y,\epsilon)+{\cal D}_+(x,y,-\epsilon)-I_{\text{Vec.}}(x,y)\,.
\end{equation}
Similarly, ${\mathbb S}^2_{-1}$ consists of two caps with flux $-\frac12$:
\begin{equation}
I_6[{\mathbb S}^2_{-1}](x,y,\epsilon)={\cal D}_-(x,y,\epsilon)+{\cal D}_-(x,y,-\epsilon)-I_{\text{Vec.}}(x,y)\,.
\end{equation}
\end{itemize}
Here both caps enter with $\lambda=0$. Indeed, for two identical caps the combination ${\cal D}_\pm(x,y,\epsilon)+{\cal D}_\pm(x,y,-\epsilon)-I_{\text{Vec.}}(x,y)$ is even in $\epsilon$, and after the shift $y\to y+\lambda\epsilon$ it remains even in $\epsilon$, as $I_6[{\mathbb S}^2_{\pm1}]$ must be, only for $\lambda=0$. These two equations determine the parts of ${\cal D}_\pm$ which are even in $\epsilon$. We write \be {\cal D}_\pm={\cal D}_\pm^{\text{even} }+{\cal D}_\pm^{\text{odd} }\,,\ee with
\begin{equation}
\begin{aligned}
{\cal D}_\pm^{\text{even} }=\frac12\Big(I_6[{\mathbb S}^2_{\pm1}]+I_{\text{Vec.}}\Big)
={}&\frac{d_\g(1-h_\g)-r_\g}{12}\,x^3+\frac{d_\g h_\g}{12}\,xy^2\pm\frac{d_\g h_\g}{12}\,x^2y-\frac{d_\g\pm K}{12}\,y^3\\
&+\frac{r_\g-d_\g}{48}\,x\,p_1+\frac{d_\g\pm r_\g}{48}\,y\,p_1\,.
\end{aligned}
\label{eq:capeven}
\end{equation}
The odd part, ${\cal D}_\pm^{\text{odd} }$, is the most general one allowed by the degree of the polynomial:
\begin{equation}
{\cal D}_\pm^{\text{odd} }=\epsilon\,\left(\nu^\pm_{xx}\,x^2+\nu^\pm_{yy}\,y^2+\nu^\pm_{xy}\,xy+\nu^\pm_\epsilon\,\epsilon^2+\nu^\pm_p\,p_1\right)\,.
\end{equation}
Note that all the coefficients here are independent of $z$ by construction.
\begin{itemize}
\item ${\mathbb S}^2_0$ is a cap with flux $+\frac12$ glued to a cap with flux $-\frac12$:
\begin{equation}
I_6[{\mathbb S}^2_0](x,y,\epsilon)=\Big[{\cal D}_+(x,y,\epsilon)+{\cal D}_-(x,y,-\epsilon)-I_{\text{Vec.}}(x,y)\Big]_{y\to y+\lambda\epsilon}\,.
\label{eq:S0caps}
\end{equation}
Note that here we allow for a shift of $y$ by some $\lambda$, since otherwise we would not be able to match the anomalies.
\end{itemize}
Matching both sides of the equation, using \eqref{eq:sphereISO} with $z=0$, we find $\lambda^2=\frac14$ (we choose $\lambda=\frac12$ without loss of generality), and we obtain
\begin{equation}
{\cal D}_\pm^{\text{odd} }= \epsilon\,\left(Q\mp\frac{d_\g h_\g}{12}\,xy\right)\,,\qquad
Q=\nu_{xx}\,x^2+\nu_{yy}\,y^2+\nu_{xy}\,xy+\nu_\epsilon\,\epsilon^2+\nu_p\,p_1\,.
\label{eq:capodd}
\end{equation}
The five constants in $Q$ are not determined by this procedure.

We now insert \eqref{eq:capeven} and \eqref{eq:capodd} into \eqref{eq:tubedirectgen} to obtain the anomalies of the tube. In its symmetric frame the tube must have no anomalies odd under the symmetry ${\cal J}\to-{\cal J}\,, {\rm L} \leftrightarrow {\rm R}$. That is, $I_6\left[{\cal T}^{(\g)}_z\right]$ must be even in $\epsilon$. This requirement implies
\begin{equation}
\lambda_{z}=\frac12\,,\qquad m_z=\frac z2\,.
\end{equation}
Thus, in its symmetric frame, the moment maps of the tube have ${\cal J}(\mu_{\rm L})=-\frac z2$ and ${\cal J}(\mu_{\rm R})=+\frac z2$. For $\g=\su_3$ this is what we found using the explicit Lagrangian construction, see~\eqref{eq:Jseam}. Finally, the anomaly polynomial of the tube is
\begin{equation}
I_6\left[{\cal T}^{(\g)}_z\right](x,y,\epsilon)=I_6[{\mathbb S}^2_z]\big(x,y-\tfrac12\epsilon,\epsilon\big)-{\cal D}_+\big(x,y-\tfrac z2\epsilon,\epsilon\big)-{\cal D}_-\big(x,y+\tfrac z2\epsilon,-\epsilon\big)\,.
\label{eq:tubedirect}
\end{equation} We can now read off all the anomalies explicitly.
The anomalies not involving ${\cal J}$ are obtained by setting $\epsilon=0$, and they coincide with \eqref{eq:tubeI6}.
By construction, all the anomalies odd in ${\cal J}$ vanish,
\be 
{\rm Tr}\,{\cal J}={\rm Tr}\,{\cal J}^3={\rm Tr}\,{\cal J}R_0^2={\rm Tr}\,{\cal J}{\cal F}^2={\rm Tr}\,{\cal J}R_0{\cal F}=0\,.
\ee Moreover, $\nu_{xx}$, $\nu_\epsilon$ and $\nu_p$ completely drop out and do not appear in the anomaly polynomial of the tube.
 The anomalies quadratic in ${\cal J}$ are
\begin{equation}
\boxed{{\rm Tr}\,{\cal J}^2R_0=2\nu_{xy}\,z\,,\qquad
{\rm Tr}\,{\cal J}^2{\cal F}=\frac{z}{12}\Big(3d_\g z-K\,z^2-2K\Big)+4\nu_{yy}\,z\,.}
\label{eq:tubeJbar}
\end{equation}
Thus these anomalies are fixed up to an ambiguity which affects only the terms linear in the flux $z$.

We still need to fix the anomalies involving the puncture symmetries,
${\rm Tr}\,{\cal J}\,\g^2_{\rm L,R}$. These are fixed by requiring that gauging the diagonal $\g$ at each cut on the sphere is free of anomalies. Following our conventions \eqref{eq:superpotentialJ}, $\Phi$ has ${\cal J}=-\lambda$, so that it contributes $-h_\g\,\lambda$ to ${\rm Tr}\,\g^2\,{\cal J}$. We parametrize the anomalies ${\rm Tr}\,{\cal J}\,\g^2$ of the caps ${\cal C}_{\pm\frac12}$, in the frames of the caps, by $\gamma_\pm$.
For ${\mathbb S}^2_0$, with $\lambda=\frac12$, the requirement that the gluing of the two caps is free of anomalies implies
\be 
\big(\gamma_++\frac{h_\g}{4}\big)-\frac{h_\g}{ 2}+\big(-\gamma_-+\frac{h_\g}{4}\big)=0\;\;\to\;\; \gamma_+=\gamma_-\equiv\gamma_\g\,.
\ee
For the tube, the two cuts have $\lambda=\frac{1-z}{2}$ and $\lambda=\frac{1+z}{2}$. Thus, in the frame of the sphere, the
absence of gauge anomalies at the two cuts gives,
\begin{equation}
{\rm Tr}\,{\cal J}\,\g_{\rm L}^2=\frac{h_\g\, (1-z)}{4}-\gamma_\g\,,\qquad
{\rm Tr}\,{\cal J}\,\g_{\rm R}^2=\frac{h_\g\, (1+z)}{4}+\gamma_\g\,.
\end{equation}
Going to the symmetric frame of the tube, ${\cal J}\to{\cal J}-\frac12{\cal F}$, subtracts $\frac12\,{\rm Tr}\,{\cal F}\,\g^2=\frac{h_\g}{4}$ from both anomalies. This gives,
\be \label{eq:gpunctureanomalies}
\boxed{{\rm Tr}\,{\cal J}\,\g_{\rm R}^2=-{\rm Tr}\,{\cal J}\,\g_{\rm L}^2=\frac{h_\g\,z}{4}+\gamma_\g\,.}
\ee
To summarize, in \eqref{eq:tubeJbar} and \eqref{eq:gpunctureanomalies} we have determined all the non-zero anomalies of the tubes with flux $z$ which involve ${\cal J}$, up to three $z$-independent coefficients $\nu_{xy}$, $\nu_{yy}$ and $\gamma_\g$.
The ${\cal J}$-independent anomalies are fixed completely, and they are summarized in \eqref{eq:nonJtubeanomalies} and \eqref{eq:punctanom}. Next, we will see that these anomalies are consistent with the claim \eqref{eq:mainclaim}. 
The comparison with \eqref{eq:mainclaim} will also fix the undetermined coefficients to
\be\label{eq:ambiguouscoefficients}
\nu_{xy}=0\,,\qquad \nu_{yy}=\frac{d_\g\, h_\g}{24}\,,\qquad 
\gamma_\g=-\frac12\,.
\ee 
Finally, in Appendix~\ref{sec:holography} we will perform a holographic computation of the anomalies of ${\cal T}^{(\g=\su_N)}_z$ and will derive exactly the values in~\eqref{eq:ambiguouscoefficients}.

\subsection{Gauging a puncture \texorpdfstring{$\g$ of the tube and $D_{h_\g\,z+1}(\g)$}{g of the tube and Dhgz+1}}\label{sec:gauging}

 Let us gauge the puncture symmetry $\g_{\rm L}$ of ${\cal T}^{(\g)}_z$ with an ${\cal N}=2$ vector multiplet whose adjoint ${\cal N}=1$ chiral superfield is denoted by $\Phi$. The gauging involves a superpotential ${\rm Tr}\,\Phi\,\mu_{\rm L}$. This is a generalization of what we did for $\g=\su_3$ in Section~\ref{sec:a2} and $\g=\su_2$ in Section~\ref{sec:a1}. Our claim~\eqref{eq:mainclaim} is that, for every simply-laced $\g$ and every $z$,
\begin{equation}
\boxed{\begin{gathered}{\cal T}^{(\g)}_z\,/\,(\g\oplus V_{{\cal N}=2})\ \longrightarrow\ D_{h_\g z+1}(\g)\ \oplus\ \{{\cal O}_{j,k}\}\,,\\
\Delta({\cal O}_{j,k})\equiv\Delta_{j,k}=\frac{d_j+h_\g k}{h_\g z+1}\,,\qquad j=1,\dots,r_\g\,,\quad k=0,\dots,z-1\,,\end{gathered}}
\label{eq:gaugingclaim}
\end{equation}
where ${\cal O}_{j,k}$ are $r_\g z$ free chiral fields. In particular, the operators with $k=0$ are the Casimirs ${\rm Tr}\,\Phi^{d_j}$ of the added vector multiplet. In a Lagrangian realization the free fields can be flipped, and then the theory flows to $D_{h_\g z+1}(\g)$ alone. We now show that \eqref{eq:gaugingclaim} is consistent with all the 't~Hooft anomalies, and that this consistency fixes the three undetermined constants $\gamma_\g$, $\nu_{xy}$ and $\nu_{yy}$ of Section~\ref{sec:tubeanomalies}.

  In the symmetric frame of the tube, ${\cal J}(\mu_{\rm L})=-\frac z2$, so by \eqref{eq:superpotentialJ} the adjoint chiral field has $(R_0,{\cal F},{\cal J})(\Phi)=(1,-1,\frac z2)$. The anomaly polynomial of the gauged theory is $I_6[{\cal T}^{(\g)}_z](x,y,\epsilon)+I_{\text{Vec.}}(x,y-\frac z2\epsilon)$, with the anomalies of the tube given in \eqref{eq:nonJtubeanomalies}, \eqref{eq:tubeJbar} and \eqref{eq:gpunctureanomalies}. At the gauged node, using \eqref{eq:punctanom},
\begin{equation}
{\rm Tr}\,\g_{\rm L}^2\,(R_0,\ {\cal F},\ {\cal J})=\Big(\frac{h_\g}{2},\ -\frac{h_\g}{2},\ \frac{h_\g z}{4}-\gamma_\g\Big)\,.
\end{equation}
Thus both $R_0$ and ${\cal F}$ are anomalous, with the non-anomalous R-symmetries being $R_0+\alpha{\cal F}+\beta{\cal J}$ such that,
\begin{equation}
\frac{h_\g}{2}\,(1-\alpha)+\beta\Big(\frac{h_\g z}{4}-\gamma_\g\Big)=0\,.
\label{eq:family}
\end{equation}
There is a single non-anomalous abelian flavor symmetry, and it necessarily involves the isometry ${\cal J}$. The remaining puncture provides the flavor symmetry $\g_{\rm R}$.

Let us compare this to $D_p(\g)$.  For $D_p(\g)$ with $\gcd(p,h_\g)=1$~\cite{Cecotti:2012jx,Cecotti:2013lda,Xie:2013jc}, the various anomalies are
\begin{equation}
a=\frac{(4p-1)(p-1)}{48\,p}\,d_\g\,,\qquad c=\frac{p-1}{12}\,d_\g\,,\qquad k_\g=\frac{2(p-1)}{p}\,h_\g\,.
\label{eq:Dpdata}
\end{equation}
In an ${\cal N}=2$ SCFT all the anomalies of the R-symmetries and of the flavor symmetry follow from $a$, $c$ and $k_\g$~\cite{Shapere:2008zf}. With $n_v=4(2a-c)$ and $n_h=4(5c-4a)$, and with $2I_3$ and $r$ normalized so that the gauginos have charge one under both,
\be 
&&{\rm Tr}\,(2I_3)={\rm Tr}\,(2I_3)^3={\rm Tr}\,(2I_3)\,r^2=0\,,\qquad
{\rm Tr}\,(2I_3)^2r=2n_v\,,\qquad {\rm Tr}\,r={\rm Tr}\,r^3=2(n_v-n_h)\,,\nonumber\\
&&{\rm Tr}\,(2I_3)\,\g^2=0\,,\qquad {\rm Tr}\,r\,\g^2=-\frac{k_\g}{2}\,.
\label{eq:N2anom}
\ee 
The fermions of the free fields have $2I_3=-1$ and $r=2\Delta_{j,k}-1$. Their contributions involve the sums $\sum_{j,k}(2\Delta_{j,k}-1)^n$, which can be evaluated in closed form using the following identities for the exponents $m_j=d_j-1$ of a simply-laced $\g$:\footnote{The first identity states that $\sum_jm_j$ is the number of positive roots. The second can be checked for $\su_n$, ${\frak{so}}_{2n}$ and ${\frak e}_{6,7,8}$. The third follows from the first two and from the symmetry $m_j+m_{r_\g+1-j}=h_\g$.}
\begin{equation}
\sum_j m_j=\frac{r_\g h_\g}{2}\,,\qquad
\sum_j m_j^2=\frac{r_\g h_\g(2h_\g-1)}{6}\,,\qquad
\sum_j m_j^3=\frac{r_\g h_\g^2(h_\g-1)}{4}\,.
\label{eq:powersums}
\end{equation}
 We set $p=h_\g z+1$ and identify the ${\cal N}=2$ R-symmetries with non-anomalous combinations of $R_0$, ${\cal F}$ and ${\cal J}$.
\begin{itemize}
\item The natural candidate for $2I_3$ is $R_0+{\cal F}$, which is $2I_3$ already in six dimensions. It is free of ABJ anomalies, $\alpha=1$ and $\beta=0$ in \eqref{eq:family}, and ${\rm Tr}\,(R_0+{\cal F})\,\g_{\rm R}^2=0$, as required by \eqref{eq:N2anom}.
\item For $r=R_0+\alpha{\cal F}+\beta{\cal J}$, \eqref{eq:family} and \eqref{eq:gpunctureanomalies} give ${\rm Tr}\,r\,\g_{\rm R}^2=\frac{h_\g z}{2}\,\beta$. Matching $k_\g$ of \eqref{eq:Dpdata}, together with \eqref{eq:family}, then fixes
\begin{equation}
r=R_0+\frac{1+4\gamma_\g}{p}\,{\cal F}-\frac{2h_\g}{p}\,{\cal J}\,.
\end{equation}
\end{itemize}
We now compare the remaining six anomalies of the gauged tube with those of $D_p(\g)$ plus the free fields. The anomalies ${\rm Tr}\,(2I_3)={\rm Tr}\,(2I_3)^3=-r_\g z$ match identically: the right number, $r_\g z$, of free fields comes out automatically. For the other four, the differences between the two sides are
\begin{equation}
\begin{aligned}
{\rm Tr}\,r:&\quad -\frac{2\,(1+2\gamma_\g)\,r_\g\,z}{p}\,,\qquad\qquad
{\rm Tr}\,(2I_3)^2r:\quad -\frac{2\,(1+2\gamma_\g)\,(2d_\g h_\g+3r_\g)\,z}{3p}\,,\\
{\rm Tr}\,(2I_3)\,r^2:&\quad -\frac{2h_\g^2z}{3p^2}\,\big(d_\g h_\g-12\,\nu_{xy}-24\,\nu_{yy}\big)\,,\\
{\rm Tr}\,r^3:&\quad \frac{2h_\g^2z}{p^3}\,\big(d_\g h_\g+12\,(1+h_\g z)\,\nu_{xy}-24\,\nu_{yy}\big)\,,
\end{aligned}
\end{equation}
where in the last two lines we already used $\gamma_\g=-\frac12$, which is required by the first line. All the anomalies thus match if and only if
\begin{equation}
\boxed{\ \gamma_\g=-\frac12\,,\qquad \nu_{xy}=0\,,\qquad \nu_{yy}=\frac{d_\g h_\g}{24}\,.\ }
\label{eq:constantsfixed}
\end{equation}
This is a nontrivial check. The three constants are local to the puncture, and thus they cannot depend on $z$, while the matching conditions depend on $z$ explicitly. Moreover, the values \eqref{eq:constantsfixed} agree with those of the explicit tubes: for $\g=\su_2$ the ${\cal J}$ anomalies of ${\cal T}^{(\su_2)}_{z=1}$ vanish, and for $\g=\su_3$ the tube ${\cal T}^{(\su_3)}_{z=1}$ of Table~\ref{tab:tube} has ${\rm Tr}\,{\cal J}^2{\cal F}=-\frac12$ and ${\rm Tr}\,{\cal J}^2R_0=0$. With these values the R-symmetry of $D_{h_\g z+1}(\g)$ is
\begin{equation}
\boxed{\ 2I_3=R_0+{\cal F}\,,\qquad r=R_0-\frac{1}{h_\g z+1}\,{\cal F}-\frac{2h_\g}{h_\g z+1}\,{\cal J}\,,\ }
\label{eq:map}
\end{equation}
which for $\g=\su_3$ is \eqref{eq:D3z1map}. Part of the ${\cal N}=2$ R-symmetry of the Argyres--Douglas theory is thus the isometry of the two-punctured sphere. As a check, ${\rm Tr}\,\Phi^{d_j}$ has $r=\frac{2d_j}{p}$, {\it i.e.}\ $\Delta=\frac{d_j}{p}=\Delta_{j,0}$.

Finally, let us comment on how the fixed point is reached by $a$-maximization~\cite{Intriligator:2003jj} in the theory without flip fields. The non-anomalous R-symmetries form the family $R_x=x\,(2I_3)+(1-x)\,r$, and, after removing the operators ${\cal O}_{j,k}$, all the anomalies are those of $D_{h_\g z+1}(\g)$. Thus
\begin{equation}
\begin{gathered}
a(x)=a_{{\cal N}=2}(x)+\sum_{j=1}^{r_\g}\sum_{k=0}^{z-1}a_\chi\Big(2\Delta_{j,k}(1-x)\Big)\,,\\
a_{{\cal N}=2}(x)=\frac{3}{16}\,(1-x)\Big[9\,n_v\,x^2+(n_v-n_h)\big(3(1-x)^2-1\big)\Big]\,,
\end{gathered}
\label{eq:atrialg}
\end{equation}
with $a_\chi(R)$ defined in \eqref{eq:atrial}. The argument we used in Section~\ref{sec:a2} for $\g=\su_3$ applies verbatim;  $a_{{\cal N}=2}(x)$ is increasing for $\frac13<x<\frac23$ and is maximized at $x=\frac23$. The operators ${\cal O}_{j,k}$ have $R<\frac23$ there, so they push the maximum to $x<\frac23$, where some of them violate the unitarity bound, while the Coulomb branch operators do not. Iterating the procedure of~\cite{Kutasov:2003iy}, which in general requires several steps, decouples exactly the operators ${\cal O}_{j,k}$ and ends at $x=\frac23$, {\it i.e.}\ at the R-symmetry \eqref{eq:map}. In practice, it is simplest to flip the whole tower from the start, as we did in Section~\ref{sec:a2}, and then a single $a$-maximization lands on the Argyres--Douglas point.
Note that this argument is not a proof but an outline of how $a$-maximization is expected to work. For example, one should show in all the cases that the operators claimed to decouple are the only ones which actually do so. This requires either a general argument, which we do not provide, or an explicit check, as can be done when a Lagrangian is available (as in the $\g=\su_2$ and $\g=\su_3$ cases we analyzed in earlier sections).

\subsection{Summary of results for \texorpdfstring{$\g=\su_N$}{g=suN}}\label{sec:suN}

Let us finally summarize the results for the case of $\g=\su_N$, for which $r_\g=N-1$, $d_\g=N^2-1$, $h_\g=N$, $K=N^3-1$, and the Casimir degrees are $d_j=j$, $j=2,\dots,N$. With the values \eqref{eq:constantsfixed} of the constants, $\gamma_\g=-\frac12$, $\nu_{xy}=0$ and $\nu_{yy}=\frac{N(N^2-1)}{24}$, the tube ${\cal T}^{(\su_N)}_z$ of flux $z$ has the anomalies
\begin{equation}
\begin{gathered}
{\rm Tr}\,R_0={\rm Tr}\,R_0^3=-(N^2-1)\,,\qquad {\rm Tr}\,R_0^2{\cal F}=\frac{N(N^2-1)\,z}{3}\,,\qquad {\rm Tr}\,R_0{\cal F}^2=0\,,\\
{\rm Tr}\,{\cal F}=N^2-1-(N-1)z\,,\qquad {\rm Tr}\,{\cal F}^3=N^2-1-(N^3-1)z\,,\\
{\rm Tr}\,\su_N^2\,(R_0,{\cal F})=\Big(-\frac N2,\,\frac N2\Big)\,,\qquad {\rm Tr}\,{\cal J}^2R_0=0\,,\\
{\rm Tr}\,{\cal J}^2{\cal F}=\frac{z}{12}\Big(3(N^2-1)z-2(N-1)-(N^3-1)z^2\Big)\,,\\
{\rm Tr}\,{\cal J}\,(\su_N)_{\rm R}^2=-{\rm Tr}\,{\cal J}\,(\su_N)_{\rm L}^2=\frac{Nz-2}{4}\,,
\end{gathered}
\end{equation}
with all anomalies odd in ${\cal J}$ vanishing. The anomalies of the tube theory which do not involve ${\cal J}$ can be derived directly in four dimensions, starting from the trinion $T_N$ and closing one of its punctures, as we did for the special cases $N=2$ and $N=3$ in Sections~\ref{sec:a1} and~\ref{sec:a2}~\cite{Gadde:2013fma,Agarwal:2014rua,Agarwal:2015vla,Gaiotto:2015usa,Razamat:2022gpm}. The ${\cal N}=2$ theory $T_N$ has $\su_N^{(1)}\times\su_N^{(2)}\times\su_N^{(3)}$ flavor symmetry associated with the three punctures, each factor with $k_{\su_N}=2N$. The conformal anomalies are encoded in the effective numbers of free vector multiplets and hypermultiplets~\cite{Gaiotto:2009gz,Chacaltana:2010ks}
\begin{equation}
n_v(T_N)=\frac{2N^3}{3}-\frac{3N^2}{2}-\frac N6+1\,,\qquad n_h(T_N)=\frac{2N^3}{3}-\frac{2N}{3}\,.
\end{equation}
In terms of $R_0$ and ${\cal F}$, {\it i.e.}\ with $2I_3=R_0+{\cal F}$ and $r=R_0-{\cal F}$ in \eqref{eq:N2anom}, the anomaly polynomial of $T_N$ is
\begin{equation}
I_6[T_N]=n_v\,u^2v+(n_v-n_h)\Big(\frac{v^3}{3}-\frac{v\,p_1}{12}\Big)\,,\qquad u=\frac{x+y}{2}\,,\quad v=\frac{x-y}{2}\,,
\end{equation}
together with ${\rm Tr}\,\su_N^{(i)\,2}\,(R_0,{\cal F})=(-\frac N2,\frac N2)$. This coincides with \eqref{eq:genred} for $\chi=-1$, $s=3$ and $z=\frac12$. To close the third puncture, we first flip its moment map $\mu_3$ with an adjoint chiral field $M$, with the superpotential ${\rm Tr}\,M\mu_3$, so that $M$ has $(R_0,{\cal F})=(1,-1)$. We then give $M$ the principal nilpotent vacuum expectation value $\langle M\rangle=\rho(\sigma^+)$. Let $T_\rho$ be the generator of the principal $\su_2$ embedding $\rho$ under which $\sigma^+$ has charge $+1$. The vacuum expectation value breaks $\su_N^{(3)}$ completely, and preserves
\begin{equation}
R_0'=R_0-T_\rho\,,\qquad {\cal F}'={\cal F}+T_\rho\,.
\end{equation}
Under $\rho$ the adjoint decomposes into spins $j=1,\dots,N-1$. The components $M_{j,m}$ with $m>-j$ decouple, and the $N-1$ singlets $M_j\equiv M_{j,-j}$ remain, with $(R_0',{\cal F}')=(j+1,-j-1)=(d_j,-d_j)$.  For example, for $N=2$, $M_1$ is the singlet flipping $\det Q$ in the tube of Section~\ref{sec:a1}. The anomalies of the resulting theory are obtained by setting the background of $T_\rho$ to $y-x$, using ${\rm Tr}\,X\,T_\rho^2=I_\rho\,{\rm Tr}\,X\,\su_N^2$ with the embedding index $I_\rho=\frac{N(N^2-1)}{6}$, and adding the singlets:
\begin{equation}
I_6\big[{\cal T}^{(\su_N)}_{z=1}\big]=I_6[T_N]+\frac{N^2(N^2-1)}{24}\,(y-x)^3+\sum_{j=1}^{N-1}\Big(\frac{(jx-(j+1)y)^3}{6}-\frac{p_1\,(jx-(j+1)y)}{24}\Big)\,.
\end{equation}
This coincides with \eqref{eq:tubeI6} at $z=1$, for all $N$, while the anomalies of the two remaining punctures are unchanged and given by \eqref{eq:punctanom}. Thus, closing a puncture removes it and shifts the flux by $+\frac12$, from $\frac12$ to $1$. In fact, the last two terms in the anomaly polynomial  are exactly ${\cal D}^{\text{even}}_+$ of \eqref{eq:capeven}: at the level of anomalies, closing a puncture is the same as gluing the cap with flux $+\frac12$ to it.\footnote{The odd part of ${\cal D}_+$ is proportional to $\epsilon$. Since the three-punctured sphere has no geometric isometry, it does not appear here.} Finally, gluing $z$ unit tubes with $z-1$ ${\cal N}=2$ vector multiplets gives \eqref{eq:tubeI6} for general $z$. On the other hand, the anomalies involving ${\cal J}$ cannot be obtained this way: $T_N$ has no geometric isometry, and ${\cal J}$ emerges only in the IR, as we have seen explicitly for $N=3$ in Section~\ref{sec:a2}.

\ 

By \eqref{eq:gaugingclaim}, gauging one of the punctures of ${\cal T}^{(\su_N)}_z$ with an ${\cal N}=2$ vector multiplet gives $D_{Nz+1}(\su_N)$ together with $(N-1)z$ free chiral fields, and by \eqref{eq:map} the ${\cal N}=2$ R-symmetry is
\begin{equation}
2I_3=R_0+{\cal F}\,,\qquad r=R_0-\frac{1}{Nz+1}\,{\cal F}-\frac{2N}{Nz+1}\,{\cal J}\,.
\end{equation}
The central charges and the flavor central charge are
\begin{equation}
a=\frac{Nz\,(4Nz+3)(N^2-1)}{48\,(Nz+1)}\,,\qquad c=\frac{Nz\,(N^2-1)}{12}\,,\qquad k_{\su_N}=\frac{2N^2z}{Nz+1}\,.
\end{equation}
The Coulomb branch operators have dimensions $\Delta=j-\frac{N}{Nz+1}\,s>1$~\cite{Cecotti:2013lda}, with $j=2,\dots,N$ and $s=1,\dots,(j-1)z$, so that the rank is $\frac{N(N-1)}{2}\,z$. The free fields have dimensions
\begin{equation}
\Delta=\frac{j+N\,k}{Nz+1}\,,\qquad j=2,\dots,N\,,\qquad k=0,\dots,z-1\,.
\end{equation}
These are the continuation of the Coulomb branch tower below the unitarity bound, $s=jz-k$, {\it i.e.}\ $s=(j-1)z+1,\dots,jz$. The members with $k=0$ are the Casimirs ${\rm Tr}\,\Phi^j$ of the gauged vector multiplet. For $N=3$ these reproduce the results of Section~\ref{sec:a2}, and for $N=2$ those of Section~\ref{sec:a1}.

\ 

This completes our argument that, starting from the $T_N$ theory, closing one of its maximal punctures and gauging the symmetry of another one, we obtain a description of the $D_{N+1}(\su_N)$ theory plus decoupled free fields.
To obtain more general values of $p$, we can either glue in tubes, thereby increasing the flux, or add matter, generalizing our discussion for $N=2$ and $N=3$. We leave the latter for general $N$ as an exercise for the enthusiastic reader.

\section{Discussion}\label{sec:discussion}

We have argued in this paper that two types of compactifications of a $6d$ $(2,0)$ theory, with regular punctures and with irregular punctures, are intimately and simply related to each other. The main statement is \eqref{eq:mainclaim}, together with its generalizations, which we discussed only for $\g=\su_2$ and $\g=\su_3$. For $\g=\su_3$, our discussion also provides novel Lagrangian constructions of the $D_p(\su_3)$ Argyres--Douglas theories, albeit in general involving the gauging of emergent symmetries.

An important ingredient of our construction explicitly mimics the construction of an irregular puncture. 
The Seiberg--Witten description of an irregular puncture involves a singularity of, in general, fractional order~\cite{Cecotti:2012jx,Cecotti:2013lda,Xie:2012hs,Wang:2015mra,Nanopoulos:2009uw,Gaiotto:2009hg,Gaiotto:2009ma}. Such a singularity is invariant neither under the $\u1$ part of
the ${\cal N}=2$ R-symmetry nor under the geometric rotation around the puncture. Only a certain combination of the two symmetries is preserved and becomes the $\u1$ R-symmetry of the Argyres--Douglas theory~\cite{Xie:2012hs,Bah:2021mzw,Bah:2021hei,Bah:2022yjf}. In our construction this is implemented by gauging the maximal puncture symmetry. Both the $\u1$ part of the R-symmetry and the geometric isometry are anomalous, with a certain combination surviving. This combination is the $\u1$ R-symmetry of the IR theory. The combination which is preserved depends on the value of the flux (and on the details of the gauging) in our construction, and on the details of the singularity in the usual construction of irregular punctures. We thus provide a dictionary between the two constructions. 
Our construction gives the Argyres--Douglas theories up to a set of $r_\g\,z$ decoupled free fields~\eqref{eq:gaugingclaim}.
All these fields are singlets of the $\su_2$ part of the R-symmetry, and their naive conformal dimension is half of their $\u1$ R-charge. They are not Coulomb branch operators. For example, for $\g=\su_N$ and $p=z\,N+1$ they violate the unitarity bound.   They have the same dimensions, $2-\Delta$, as the couplings of the relevant Coulomb branch operators. As a result, they correspond naturally to the couplings of the associated deformations.

Our construction assumes $z\neq 0$. However, it can be formally extended to $z=0$, where it connects to a known construction of a certain irregular singularity. A tube with $z=0$ is an empty theory: gluing it to a theory does not change the associated geometry, and thus leaves the theory unchanged. This is also the only value of the flux which preserves ${\cal N}=2$ supersymmetry. This theory can be thought of as a sort of ``$\delta$-function'' theory~\cite{Gaiotto:2011xs}. Gauging the $\g$ symmetries of both of its maximal punctures, each with an ${\cal N}=2$ vector multiplet, then simply gives the pure ${\cal N}=2$ $\g$ SYM theory. This theory has a description as a compactification with two particular irregular punctures. For $\g=\su_N$ this puncture is usually denoted by $I_{N,1-N}$~\cite{Xie:2012hs,Xie:2013jc,Wang:2015mra,Cecotti:2013lda,Song:2017oew,Buican:2015ina,Kimura:2020krd}. In this notation, our construction generalizes it to $I_{N,p-N}$ with $p=Nz+1$. An application of this statement, due to Gaiotto~\cite{Gaiotto:2009ma}, is the construction of the so-called Gaiotto--Whittaker state in the framework of the AGT correspondence~\cite{Alday:2009aq}. Similar ideas were used to describe some irregular singularities in the superconformal index~\cite{Song:2015wta,Buican:2015ina,Buican:2015tda,Buican:2017uka}, using its  limits~\cite{Gadde:2011ik,Gadde:2011uv}. It would be very interesting to understand how to introduce a fugacity for the geometric isometry into the TQFT expression~\cite{Gadde:2009kb,Gadde:2011uv} for the index, extending the ${\cal N}=1$ discussion of~\cite{Beem:2012yn}. It would also be interesting to relate the construction of irregular punctures with flux $z$ discussed here to their construction by collision of punctures in the framework of AGT~\cite{Gaiotto:2012sf,Bonelli:2011aa}.

Related to the above, for $\g=\su_N$, our construction can be viewed more generally as follows. Start with an Argyres--Douglas theory $D_p(\su_N)$, which corresponds to a sphere with one maximal and one irregular puncture, and glue a tube of flux $z$ to it. Up to decoupled free fields, the resulting theory is $D_{zN+p}(\su_N)$. For example, $D_1(\su_N)$ is the empty theory, and gluing a tube of flux $z$ to it, {\it i.e.} gauging one of the punctures of the tube, produces $D_{zN+1}(\su_N)$: this is the statement~\eqref{eq:mainclaim}. More generally, for $N=m\,p_0$ with $p_0,\,m\in{\mathbb N}$, the theory $D_{p_0}(\su_N)$ is a superconformal tail of Gaiotto~\cite{Gaiotto:2009we}, namely the linear quiver with gauge groups $\su_{km}$, $k=1,\dots,p_0-1$, and an $\su_N$ flavor symmetry at its end~\cite{Cecotti:2013lda}. Gluing a tube of flux $z$ to it, we obtain the $D_{zN+p_0}(\su_N)$ theory.
We have discussed this explicitly for $\g=\su_2$ (following~\cite{ChoSong:2026}) and $\g=\su_3$, and we expect the statement to hold more generally.  

Another possible extension is the $3d$ reduction of Argyres--Douglas theories. Unlike in $4d$, where these theories lack an ${\cal N}=2$ Lagrangian description, in $3d$ ${\cal N}=4$ Lagrangian descriptions are abundant~\cite{Buican:2015hsa,Benvenuti:2017kud,Agarwal:2018oxb}. For example, the $3d$ ${\cal N}=4$ Lagrangian for $D_p(\su_N)$ is known~\cite{Carta:2021whq,Giacomelli:2020gee}.  For $\gcd(p,N)=1$ and
$p>N$ it is the $T[\su_N]$ theory~\cite{Gaiotto:2008ak} together with $\frac12(N-1)(p-N-1)$ free twisted
hypermultiplets~\cite{Dedushenko:2019yiw,Giacomelli:2020gee}. On the other hand, the circle reduction of the ${\cal N}=1$ tubes with flux for $\g=\su_N$ has a known
dual description in $3d$. It is given by gluing two copies of $T[\su_N]$, with a number of adjoint chiral superfields which depends on $z$~\cite{Razamat:2019sea}. Our construction predicts that gauging, in $3d$, one of the two $\su_N$ symmetries of this quiver gives the theory of~\cite{Dedushenko:2019yiw,Giacomelli:2020gee}. To understand this statement precisely, one might need to employ the recent understanding of {\it bad} quivers developed, {\it e.g.}, in~\cite{Giacomelli:2023zkk,Giacomelli:2024laq,Comi:2025zwu}.

It would also be interesting to relate our results directly to holographic constructions of Argyres--Douglas theories. See~\cite{Aharony:1998xz} for an older construction and~\cite{Bah:2021mzw,Bah:2021hei,Couzens:2022yjl,Bah:2022yjf,Bomans:2023ouw,Couzens:2023kyf,Bomans:2024mrf} for modern approaches. Argyres--Douglas theories can also be engineered from type IIB string theory on canonical threefold singularities~\cite{Shapere:1999xr,Cecotti:2010fi,Cecotti:2012jx,Xie:2015rpa}, see {\it e.g.}~\cite{Closset:2020scj,Closset:2021lwy,Closset:2020afy}. 
Holography has been extremely useful in understanding these strongly coupled theories, and in fact the first computation of the anomalies of some of the Argyres--Douglas theories was holographic~\cite{Aharony:2007dj}.
For example, we have derived in Appendix~\ref{sec:holography}, by performing a holographic computation, the three coefficients in the anomaly polynomial ($\nu_{xy}$, $\nu_{yy}$, and $\gamma_\g$) that could not be fixed by the field-theoretic procedure of cutting a sphere into two caps and a tube. 

The compactification of $(2,0)$ theories to $4d$ is just one example of a large class of constructions of $4d$ SCFTs starting from general $(1,0)$ $6d$ SCFTs. See~\cite{Razamat:2022gpm} for a review. In that context, we can consider the compactification of a $(1,0)$ theory on a sphere with two punctures and flux and then gauge the puncture symmetry. Without taking into account the isometry symmetry, this would in general break the R-symmetry and/or part of the Cartan generators of the $6d$ symmetry group. However, as in this paper, because of the isometry symmetry, all these can be preserved. 
We can thus think of this construction as a new {\it ``irregular''} puncture of the $(1,0)$ compactifications. It would be interesting to understand whether there is a $5d$ picture of such a construction and whether, like the Argyres--Douglas theories with ${\cal N}=2$ supersymmetry, these ${\cal N}=1$ theories have interesting novel properties.

Finally, let us stress that other constructions of, and connections between, class ${\cal S}$ theories with and without irregular punctures, and with or without manifest ${\cal N}=2$ supersymmetry, exist. For the former see, for example,~\cite{Giacomelli:2025reg,Giacomelli:2024ycb,Beem:2020pry}. 
For the latter, many ${\cal N}=1$ Lagrangians are known for a variety of Argyres--Douglas theories~\cite{Maruyoshi:2016tqk,Maruyoshi:2016aim,Agarwal:2016pjo,Agarwal:2017roi,Benvenuti:2017bpg,Maruyoshi:2018nod,Kang:2023dsa,Maruyoshi:2023mnv}, and some geometric understanding of them has been developed~\cite{Giacomelli:2017ckh,Carta:2018qke,Carta:2019hbi}. Some of these constructions overlap with~\cite{ChoSong:2026}, which is both the main motivation for, and a particular example of, our construction. Argyres--Douglas theories have also been used as building blocks of other interesting ${\cal N}=2$ theories, glued together by ${\cal N}=2$ vector multiplets~\cite{Buican:2014hfa,Xie:2017vaf,Kang:2021lic}.

It seems that some of the structures of the ${\cal N}=2$ world become more apparent once we leave the comfort of ${\cal N}=2$ supersymmetry and venture into the wild world of ${\cal N}=1$ supersymmetry. It will be very interesting to explore this further. 

\

\noindent{\bf Acknowledgments:}~We are grateful to Minseok Cho for discussions.   
We would like to thank Sergio Benvenuti and Sara Pasquetti for organizing the workshop {\it Quivers, Symmetries, and SCFTs} in Trieste where this work was initiated. 
 The research of SSR  is supported in part by the Planning and Budgeting Committee, by the Israel Science Foundation under grant no.~2683/26, and by BSF-NSF grant no.~2023769.
The research of JS is supported by the National Research Foundation of Korea (NRF) grants RS-2024-00405629 and RS-2026-25482546, and the KAIST-KIAS collaboration program. The research of TW is supported at the Technion by a Zuckerman Fellowship. The research of KHL is supported by ISF grant no.~759/23.

\

\noindent {\bf AI statement:
}~We have used Claude Code (Anthropic) and  Codex (OpenAI) for discussions, to assist with computations and simplifying formulae, preparing figures, and proofreading the text. Core physics ideas and computations came from, and all the computations were verified by, the authors.

\appendix 

\section{Anomalies of the \texorpdfstring{$\su_N$}{suN} tube from M5-brane inflow}\label{sec:holography}

In Section \ref{sec:g}, we computed the anomalies of the $\mathcal{T}_z^{(\frak{g})}$ theories in order to compare them to the known results for the $D_p(\frak{su}_N)$ theories.
The former we could fix using a purely field-theoretic technique only up to an ambiguity parametrized by three coefficients $\nu_{xy}$, $\nu _{yy},$ and $\gamma_{\frak{g}}$.
We found the anomalies to match
once the undetermined coefficients were fixed to particular values. In this appendix, we offer an alternative derivation of the anomaly coefficients for $\mathcal{T}_z^{(\frak{su}_N)}$ through a holographic realization of the class $\mathcal{S}$ construction in M-theory. Through anomaly inflow on a stack of $N$ M5-branes wrapping a sphere with two maximal punctures, we recover all the relevant anomalies of the 4d theory, and thus fix the three anomaly coefficients. 

We start by reviewing the formalism of anomaly inflow in M-theory, before specializing to the particular geometry with which we are concerned. After collecting the necessary pieces, we perform the explicit computation in Section~\ref{sec:inflow_polynomial}. The goal of this appendix is not to be completely self-contained. While we will review the relevant concepts and constructions, in-depth details and analysis can be found in the cited references.

\subsection{Anomaly inflow from M5-branes}\label{sec:inflow_general}

Our starting point is a stack of $N$ M5-branes with worldvolume denoted $W^6$. When restricting to the worldvolume, the tangent bundle of the ambient space $M^{11}$ decomposes accordingly.
\begin{equation}
    TM^{11}\big|_{W^6} = TW^6 \oplus NW^6
\end{equation}
The two factors, respectively, denote the tangent bundle and normal bundle for the worldvolume of the M5-branes. The latter has an $\frak{so}_5$ structure group coming from rotations in the five directions transverse to the stack. This structure group holographically encodes the data of the $\frak{so}_5$ R-symmetry of the 6d theory living on $W^6$.\footnote{For flat M5-branes, this theory is the 6d $\mathcal{N} = (2,0)$ theory.}

The worldvolume theory of the M5-branes, and by extension any theory obtained from this one through dimensional reduction, can have 't Hooft anomalies for its global symmetries. Since the full 11d M-theory is anomaly free, any anomalies of the lower-dimensional QFTs must be canceled via anomaly inflow from the M-theoretic bulk, as originally discussed in \cite{Freed:1998tg, Harvey:1998bx}. The stack of M5-branes acts as a magnetic source for the $G_4$ flux of M-theory, so that a proper analysis in supergravity requires excising a small tubular neighborhood $B^{11}_\varepsilon$ surrounding the stack. This introduces a boundary $M^{10} \coloneq \partial(M^{11}\setminus B^{11}_\varepsilon)$, causing the M-theory effective action to no longer be invariant under the gauge transformations of the theory (including diffeomorphisms). The classical variation in the action then gives rise, by standard descent procedures, to a formal anomaly 12-form characteristic class.
\begin{equation}
    \begin{gathered}
        \frac{\delta S_{\rm M}}{2\pi} = \int_{M^{10}}\mathcal{I}_{10}^{(1)}\,,\quad d\mathcal{I}^{(1)}_{10} = \delta\mathcal{I}_{11}^{(0)}\,,\quad d\mathcal{I}_{11}^{(0)} = \mathcal{I}_{12}\\
        \mathcal{I}_{12} = -\frac{1}{6}E_4\wedge E_4\wedge E_4 - E_4\wedge X_8
    \end{gathered}
\end{equation}
In the above expression, $E_4$ is the M-theory 4-form flux, but we have turned on gauge fields associated to isometries of the spacetime \cite{Bah:2018gwc, Bah:2019rgq}. That is, we have taken $G_4$ and turned it into a 4-form that is both closed and invariant under the action of our structure group. For reductions to theories of dimension $d<6$, this also accounts for isometries of the compactification surface, as will be relevant for our case of interest. The form $X_8$ is given by
\begin{equation}\label{eq:X8}
    X_8 = \frac{1}{192}\bigg(p_1(TM^{11})^2 - 4p_2(TM^{11})\bigg)\,,
\end{equation}
where $p_i(TM^{11})$ denotes the $i^{\rm th}$ Pontryagin class of the tangent bundle of $M^{11}$, implicitly pulled back to $M^{10}$.

By integrating $\mathcal{I}_{12}$ over the $S^4$ transverse to the M5-brane stack, we obtain the inflow anomaly polynomial for the theory living on $W^6$.
\begin{equation}
    I^{\rm inflow}_8 = \int_{S^4}\mathcal{I}_{12}
\end{equation}
This anomaly polynomial cancels against the anomalies of chiral degrees of freedom living on the M5-branes. There are two contributions to this: those coming from the type $A_{N-1}$ CFT on the stack, and a decoupled 6d $(2,0)$ tensor multiplet understood as the center-of-mass
mode for the $N$ M5-branes. Cancellation of all anomalies thus means the following.
\begin{equation}
    I_8^{\rm inflow} + I_8^{\rm CFT} + I_8^{\rm dec.} = 0
\end{equation}
In the case of the M5-brane worldvolume having compact dimensions, and thus producing a theory on $W^d$ for $d<6$, the same argument yields an inflow anomaly polynomial $I_{d+2}^{\rm inflow}$ that must cancel against the CFT and decoupled modes. We are interested in the case $d=4$, where the M5-brane worldvolume includes a Riemann surface $\Sigma_{g,n}$ with genus $g$ and $n$ punctures. We now turn to describing the geometry of this space and how it will affect the construction of $I_6^{\rm inflow}$.

\subsection{M-theory geometry and isometries}\label{sec:mtheory_geometry}

We follow the analysis of \cite{Bah:2019jts} closely (see also~\cite{Bah:2018jrv}). For the 4d theories of class $\mathcal{S}$, the holographic description comes from a stack of $N$ M5-branes wrapping $W^6 = W^4 \times \Sigma_{g,n}$. We may describe the total $M^{10}$ as a fibration, 
\begin{equation}
    M^6 \hookrightarrow M^{10} \to W^4\,,
\end{equation}
where $M^6$ itself is a non-trivial fibration of the 4-sphere transverse to the stack over the wrapped Riemann surface,
\begin{equation}\label{eq:M6_fibration}
    S^4 \hookrightarrow M^6 \to \Sigma_{g,n}\,.
\end{equation}

The compact space $M^6$ can be decomposed into a ``bulk'' space and small neighborhoods of the individual punctures, which we index by $\alpha$.
\begin{equation}\label{eq:M6_decomp}
    M^6 = M^6_{\rm bulk}\cup \bigcup_{\alpha} X_\alpha^6
\end{equation}
We discuss the exact form of the spaces $X^6_\alpha$ more below. For now, let us emphasize that the decomposition \eqref{eq:M6_decomp} causes the inflow anomaly polynomial to have multiple contributions: one from $M^6_{\rm bulk}$, and one for each puncture $X^6_\alpha$.
\begin{equation}
    \begin{gathered}
        I_6^{\rm inflow} = I_{6,\,\rm bulk}^{\rm inflow} + \sum_\alpha I_{6,\,\alpha}^{\rm inflow}\,,\\
        I_{6,\,\rm bulk}^{\rm inflow} = \int_{M^6_{\rm bulk}}\mathcal{I}_{12}\,,\quad I^{\rm inflow}_{6,\,\alpha} = \int_{X^6_\alpha}\mathcal{I}_{12}\,.
    \end{gathered}
\end{equation}
The total inflow 6-form is given by the sum of each of the constituent contributions.

As a consequence of the decomposition of $W^6 = W^4 \times \Sigma_{g,n}$, the tangent space of the worldvolume manifold similarly decomposes,
\begin{equation}
    TW^6 = TW^4\oplus T\Sigma_{g,n}\,.
\end{equation}
In the fibration \eqref{eq:M6_fibration}, a topological twist is performed that identifies the $\frak{so}_2 \subset \frak{so}_5$ action on $NW^6$ with the $\frak{so}_2^\Sigma$ action on the tangent bundle of the Riemann surface $T\Sigma_{g,n}$. This is done to preserve $\mathcal{N}=2$ supersymmetry in four dimensions, and reduces the structure group of $NW^6$ to a subgroup $\frak{so}_2^\phi\times\frak{so}_3^\Omega \subset \frak{so}_5$. As a result, the normal bundle of $W^6$ also decomposes,
\begin{equation}\label{eq:N6_decomp}
    NW^6 = N_{\frak{so}_2}\oplus N_{\frak{so}_3}\,.
\end{equation}
If the Chern roots of these bundles are $\hat{n}_1$ and $\hat{n}_2$ respectively, the aforementioned topological twist is implemented via\footnote{This is in the convention of \cite{Bah:2019jts}, which differs from \eqref{eq:twist} by a shift of the $\mathcal{F}$ flux and frame.}
\begin{equation}
    \hat{n}_1 = -\hat{t} + \hat{n}^{\rm ext}_1\,,
\end{equation}
where $\hat{n}^{\rm ext}_1$ denotes the component of $\hat{n}_1$ along $W^4$, or the Chern root of $N_{\frak{so}_2}$ in the absence of the twist, and $\hat{t}$ denotes the Chern root of $T\Sigma_{g,n}$.

\paragraph{\texorpdfstring{$M^6_{\rm bulk}$}{M6bulk} Bulk Geometry}

Away from the punctures, the space $M^6_{\rm bulk}$ can be viewed as an $S^4$ fibration over $\Sigma_{g,n}$ with disks around the punctures excised. The decomposition of the normal bundle to the branes \eqref{eq:N6_decomp} motivates viewing the $S^4$ itself as an  $S^1_\phi\times S^2_\Omega$ fibration over an interval parametrized by $\mu \in [0,1]$, described by the following line element.
\begin{equation}\label{eq:S4_metric}
    ds^2(S^4) = \frac{d\mu^2}{1-\mu^2} + (1-\mu^2)d\phi^2 + \mu^2 ds^2(S^2_\Omega)
\end{equation}
The isometries of $S^1_\phi$ and $S^2_\Omega$ respectively correspond to  the $\frak{u}_1$ and $\frak{su}_2$ R-symmetries of the dual 4d $\mathcal{N}=2$ theory. For later use, we denote the Killing vectors associated to these isometries as $\partial_\phi$ and $\partial_\Omega^{(i)}$ respectively.

\paragraph{\texorpdfstring{$X^6_{\alpha}$}{X6} Puncture Geometry}

We are now interested in describing the spaces $X^6_\alpha$, which contain the $S^4$ above as well as the excised disks. Let the excised disks near each puncture be parameterized with coordinates $r \in [0,r_0]$ and $\beta\in [0,2\pi)$. As described in \cite{Gaiotto:2009gz, Bah:2019jts}, the geometry near each puncture can be modeled as a fibration of $S^2_\Omega$ over a multi-center Gibbons--Hawking space with $q$ monopoles dictated by the puncture behavior. Near each monopole, the metric can be written as a Taub-NUT space. For a monopole of charge $k_a$ located at $\eta_a$, the line element is
\begin{equation}
    \begin{gathered}
        ds^2({\rm TN}_{k_a}) = V_a^{-1}D\beta^2 + V_a(d\rho^2 + d\eta^2 + \rho^2d\chi^2)\\
        dD\beta = -\star_{\mathbb{R}^3}dV_a,\quad V_a = \frac{k_a/2}{\sqrt{\rho^2 + (\eta-\eta_{a})^2}}\,,
    \end{gathered}
\end{equation}
where $\rho$, $\eta$, and $\chi$ are defined by\footnote{To account for the $\mathbb{Z}_{k_a}$ orbifold action, the covering angles are defined as $\tilde{\phi} = \phi + \frac{k_a-1}{k_a}\beta$ and $\tilde{\beta} = \frac{1}{k_a}\beta$. Their sum, $\chi$, is $\mathbb{Z}_{k_a}$-invariant.}
\begin{equation}
    \rho = \frac{r^{1/k_a}}{k_a}\sqrt{1-\mu^2}\,,\quad \eta = \eta_{a} + \frac{1}{2k_a}\left(r^{2/k_a} - 1+ \mu^2\right)\,,\quad \chi = \tilde{\phi}+\tilde{\beta} = \phi+\beta\,,
\end{equation}
and $D\beta \coloneq d\beta - L(\rho,\eta) d\chi$. The function $L(\rho,\eta)$ is piecewise continuous along the $\eta$-axis, and has discontinuities only at $\eta_a$ where the monopole charges are located. Its form encodes the profile of the regular puncture modeled by the $q$ monopoles. 

Near each monopole location, where the metric is modeled by that of TN$_{k_a}$, there is an $\mathbb{R}^4/\mathbb{Z}_{k_a}$ orbifold singularity. We can resolve this singularity with $k_a-1$ 2-cycles; let us denote the resolved Taub-NUT space as $\widetilde{\rm TN}_{k_a}$. These 2-cycles have specific intersection numbers between them. In particular, if $\hat{\omega}_2^{a,I}$ denotes the Poincar\'e dual of the $I^{\rm th}$ 2-cycle in $\widetilde{\rm TN}_{k_a}$, the intersection pairing is given by
\begin{equation}
    \int_{\widetilde{\rm TN}_{k_a}}\hat{\omega}_2^{a,I}\wedge\hat{\omega}_2^{b,J} = -\delta^{ab}C^{IJ}_{\frak{su}_{k_a}}\,,
\end{equation}
where $C_{\frak{su}_{k_a}}^{IJ}$ denotes the Cartan matrix of $\frak{su}_{k_a}$. For our case of interest, a maximal puncture, we have only one monopole of charge $N$. The approach for general regular punctures, and how the data of $L(\rho,\eta)$ maps onto the associated Young tableaux, can be found in \cite{Bah:2019jts}. 

\paragraph{Introducing Flux and Gauging the Isometries}

Before computing the inflow anomaly polynomial, there are two additional details we must add to the above discussion: the non-trivial flux $z$ through $\Sigma_{g,n}$, and the gauging of the isometry directions. To assist with both of these, it is useful to recombine the isometry directions mentioned previously into quantities more relevant to the final desired $\mathcal{N} = 1$ SCFT.

As mentioned in the discussion surrounding \eqref{eq:S4_metric}, the isometries generated by $\partial_\phi$ and $\partial_\Omega^{(i)}$ correspond to the $\frak{u}_1$ and $\frak{su}_2$ R-symmetries of the $\mathcal{N}=2$ theory. When viewing this theory as one with only $\mathcal{N}=1$ supersymmetry, we can identify certain linear combinations as generating the $\frak{u}_1$ R-symmetry $R_0$ and $\frak{u}_1$ flavor symmetry $\mathcal{F}$. With a slight abuse of notation, we write these generators as follows.
\begin{equation}\label{eq:isometry_redef}
    \begin{gathered}
        R_0 = \partial_\Omega^{(3)} + \partial_\phi\,,\quad \mathcal{F} = \partial_\Omega^{(3)} - \partial_\phi\\
        \partial_\Omega^{(3)} = \frac{1}{2}(R_0 + \mathcal{F})\,,\quad \partial_\phi = \frac{1}{2}(R_0 - \mathcal{F})
    \end{gathered}
\end{equation}
This gives a relationship between the Chern classes associated to the $\partial_\phi$ and $\partial_\Omega^{(3)}$ isometries and those of the $(\frak{u}_1)_{R_0}$ and $(\frak{u}_1)_\mathcal{F}$ bundles. 
\begin{equation}\label{eq:chern_redef}
    \begin{split}
        c_1(\Omega^{(3)}) = c_1^\Omega &= c_1(R_0) + c_1(\mathcal{F}) = x+y\\
        c_1(\phi) = c_1^\phi &= c_1(R_0) - c_1(\mathcal{F}) = x-y
    \end{split}
\end{equation}
We are now in a position to implement both the flux $z$ and the gauging of the isometries. From here on we specialize our Riemann surface to the case of interest: $\Sigma_{0,2}$ with two maximal punctures.

By adding flux $z$ for $\mathcal{F}$ through $\Sigma_{0,2}$, we are shifting the first Chern class of the $(\frak{u}_1)_{\mathcal{F}}$ bundle by a 2-form $f$ such that
\begin{equation}\label{eq:flavor_flux_shift}
    y \to y - f\,,\qquad \int_{\Sigma_{0,2}}f = z\,,
\end{equation}
where we have used the notation in \eqref{eq:twist} in the main text. This flux, containing only the Cartan of $\frak{su}_2$, breaks the $\mathcal{N}=2$ R-symmetry down to its subgroup $(\frak{u}_1)_{R_0}\times (\frak{u}_1)_\mathcal{F} \subset \frak{u}_1\times\frak{su}_2$, and thus breaks the supersymmetry down to $\mathcal{N}=1$.

As we are evaluating integrals over the tube where neighborhoods of the punctures have been excised, it is useful to have an explicit expression for $f$. We parameterize the tube in terms of $s\in [-1,1]$ along its length and $\beta \in [0,2\pi)$ around it. In terms of these coordinates, an axially symmetric representative for $f$ can be written as
\begin{equation}
    f = d\left(\xi(s) \frac{d\beta}{2\pi}\right)\,.
\end{equation}
Here $\xi(s)$ is some smooth function that takes values $\pm\frac{z}{2}$ at $s=\pm 1$ respectively. With regard to the Chern roots $\hat{n}_1$ and $\hat{n}_2$ of $N_{\frak{so}_2}$ and $N_{\frak{so}_3}$, the relations \eqref{eq:chern_redef} and \eqref{eq:flavor_flux_shift} yield the following twisted Chern roots.
\begin{equation}\label{eq:chern_root_flux}
    n_1 = \hat{n}_1 + d\left(\xi(s)\frac{d\beta}{2\pi}\right)\,,\quad n_2 = \hat{n}_2 - d\left(\xi(s)\frac{d\beta}{2\pi}\right)
\end{equation}

The remaining step is to turn on gauge fields for the isometries of the background geometry, of which there are three. The first two correspond to the isometries already discussed: those generated by $\partial_\Omega^{(3)}$ and $\partial_\phi$. Gauging these amounts to turning on connection 1-forms as follows.
\begin{equation}
    d\phi \to D\phi = d\phi - \xi(s)d\beta + A_1^{\Sigma} - A_1^\phi\,,\quad d\Omega^{(3)} \to D\Omega^{(3)} = d\Omega^{(3)} + \xi(s)d\beta - A_1^\Omega
\end{equation}
The presence of $A_1^\Sigma$ comes from the topological twisting between the actions on $N_{\frak{so}_2}$ and $T\Sigma_{0,2}$ mentioned previously. An explicit expression for $A_1^\Sigma$ can be written similarly to $f$, but with a different smooth function $\xi_\Sigma(s)$ defined such that $\xi_\Sigma(\pm 1) = 0$. That is, $A_1^\Sigma = \xi_\Sigma(s)d\beta$. We have defined $D\phi$ and $D\Omega^{(3)}$ so that
\begin{equation}
    d\left(\frac{D\phi}{2\pi}\right) = -{n}_1\,,\quad d\left(\frac{D\Omega^{(3)}}{2\pi}\right) = -{n}_2\,.
\end{equation}
Near the punctures the $(\phi,\beta)$ coordinates are replaced with $(\chi, \beta)$, but the gauging procedure holds identically. We introduce a connection $A_1^\chi$ so that
\begin{equation}\label{eq:Dchi_def}
    d\chi \to D\chi = d\chi - A_1^\chi\,,\quad d\left(\frac{D\chi}{2\pi}\right) = -{n}_1\,.
\end{equation}
Importantly, the definition of $A_1^\chi$ contains the information of the flux shift previously in $D\phi$, so that indeed the above relation holds. The final isometry comes from that of the tube $\Sigma_{0,2}$ itself, corresponding to rotations in the $\beta$ direction. We denote this isometry by $\mathcal{J}$ in order to connect with the main text. We gauge this isometry through another connection 1-form.
\begin{equation}
    d\beta \to d\beta^g = d\beta - A_1^J
\end{equation}
Crucially, this also affects the Chern roots in the presence of the flux, as can be seen from \eqref{eq:chern_root_flux}. As a result of this, we see that $\mathcal{J}$ should not be understood as a rotation in $S^1_\beta$ alone, but as an isometry of the larger space $M^6$.\footnote{At a fixed fiber angle, $\mathcal{J}|_{\rm fixed} = \partial_\beta$, but generically this isometry is given by a linear combination
\begin{equation*}
    \mathcal{J} = \partial_\beta + a\partial_\phi + b\partial_\Omega^{(3)}\,.
\end{equation*}
The coefficients $a$ and $b$ are non-trivial with non-zero flux. In a symmetric frame, they take values $\pm\frac{z}{2}$ near the punctures, with $a=-b$. From one puncture to the other, they swap signs.} We use a superscript $g$ to denote forms containing the gauged $d\beta^g$. For example,
\begin{equation}
    n_2^g = \hat{n}_2 - d\left(\xi(s)\frac{d\beta^g}{2\pi}\right)\,.
\end{equation}

\subsection{Computing \texorpdfstring{$I_6$}{I6}}\label{sec:inflow_polynomial}

Now that we have all the pieces, we may compute the desired anomaly polynomial. Recall that there are three sources we must account for: the bulk inflow contributions $I_{6,{\rm bulk}}^{\rm inflow}$, the puncture contributions $I_{6,\alpha}^{\rm inflow}$, and the contributions from the decoupled center-of-mass modes $I_6^{\rm dec.}$. We work through these systematically.

\paragraph{Bulk Contributions}

To begin, we compute $-I_{6,{\rm bulk}}^{\rm inflow}$. Away from the location of the punctures, $E_4$ takes the form
\begin{equation}
    E_4 = \mathcal{E}_2^g\wedge e_2^\Omega + \dots\,,
\end{equation}
where $e_2^{\Omega}$ is the global angular form for the $N_{\frak{so}_3}$ bundle, though here we are restricting only to the $\frak{u}_1 \subset \frak{so}_3$ Cartan. The 2-form $\mathcal{E}_2^g$ is closed, gauge-invariant, and can be written as follows.
\begin{equation}
    \mathcal{E}_2^g = Nd\left[\gamma(\mu)\frac{D\phi^g}{2\pi}\right]
\end{equation}
The function $\gamma(\mu)$ is fixed by regularity so that $\gamma(0) = 0$ and $\gamma(1) = 1$.

We first look at the $E_4^3$ term of $\mathcal{I}_{12}$.
\begin{align}\label{eq:bulk_E43}
    \frac{1}{6}\int_{M^6_{\rm bulk}}E_4^3 &= \frac{1}{24}\int_{S^1_\phi\times [\mu]\times S^1_\beta\times [s]}({n}_2^g)^2 \left(\mathcal{E}_2^g\right)^3\nonumber\\
    &= \frac{N^3}{24}\int_{[\mu]}d(\gamma^3)\int_{S^1_\beta\times[s]}\left(c_1^\Omega -d\xi\wedge\frac{d\beta^g}{2\pi} + \xi\epsilon\right)^2\left(c_1^\phi +d\xi\wedge\frac{d\beta^g}{2\pi} - \xi\epsilon\right)^2\nonumber\\
    &=-\frac{N^3}{24}\frac{1}{\epsilon}\int_{-1}^1 ds\, \partial_s\left[(c_1^\Omega + \xi(s)\,\epsilon)^2\wedge(c_1^\phi - \xi(s)\,\epsilon)^2\right]\nonumber\\
    &= \frac{N^3z}{6}\left(x^2y - y^3 - \frac{z^2}{4}y\epsilon^2\right)
\end{align}
In the first line, we have applied the Bott--Cattaneo formula \eqref{eq:BC} to the present setup, with $e_2^\Omega = \frac{1}{2}e$ defined there, and we have used in the final line the values of $\xi(s)$ at the punctures.

To compute the $E_4\wedge X_8$ term, we make use of the decomposition of $TM^{11}\big|_{W^6}$ in this region. From the discussion above,
\begin{equation}
    TM^{11}\big|_{W^6} = TW^4 \oplus T\Sigma_{0,2}\oplus N_{\frak{so}_2}\oplus N_{\frak{so}_3}\,.
\end{equation}
Then, from the general properties of Pontryagin classes, we can decompose $p_i(TM^{11}\big|_{W^6})$ into Pontryagin classes over the constituent bundles.\footnote{For a direct sum bundle $E\oplus F$, the first and second Pontryagin classes obey the following relations.
\begin{equation*}
    p_1(E\oplus F) = p_1(E)+p_1(F)\,,\quad p_2(E\oplus F) = p_2(E) + p_2(F) + p_1(E)p_1(F)
\end{equation*}}

\begin{equation}\label{eq:bulk_pontryagin}
    \begin{split}
        p_1(TM^{11}\big|_{W^6}) &= p_1(TW^4) + (n_1^g)^2 + (n_2^g)^2\\
        p_2(TM^{11}\big|_{W^6}) &= p_2(TW^4) + (n_1^g)^2(n_2^g)^2 + p_1(TW^4)\left((n_1^g)^2 + (n_2^g)^2\right)
    \end{split}
\end{equation}
In writing these, we have omitted terms depending on $\hat{t}$, the Chern root of $T\Sigma_{0,2}$, as its contribution vanishes due to the requirement $\xi_\Sigma(\pm 1) = 0$. Using \eqref{eq:bulk_pontryagin}, the 8-form \eqref{eq:X8} contains the following terms.
\begin{equation}
    X_8 \supset \frac{1}{24}(p_1(TW^4) - 4x^2)\wedge y\wedge d\left(\xi(s)\frac{d\beta^g}{2\pi}\right)
\end{equation}
These are the only terms relevant to the desired integral, due to the form of $E_4$ away from the punctures.
\begin{align}\label{eq:bulk_E4X8}
    \int_{M^6_{\rm bulk}}E_4\wedge X_8 &= N\int_{\Sigma_{0,2}}X_8\nonumber\\
    &= \frac{N}{24}(p_1(TW^4) - 4x^2)y\int_{\Sigma_{0,2}}d\left(\xi(s)\frac{d\beta^g}{2\pi}\right)\nonumber\\
    &= \frac{Nz}{24}(p_1(TW^4)y - 4x^2y)
\end{align}

\paragraph{Puncture Contributions}

Next, we compute the contributions from $-I_{6,\alpha}^{\rm inflow}$. Near the punctures, the expression for $E_4$ contains additional terms corresponding to the local geometry of the Gibbons--Hawking monopoles induced by the puncture structure. For a maximal puncture, we have the following.
\begin{equation}
    E_4 = \mathcal{E}_2^g\wedge e_2^\Omega + \sum_{I=1}^{N-1}\frac{\hat{F}^{I}_2}{2\pi}\wedge (\hat{\omega}^g_2)^{I}
\end{equation}
Here, the forms $(\hat{\omega}_2^g)^{I}$ are the equivariant completion of the forms $\hat{\omega}_2^{I}$ Poincar\'e dual to the 2-cycles in the resolved Taub-NUT space $\widetilde{\rm TN}_N$, and $\mathcal{E}_2^g$ is written in the puncture coordinates as
\begin{equation}
    \mathcal{E}_2^g = d\left(Y\frac{D\chi^g}{2\pi} - W\frac{D\beta^g}{2\pi}\right)\,,
\end{equation}
where $Y(\rho,\eta)$ and $W(\rho,\eta)$ are piecewise continuous functions along the $\eta$-axis, whose explicit properties are detailed in \cite{Bah:2019jts} for general punctures. For our purposes, we need only the following properties.
\begin{equation}
    \begin{gathered}
        W(\rho,0) = Y(\rho,0) = 0\,,\\
        W(0,\eta < \eta_{\rm max}) = 0\,,\quad W(0,\eta > \eta_{\rm max}) = 1\,,\\
        Y(0,\eta < \eta_{\rm max}) = 0\,,\quad Y(0,\eta > \eta_{\rm max}) = N
    \end{gathered}
\end{equation}
The parameter $\eta_{\rm max}$ denotes the location of the maximal puncture viewed as a charge $N$ monopole along the $\eta$-axis.

We can now compute the anomalies from each puncture. We start with the puncture located at $s=-1$ in the Riemann surface $\Sigma_{0,2}$. From the $E_4^3$ component, there are two terms.
\begin{equation}\label{eq:puncture_E43_terms}
    \int_{X^6_\alpha}E_4^3 = \frac{1}{4}(n_2^g)^2\int_{\tilde{X}^4}(\mathcal{E}_2^g)^3 + 3\int_{\tilde{X}^4}\mathcal{E}^g_2\wedge \left(\sum_{I=1}^{N-1}\frac{\hat{F}^I_2}{2\pi}\wedge(\hat{\omega}^g_2)^I\right)^2
\end{equation}
We approach them one at a time. The first term is evaluated as follows.
\begin{align}
    \int_{\tilde{X}^4}(\mathcal{E}^g_2)^3 &= 3\left(x-y+\frac{z}{2}\epsilon\right)\int_{(\rho,\eta)} d[(Y+WL)^2]\wedge dW\nonumber\\
    &\qquad +6\epsilon\int_{(\rho,\eta)}(dY + WdL)\wedge dW\nonumber\\
    &= -2N^2\left(x-y+\frac{z}{2}\epsilon\right) + N\epsilon
\end{align}
In evaluating this integral we have used that we are looking near the puncture at $s = -1$, so that $\xi(s) \to -\frac{z}{2}$ and thus $n_{1,2}^g\to x\mp y\pm\frac{z}{2}\epsilon$. Moving to the second term, we find
\begin{align}
    \int_{\tilde{X}^4}\mathcal{E}_2^g\wedge\left(\sum_{I=1}^{N-1}\frac{\hat{F}^I_2}{2\pi}\wedge(\hat{\omega}^g_2)^I\right)^2&= -\left(x-y + \frac{z}{2}\epsilon\right)\sum_{I,J=1}^{N-1} \frac{\hat{F}^I_2}{2\pi}\wedge\frac{\hat{F}^J_2}{2\pi}\int_{\tilde{X}^4}(Y+WL)\hat{\omega}^I_2\wedge\hat{\omega}_2^J\nonumber\\
    &\qquad +\epsilon\sum_{I,J=1}^{N-1} \frac{\hat{F}^I_2}{2\pi}\wedge\frac{\hat{F}^J_2}{2\pi}\int_{\tilde{X}^4}W\,\hat{\omega}^I_2\wedge\hat{\omega}_2^J\nonumber\\
    &= \left(N(x-y) - \left(1 - \frac{Nz}{2}\right)\epsilon\right)\sum_{I,J=1}^{N-1}C^{IJ}_{\frak{su}_N}\frac{\hat{F}^I_2}{2\pi}\wedge\frac{\hat{F}^J_2}{2\pi}
\end{align}
The sum in the final expression can be equivalently written as the second Chern class of the $\frak{su}_N$ flavor symmetry at the puncture. That is, the sum can be written as $2c_2(\frak{g}_L)$ with our chosen normalization.

By combining these two integrals with the coefficients in \eqref{eq:puncture_E43_terms}, we can find the contribution of the $E_4^3$ component for the maximal puncture located at $s=-1$, or L. For the maximal puncture at $s=+1$, or R, the above computation is identical save the sign of $\epsilon$ wherever it arises. When computing the total puncture contribution to $I_6^{\rm inflow}$, the terms odd in $\epsilon$ thus cancel aside from those involving the puncture flavor symmetries, and we are left with the remaining terms.
\begin{align}\label{eq:puncture_E43}
    \frac{1}{6}\sum_\alpha\int_{X^6_\alpha}E_4^3 &= -\frac{N^2}{6} x^3 - \frac{N^2}{6} x^2 y + \frac{N^2}{6} xy^2 + \frac{N^2}{6} y^3 \\
    &\quad + \frac{Nz(Nz-2)}{24}x\epsilon^2 + \frac{Nz(3Nz - 2)}{24}y\epsilon^2\nonumber\\
    &\quad +\frac{Nz-2}{2}\epsilon(c_2(\frak{g}_{\rm L})-c_2(\frak{g}_{\rm R}))\nonumber\\
    &\quad +N(x-y)(c_2(\frak{g}_{\rm L})+c_2(\frak{g}_{\rm R}))\nonumber
\end{align}

Near the punctures, the tangent bundle of the total eleven-dimensional space has a different decomposition, owing to the topological twist between $N_{\frak{so}_2}$ and $T\Sigma_{0,2}$.
\begin{equation}
    TM^{11} = TW^4 \oplus T\tilde{X}^4\oplus N_{\frak{so}_3}
\end{equation}
The bundle $T\tilde{X}^4$ {\it a priori} has two distinct Chern roots $\hat{\lambda}_1$ and $\hat{\lambda}_2$. However, as $\tilde{X}^4$ is a resolved Taub-NUT space, its first Chern class vanishes and the two Chern roots should be opposite one another $\hat{\lambda}_1 = -\hat{\lambda}_2 \coloneq \hat{\lambda}$. The topological twist and gauging $\mathcal{J}$ introduce shifts to the roots.\footnote{The $\hat{t}^g$ component of $n^g_1$ does not contribute here as $\xi_\Sigma(\pm1) = 0$, hence $n_1^g = -d\left(D\chi^g/2\pi\right)$ as in \eqref{eq:Dchi_def}.}
\begin{equation}
    \lambda_1 = \hat{\lambda} + \frac{1}{2} n_1^g\,,\quad \lambda_2 = -\hat{\lambda} + \frac{1}{2}n_1^g
\end{equation}
Thus, the total Pontryagin classes of $\tilde{X}^4$ are modified to
\begin{equation}
    \begin{split}
        p_1^g(T\tilde{X}^4) &= \lambda_1^2 + \lambda_2^2 = p_1(T\tilde{X}^4) + \frac{1}{2}(n_1^g)^2\\
        p_2^g(T\tilde{X}^4) &= \lambda_1^2\lambda_2^2 = p_2(T\tilde{X}^4) -\frac{1}{4}p_1(T\tilde{X}^4)(n_1^g)^2 + \frac{1}{16}(n_1^g)^4\,,
    \end{split}
\end{equation}
where $p_i^g(T\tilde{X}^4)$ denotes the Pontryagin classes with gauged isometries. The 8-form $X_8$ can then be expanded in terms of the Pontryagin classes and Chern roots of the bundles comprising $TM^{11}$. For our purposes, the relevant terms are collected below.
\begin{equation}
    X_8 \supset -\frac{1}{96}p_1(T\tilde{X}^4)\left(p_1(TW^4) - (n_1^g)^2 + (n_2^g)^2\right)
\end{equation}
With this, we can find the $E_4\wedge X_8$ component of the puncture contributions. Specializing to maximal punctures, we find
\begin{align}
    \int_{X^6_\alpha}E_4\wedge X_8 &= \int_{\tilde{X}^4}\mathcal{E}_2^g\wedge X_8\nonumber\\
    &= \frac{1}{96}\left(p_1(TW^4) - (n_1^g)^2 + (n_2^g)^2\right)\int_{\tilde{X}^4}p_1(T\tilde{X}^4)\left((Y+WL)n_1^g - W\epsilon\right)\nonumber\\
    &= \frac{N}{48}\left(p_1(TW^4) + 4x\left(y-\frac{z}{2}\epsilon\right)\right)
    \left(N(x-y)+\left(\frac{Nz}{2}-1\right)\epsilon\right)
\end{align}
The final step is to sum over the individual punctures, keeping in mind the flip $\epsilon\to -\epsilon$ from $L$ to $R$.
\begin{align}\label{eq:puncture_E4X8}
    \sum_\alpha \int_{X^6_\alpha}E_4\wedge X_8 &= \frac{N^2}{24}p_1(TW^4)(x-y) + \frac{N^2}{6}(x^2 y - xy^2) - \frac{Nz(Nz-2)}{24}x\epsilon^2
\end{align}

\paragraph{Decoupled Modes Contribution}

The final piece needed for the total anomaly polynomial comes from the decoupled center-of-mass modes for the brane stack, computed by reducing $\mathcal{I}_{12}$ over $M^6$ without any punctures, with flux $z$ threading the sphere $\Sigma_{0,0}$, and with $N=1$. Reducing over the $S^4$ directly yields the 6d $(2,0)$ anomaly polynomial of \eqref{eq:I8}, which we rewrite here for $\frak{g} = \frak{u}_1$.
\begin{equation}
    I_8^{\rm dec.} = \frac{1}{48}\left[p_2(NW^6) - p_2(TW^6) + \frac{1}{4}\left(p_1(NW^6) - p_1(TW^6)\right)^2\right]
\end{equation}
As before, we can write this in terms of the Chern roots and Pontryagin classes of the constituent bundles, which yields
\begin{align}
    I_8^{\rm dec.} &= \frac{1}{48}\left[(n_1^g)^2(n_2^g)^2 - (\hat{t}^g)^2p_1(TW^4) +\frac{1}{4}\left((n_1^g)^2+(n_2^g)^2 - p_1(TW^4) - (\hat{t}^g)^2\right)^2\right]
\end{align}
From here, we may isolate the terms saturating the integral over the Riemann surface $\Sigma_{0,0}$, and perform the same trick we did in computing \eqref{eq:bulk_E43} to evaluate the integrand only at $s=\pm 1$. What remains is our desired $I_6^{\rm dec.}$.
\begin{align}\label{eq:decoupled_contributions}
    -\int_{\Sigma_{0,0}}I_8^{\rm dec.} &= -\frac{1}{24}p_1(TW^4)x -\frac{z-1}{24}p_1(TW^4)y\\
    &\quad + \frac{1}{6}x^3 + \frac{z-1}{6}y^3 + \frac{z(z-1)(z-2)}{24}y\,\epsilon^2\nonumber
\end{align}

\paragraph{Combining} With all the contributions computed in this appendix, we may now combine them to obtain the total anomaly polynomial for the 4d theory on $W^4$. Recall that the anomaly polynomial is given by
\begin{equation}
    I_6^{\rm CFT} = -I_{6,\,{\rm bulk}}^{\rm inflow} - \sum_\alpha I_{6,\,\alpha}^{\rm inflow} - I_6^{\rm dec.}\,.
\end{equation}
Hence, we sum all the terms in \eqref{eq:bulk_E43}, \eqref{eq:bulk_E4X8}, \eqref{eq:puncture_E43}, \eqref{eq:puncture_E4X8}, and \eqref{eq:decoupled_contributions} to arrive at the final anomaly polynomial.
\begin{align}
    I_6^{\rm CFT} &= \frac{N^2-1}{24}p_1(TW^4)\,x - \frac{N^2-1-(N-1)z}{24}p_1(TW^4)\,y\\
    &\quad - \frac{N^2-1}{6}x^3 +\frac{N^2-1-(N^3-1)z}{6}y^3 + \frac{N(N^2-1)z}{6}x^2 y\nonumber\\
    &\quad + \frac{z\left(3(N^2-1)z -2(N-1) - (N^3-1)z^2\right)}{24}y\,\epsilon^2\nonumber\\
    &\quad +\frac{Nz-2}{2}\epsilon\left(c_2(\frak{g}_{\rm L})-c_2(\frak{g}_{\rm R})\right)+N(x-y)(c_2(\frak{g}_{\rm L})+c_2(\frak{g}_{\rm R}))\nonumber
\end{align}
From here one may read off all the anomaly coefficients, and thus recover the results of Section \ref{sec:tubeanomalies}. In particular, we find
\begin{equation}
    \begin{gathered}
        \Tr \mathcal{J}^2 R_0 = 0\,,\quad \Tr\mathcal{J} \frak{g}_{\rm R}^2 = -\Tr\mathcal{J}\frak{g}_{\rm L}^2 = \frac{Nz}{4}-\frac{1}{2}\,,\\
        \Tr\mathcal{J}^2F = \frac{z}{12}\left(3(N^2-1)z-(N^3-1)z^2-2(N-1)\right)
    \end{gathered}
\end{equation}
In comparing with the anomalies derived from the field theoretic techniques in \eqref{eq:tubeJbar} and \eqref{eq:gpunctureanomalies} for $\frak{su}_N$, we find the undetermined coefficients to be
\begin{equation}
    \nu_{xy} = 0\,,\quad \nu_{yy} = \frac{N(N^2-1)}{24}\,,\quad \gamma_{\frak{g}} = -\frac{1}{2}\,.
\end{equation}
This agrees directly with the results of the main text for the $\mathcal{T}_z^{(\frak{su}_N)}$ theory.

\section{The \texorpdfstring{$\su_3$ ${\cal N}=4$ SYM and $\mathcal{T}^{(\su_3)}_1$}{su3 N=4 SYM and Tsu31}}\label{sec:N4}

A test of whether the Lagrangian we have found for ${{\cal T}^{(\su_3)}_{z=1}}$ is the correct one is to use it to build a theory whose detailed properties are known independently. In this appendix we discuss such a test.\footnote{
The content of this appendix was developed in an unpublished work by Elli Pomoni and SSR.
}

Consider the compactification of a type $\su_N$ $(2,0)$ theory on a torus with one minimal puncture and flux
$+\frac12$. It can be decomposed into $\mathcal{T}^{(\su_N)}_{z=1}$, a sphere with two positive punctures and flux $+1$, and a three-punctured sphere with 
two negative maximal punctures, one positive minimal puncture, and flux $-\frac12$~\cite{Bah:2012dg}. When the two surfaces are glued along both
maximal punctures with ${\cal N}=1$ vector multiplets, the fluxes add up~\cite{Razamat:2022gpm}, giving the desired surface. On the other hand, one can construct the same surface by taking a three-punctured sphere with two maximal punctures and a minimal one, all positive: this is the free trinion, a bifundamental hypermultiplet of the two puncture $\su_N$ symmetries. Then, gluing the two maximal punctures together with an ${\cal N}=2$ vector multiplet, we obtain explicitly
the ${\cal N}=4$ $\su_N$ SYM plus a decoupled free hypermultiplet~\cite{Gaiotto:2009we}.

Thus we have a dual ${\cal N}=1$ description of the ${\cal N}=4$ $\su_N$ SYM. For $N=2$ this is well known. One can view ${\cal N}=4$ $\su_2$ SYM as an $\frak{so}_3$ gauge theory with three fields in the vector representation and perform the $\frak{so}$ Intriligator--Seiberg duality~\cite{Intriligator:1995id}. The resulting theory is an $\frak{so}_4\cong \su_2\times \su_2$ gauge theory with three bifundamental fields ($\frak{so}_4$ vectors) and gauge singlets. This is precisely the dual discussed above using the two-punctured sphere of Section~\ref{sec:a1}. Here we will thus use the two-punctured sphere of Section~\ref{sec:a2}, and follow a similar path to derive the sphere with two maximal punctures and one negative puncture of $\su_3$ class ${\cal S}$, in order to check the duality in the $\su_3$ case.

\ 

\begin{figure}[t]
\centering
\begin{tikzpicture}[>=Stealth,scale=0.85,transform shape,
 blk/.style={rectangle,draw,very thick,rounded corners=5pt,minimum width=2.6cm,minimum height=1.6cm,align=center},
 g/.style={circle,draw,very thick,minimum size=11mm,fill=blue!6}]
\node[blk,fill=green!8] (T) at (-3.4,0) {${{\cal T}^{(\su_3)}_{z=1}}$};
\node[blk,fill=orange!10] (S) at (3.4,0) {$S^{(\su_3)}_{-\frac12}$};
\node[g] (gC) at (0,1.1) {$\su_3$};
\node[g] (gB) at (0,-1.1) {$\su_3$};
\draw[very thick] (T.north east) -- node[above,font=\footnotesize]{$C$} (gC) -- node[above,font=\footnotesize]{$C$} (S.north west);
\draw[very thick,dashed] (T.south east) -- node[below,font=\footnotesize]{$B$} (gB) -- node[below,font=\footnotesize]{$B$} (S.south west);
\draw[very thick] (S.east) -- ++(0.8,0) node[right]{$\u1$};
\draw[->,very thick] (6.55,0) -- (7.55,0);
\node[right,align=left] at (7.65,0) {${\cal N}=4$ $\su_3$ SYM\\ $+$ free hyper};
\end{tikzpicture}
\caption{The torus with one minimal puncture and flux $\frac12$, built from
${{\cal T}^{(\su_3)}_{z=1}}$ and $S^{(\su_3)}_{-\frac12}$ glued with ${\cal N}=1$ vectors. The dashed line is the gluing
of the emergent punctures.}
\label{fig:N4}
\end{figure}
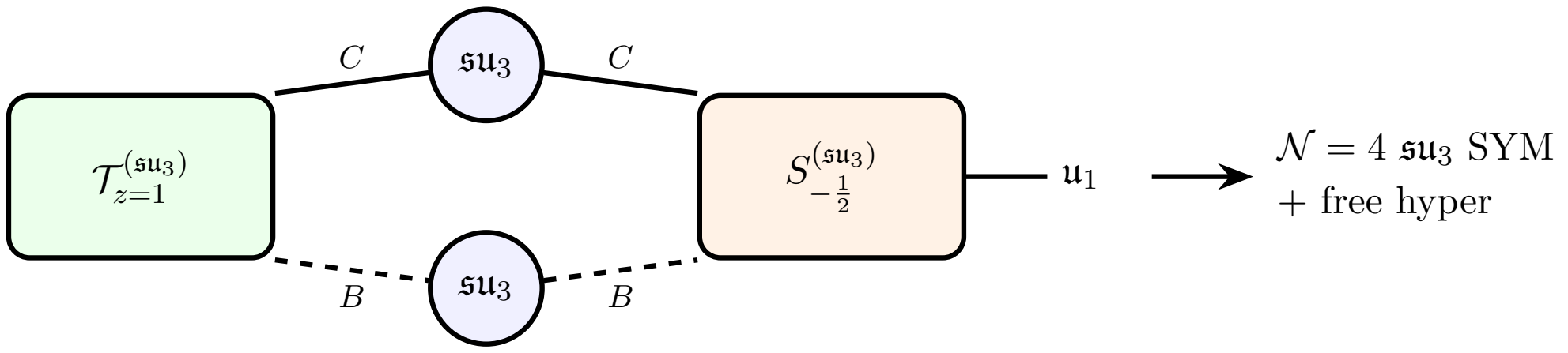

Let us start by deriving the sphere with two maximal punctures and one negative puncture of $\su_3$ class ${\cal S}$. We denote this theory by 
$S^{(\su_3)}_{-\frac12}$. It is obtained exactly like ${{\cal T}^{(\su_3)}_{z=1}}$: we flip the moment map $\mu_A$ of
$T_{\su_3}$, but now give the flip field the \emph{subregular} vacuum expectation
value, of type $[2,1]$~\cite{Agarwal:2015vla}. Closing a maximal puncture without the flip would result in a free
bifundamental hypermultiplet. However, with the flip we obtain an interacting theory. Let us discuss the matter content of the resulting theory. Four singlets survive the vacuum expectation value deformation: $S_1$,
$X_\pm$, and $S_2$, which, as before, pairs up with $Z_3$ to form a massive field. The same Higgsing
$(\su_3)_g\to (\su_2)_L$ occurs. The full matter content is summarized in Table~\ref{tab:tri}. The superpotential is
\begin{equation}
W_{S^{(\su_3)}_{-\frac12}}=A_2Y_m\Phi^m+A_2Y_4\varphi+A_1\tilde Y\tilde\Phi+\chi A_2Z+\bar\phi\det A_1
+X_-Y_4Z+X_+\chi\varphi+S_1\widehat\mu\,.
\end{equation}
Here $\widehat\mu$ is the component of the flipped moment map $\mu_A$ along the $\u1$ left unbroken by the vacuum expectation value. In terms of the fields of
Table~\ref{tab:tri}, it is a linear combination of the gauge-invariant operators with
$(R_0,{\cal F}_{S^{(\su_3)}_{-\frac12}},u,B_1,B_2)=(1,1,0,0,0)$, $\widehat\mu=c\,\bar\phi+\widehat\mu'$, where $\widehat\mu'$ is a combination of $\Phi^mA_1Y_m$, $\varphi A_1Y_4$ and
$\chi A_1Z$. We expect based on charges that the coefficient $c$ does not vanish: $\bar\phi$ is the moment map of $(\u1)_z$~\cite{GarciaEtxebarria:2021deconf}, and the generator of the unbroken $\u1$ has a non-zero $(\u1)_z$ component. Thus $S_1\bar\phi$ is a mass term, and the F-terms of $S_1$ and $\bar\phi$ set $\bar\phi=-\widehat\mu'/c$ and $S_1=-\det A_1/c$. Integrating out this vector-like pair, which does not affect the anomalies, replaces $\bar\phi\det A_1+S_1\widehat\mu$ in the superpotential by $-\frac1c\det A_1\,\widehat\mu'$. The moment map of the $\u1$ symmetry of the minimal puncture is then $\det A_1$, with $(R_0,{\cal F}_{S^{(\su_3)}_{-\frac12}},u)=(1,-1,0)$. As this puncture has the opposite sign to the maximal ones, these are the charges of the flip field $S_1$ rather than those of $\widehat\mu$. For convenience we keep $S_1$ and $\bar\phi$ in Table~\ref{tab:tri}.  

\begin{table}[h]
\centering
\renewcommand{\arraystretch}{1.25}
\begin{tabular}{|c|ccc|cc|c|cc|}
\hline
field & $(\su_2)_L$ & $(\su_2)_R$ & $(\su_3)_C$ & $B_1$ & $B_2$ & $\u1$ & $R_0$ & ${\cal F}_{S^{(\su_3)}_{-\frac12}}$\\
\hline
$A_1$ & ${\bf 2}$ & ${\bf 2}$ & ${\bf 1}$ & $\phantom{-}0$ & $\phantom{-}0$ & $\phantom{-}0$ & $\phantom{-}\frac12$ & $-\frac12$\\
$A_2$ & ${\bf 2}$ & ${\bf 2}$ & ${\bf 1}$ & $\phantom{-}0$ & $\phantom{-}0$ & $\phantom{-}0$ & $\phantom{-}\frac32$ & $-\frac32$\\
$Y_m$ & ${\bf 2}$ & ${\bf 1}$ & $\bar{\bf 3}$ & $\phantom{-}\frac12$ & $\phantom{-}\frac12$ & $-\frac16$ & $\phantom{-}\frac14$ & $\phantom{-}\frac34$\\
$Y_4$ & ${\bf 2}$ & ${\bf 1}$ & ${\bf 1}$ & $-\frac12$ & $\phantom{-}\frac12$ & $\phantom{-}\frac12$ & $\phantom{-}\frac14$ & $\phantom{-}\frac34$\\
$Z$ & ${\bf 2}$ & ${\bf 1}$ & ${\bf 1}$ & $\phantom{-}\frac12$ & $-\frac12$ & $\phantom{-}\frac12$ & $\phantom{-}\frac14$ & $\phantom{-}\frac34$\\
$\tilde Y$ & ${\bf 2}$ & ${\bf 1}$ & ${\bf 1}$ & $-\frac32$ & $-\frac32$ & $-\frac12$ & $\phantom{-}\frac34$ & $\phantom{-}\frac14$\\
$\Phi^m$ & ${\bf 1}$ & ${\bf 2}$ & ${\bf 3}$ & $-\frac12$ & $-\frac12$ & $\phantom{-}\frac16$ & $\phantom{-}\frac14$ & $\phantom{-}\frac34$\\
$\varphi$ & ${\bf 1}$ & ${\bf 2}$ & ${\bf 1}$ & $\phantom{-}\frac12$ & $-\frac12$ & $-\frac12$ & $\phantom{-}\frac14$ & $\phantom{-}\frac34$\\
$\chi$ & ${\bf 1}$ & ${\bf 2}$ & ${\bf 1}$ & $-\frac12$ & $\phantom{-}\frac12$ & $-\frac12$ & $\phantom{-}\frac14$ & $\phantom{-}\frac34$\\
$\tilde\Phi$ & ${\bf 1}$ & ${\bf 2}$ & ${\bf 1}$ & $\phantom{-}\frac32$ & $\phantom{-}\frac32$ & $\phantom{-}\frac12$ & $\phantom{-}\frac34$ & $\phantom{-}\frac14$\\
$\bar\phi$ & ${\bf 1}$ & ${\bf 1}$ & ${\bf 1}$ & $\phantom{-}0$ & $\phantom{-}0$ & $\phantom{-}0$ & $\phantom{-}1$ & $\phantom{-}1$\\
$X_\pm$ & ${\bf 1}$ & ${\bf 1}$ & ${\bf 1}$ & $\phantom{-}0$ & $\phantom{-}0$ & $\pm1$ & $\phantom{-}\frac32$ & $-\frac32$\\
$S_1$ & ${\bf 1}$ & ${\bf 1}$ & ${\bf 1}$ & $\phantom{-}0$ & $\phantom{-}0$ & $\phantom{-}0$ & $\phantom{-}1$ & $-1$\\
\hline
\end{tabular}
\caption{The theory $S^{(\su_3)}_{-\frac12}$ with two maximal punctures ($(\su_3)_C$ manifest,
$(\su_3)_B$ emergent) and one minimal puncture $\u1$.}
\label{tab:tri}
\end{table}

Having derived the two ingredients, we now glue them together into a torus with one puncture. We gauge the diagonal of each pair of maximal
punctures with an ${\cal N}=1$ vector multiplet and add a superpotential ${\rm Tr}\,\mu_{S^{(\su_3)}_{-\frac12}}\mu_{{{\cal T}^{(\su_3)}_{1}}}$ for
each pair of punctures~\cite{Gaiotto:2015usa,Razamat:2016dpl,Gadde:2013fma,Bah:2012dg} (see Figure~\ref{fig:N4}). This preserves $R_0$ and
${\cal F}={\cal F}_{{{\cal T}^{(\su_3)}_{1}}}-{\cal F}_{S^{(\su_3)}_{-\frac12}}$.  Both $\su_3$ gauge anomalies vanish:
${\rm Tr}\,{(\su_3)}^2\,R_0=3-\frac32-\frac32=0$ and
${\rm Tr}\,{(\su_3)}^2\,{\cal F}=\frac32-\frac32=0$. The gluing along puncture $C$ is a
Lagrangian one, while the gluing along $B$ gauges an emergent symmetry. Table~\ref{tab:N4} lists the anomalies of the different components. Every abelian anomaly of the glued theory coincides with that of
${\cal N}=4$ $\su_3$ SYM plus a free hypermultiplet charged under
the minimal puncture $\u1$ symmetry. Maximizing $a$~\cite{Intriligator:2003jj} over
$R_0+\omega{\cal F}$ gives $\omega=\frac13$, the ${\cal N}=2$ value, and
\begin{equation}
a=\frac{49}{24}=2+\frac1{24}\,,\qquad c=\frac{25}{12}=2+\frac1{12}\,,
\end{equation}
which are the conformal anomalies of the ${\cal N}=4$ $\su_3$ SYM plus a free hypermultiplet.\footnote{One can perform the analysis of this duality at the level of anomalies for any $\g=\su_N$.}

\begin{table}[h]
\centering
\renewcommand{\arraystretch}{1.25}
\begin{tabular}{|l|cccccc|}
\hline
 & ${\rm Tr}\,R_0$ & ${\rm Tr}\,R_0^3$ & ${\rm Tr}\,{\cal F}$ & ${\rm Tr}\,{\cal F}^3$ & ${\rm Tr}\,R_0^2\,{\cal F}$ & ${\rm Tr}\,R_0\,{\cal F}^2$\\
\hline
${{\cal T}^{(\su_3)}_{1}}$ & \ooalign{\hphantom{\ooalign{$-1$\cr$-8$\cr$-9$\cr$0$\cr$16$}}\cr\hfil$-8$\cr} & \ooalign{\hphantom{\ooalign{$-8$\cr$-\frac14$\cr$-\frac94$\cr$16$\cr$6$\cr$\frac{23}4$}}\cr\hfil$-8$\cr} & \ooalign{\hphantom{\ooalign{$-5$\cr$0$\cr$1$\cr$6$}}\cr\hfil$6$\cr} & \ooalign{\hphantom{\ooalign{$-18$\cr$-6$\cr$-\frac{23}4$\cr$0$\cr$\frac14$\cr$\frac{49}4$}}\cr\hfil$-18$\cr} & \ooalign{\hphantom{\ooalign{$-\frac{23}4$\cr$0$\cr$2$\cr$8$\cr$\frac14$\cr$\frac94$}}\cr\hfil$8$\cr} & \ooalign{\hphantom{\ooalign{$-2$\cr$-\frac14$\cr$-\frac94$\cr$0$}}\cr\hfil$0$\cr}\\
$S^{(\su_3)}_{-\frac12}$ (with ${\cal F}=-{\cal F}_{S^{(\su_3)}_{-\frac12}}$) & \ooalign{\hphantom{\ooalign{$-1$\cr$-8$\cr$-9$\cr$0$\cr$16$}}\cr\hfil$-9$\cr} & \ooalign{\hphantom{\ooalign{$-8$\cr$-\frac14$\cr$-\frac94$\cr$16$\cr$6$\cr$\frac{23}4$}}\cr\hfil$-\frac94$\cr} & \ooalign{\hphantom{\ooalign{$-5$\cr$0$\cr$1$\cr$6$}}\cr\hfil$-5$\cr} & \ooalign{\hphantom{\ooalign{$-18$\cr$-6$\cr$-\frac{23}4$\cr$0$\cr$\frac14$\cr$\frac{49}4$}}\cr\hfil$\frac{49}4$\cr} & \ooalign{\hphantom{\ooalign{$-\frac{23}4$\cr$0$\cr$2$\cr$8$\cr$\frac14$\cr$\frac94$}}\cr\hfil$-\frac{23}4$\cr} & \ooalign{\hphantom{\ooalign{$-2$\cr$-\frac14$\cr$-\frac94$\cr$0$}}\cr\hfil$-\frac94$\cr}\\
Two ${\cal N}=1$ vectors & \ooalign{\hphantom{\ooalign{$-1$\cr$-8$\cr$-9$\cr$0$\cr$16$}}\cr\hfil$16$\cr} & \ooalign{\hphantom{\ooalign{$-8$\cr$-\frac14$\cr$-\frac94$\cr$16$\cr$6$\cr$\frac{23}4$}}\cr\hfil$16$\cr} & \ooalign{\hphantom{\ooalign{$-5$\cr$0$\cr$1$\cr$6$}}\cr\hfil$0$\cr} & \ooalign{\hphantom{\ooalign{$-18$\cr$-6$\cr$-\frac{23}4$\cr$0$\cr$\frac14$\cr$\frac{49}4$}}\cr\hfil$0$\cr} & \ooalign{\hphantom{\ooalign{$-\frac{23}4$\cr$0$\cr$2$\cr$8$\cr$\frac14$\cr$\frac94$}}\cr\hfil$0$\cr} & \ooalign{\hphantom{\ooalign{$-2$\cr$-\frac14$\cr$-\frac94$\cr$0$}}\cr\hfil$0$\cr}\\
\hline
Total & \ooalign{\hphantom{\ooalign{$-1$\cr$-8$\cr$-9$\cr$0$\cr$16$}}\cr\hfil$-1$\cr} & \ooalign{\hphantom{\ooalign{$-8$\cr$-\frac14$\cr$-\frac94$\cr$16$\cr$6$\cr$\frac{23}4$}}\cr\hfil$\frac{23}4$\cr} & \ooalign{\hphantom{\ooalign{$-5$\cr$0$\cr$1$\cr$6$}}\cr\hfil$1$\cr} & \ooalign{\hphantom{\ooalign{$-18$\cr$-6$\cr$-\frac{23}4$\cr$0$\cr$\frac14$\cr$\frac{49}4$}}\cr\hfil$-\frac{23}4$\cr} & \ooalign{\hphantom{\ooalign{$-\frac{23}4$\cr$0$\cr$2$\cr$8$\cr$\frac14$\cr$\frac94$}}\cr\hfil$\frac94$\cr} & \ooalign{\hphantom{\ooalign{$-2$\cr$-\frac14$\cr$-\frac94$\cr$0$}}\cr\hfil$-\frac94$\cr}\\
\hline
${\cal N}=4$ $\su_3$ SYM & \ooalign{\hphantom{\ooalign{$-1$\cr$-8$\cr$-9$\cr$0$\cr$16$}}\cr\hfil$0$\cr} & \ooalign{\hphantom{\ooalign{$-8$\cr$-\frac14$\cr$-\frac94$\cr$16$\cr$6$\cr$\frac{23}4$}}\cr\hfil$6$\cr} & \ooalign{\hphantom{\ooalign{$-5$\cr$0$\cr$1$\cr$6$}}\cr\hfil$0$\cr} & \ooalign{\hphantom{\ooalign{$-18$\cr$-6$\cr$-\frac{23}4$\cr$0$\cr$\frac14$\cr$\frac{49}4$}}\cr\hfil$-6$\cr} & \ooalign{\hphantom{\ooalign{$-\frac{23}4$\cr$0$\cr$2$\cr$8$\cr$\frac14$\cr$\frac94$}}\cr\hfil$2$\cr} & \ooalign{\hphantom{\ooalign{$-2$\cr$-\frac14$\cr$-\frac94$\cr$0$}}\cr\hfil$-2$\cr}\\
Free hyper & \ooalign{\hphantom{\ooalign{$-1$\cr$-8$\cr$-9$\cr$0$\cr$16$}}\cr\hfil$-1$\cr} & \ooalign{\hphantom{\ooalign{$-8$\cr$-\frac14$\cr$-\frac94$\cr$16$\cr$6$\cr$\frac{23}4$}}\cr\hfil$-\frac14$\cr} & \ooalign{\hphantom{\ooalign{$-5$\cr$0$\cr$1$\cr$6$}}\cr\hfil$1$\cr} & \ooalign{\hphantom{\ooalign{$-18$\cr$-6$\cr$-\frac{23}4$\cr$0$\cr$\frac14$\cr$\frac{49}4$}}\cr\hfil$\frac14$\cr} & \ooalign{\hphantom{\ooalign{$-\frac{23}4$\cr$0$\cr$2$\cr$8$\cr$\frac14$\cr$\frac94$}}\cr\hfil$\frac14$\cr} & \ooalign{\hphantom{\ooalign{$-2$\cr$-\frac14$\cr$-\frac94$\cr$0$}}\cr\hfil$-\frac14$\cr}\\
\hline
\end{tabular}
\caption{The abelian anomalies of the glued theory of Figure~\ref{fig:N4}
compared with ${\cal N}=4$ $\su_3$ SYM plus a free hypermultiplet. Note that the geometric isometry symmetry of the two-punctured sphere ${\cal J}$ becomes anomalous due to the gauging of the maximal puncture symmetries. }
\label{tab:N4}
\end{table}

    \bibliographystyle{jhep}

	\bibliography{refs}

\end{document}